\documentclass[a4paper]{article}
\usepackage[margin=2.5cm]{geometry}
\usepackage{iftex}
\ifPDFTeX
  \usepackage[T1]{fontenc}
  \usepackage[utf8]{inputenc}
  \DeclareUnicodeCharacter{03A3}{\ensuremath{\Sigma}}
  \DeclareUnicodeCharacter{0393}{\ensuremath{\Gamma}}
  \DeclareUnicodeCharacter{03C3}{\ensuremath{\sigma}}
  \DeclareUnicodeCharacter{03C1}{\ensuremath{\rho}}
  \DeclareUnicodeCharacter{03B5}{\ensuremath{\varepsilon}}
  \DeclareUnicodeCharacter{03BB}{\ensuremath{\lambda}}
  \DeclareUnicodeCharacter{03C7}{\ensuremath{\chi}}
  \DeclareUnicodeCharacter{03BE}{\ensuremath{\xi}}
  \DeclareUnicodeCharacter{2020}{\ensuremath{\dagger}}
  \DeclareUnicodeCharacter{00F7}{\ensuremath{\div}}
  \DeclareUnicodeCharacter{2297}{\ensuremath{\otimes}}
  \DeclareUnicodeCharacter{2191}{\ensuremath{\uparrow}}
  \DeclareUnicodeCharacter{2193}{\ensuremath{\downarrow}}
  \DeclareUnicodeCharacter{1DBB}{\ensuremath{^z}}
  \DeclareUnicodeCharacter{1D62}{\ensuremath{_i}}
  \DeclareUnicodeCharacter{207B}{\ensuremath{^-}}
  \DeclareUnicodeCharacter{207A}{\ensuremath{^+}}
\else
  \usepackage{newunicodechar}
  \newunicodechar{Σ}{\ensuremath{\Sigma}}
  \newunicodechar{Γ}{\ensuremath{\Gamma}}
  \newunicodechar{σ}{\ensuremath{\sigma}}
  \newunicodechar{ρ}{\ensuremath{\rho}}
  \newunicodechar{ε}{\ensuremath{\varepsilon}}
  \newunicodechar{λ}{\ensuremath{\lambda}}
  \newunicodechar{χ}{\ensuremath{\chi}}
  \newunicodechar{ξ}{\ensuremath{\xi}}
  \newunicodechar{†}{\ensuremath{\dagger}}
  \newunicodechar{÷}{\ensuremath{\div}}
  \newunicodechar{⊗}{\ensuremath{\otimes}}
  \newunicodechar{↑}{\ensuremath{\uparrow}}
  \newunicodechar{↓}{\ensuremath{\downarrow}}
  \newunicodechar{ᶻ}{\ensuremath{^z}}
  \newunicodechar{ᵢ}{\ensuremath{_i}}
  \newunicodechar{⁻}{\ensuremath{^-}}
\newunicodechar{⁺}{\ensuremath{^+}}
\fi
\usepackage{amsmath}
\usepackage{amssymb}
\usepackage{bbm}
\usepackage[
    colorlinks=true,
    linkcolor=blue,
    citecolor=blue,
    urlcolor=blue
  ]{hyperref}
\usepackage{orcidlink}
\usepackage{graphicx}
\graphicspath{{../}{./}}
\usepackage{xcolor}
\usepackage{authblk}
\usepackage[
	backend=biber, citestyle=numeric-comp,
	sorting=none, sortcase=false, sortcites=true,
	giveninits=true,
	minnames=10,
	maxnames=10,
	date=year
]{biblatex}

\DeclareSourcemap{
  \maps[datatype=bibtex]{
    \map{
      \step[fieldsource=doi, match=\regexp{\A10\.48550/arXiv\.}, final]
      \step[fieldset=doi, null]
    }
    \map{
      \step[fieldsource=urldate, final]
      \step[fieldset=urldate, null]
    }
    \map{
      \step[fieldsource=journal, match={Physical Review Letters}, replace={Phys. Rev. Lett.}]
      \step[fieldsource=journal, match={Physical Review A}, replace={Phys. Rev. A}]
      \step[fieldsource=journal, match={Physical Review B}, replace={Phys. Rev. B}]
      \step[fieldsource=journal, match={Physical Review E}, replace={Phys. Rev. E}]
      \step[fieldsource=journal, match={Physical Review X}, replace={Phys. Rev. X}]
      \step[fieldsource=journal, match={Physical Review Research}, replace={Phys. Rev. Res.}]
      \step[fieldsource=journal, match={Reviews of Modern Physics}, replace={Rev. Mod. Phys.}]
      \step[fieldsource=journal, match={Annual Review of Condensed Matter Physics}, replace={Annu. Rev. Condens. Matter Phys.}]
      \step[fieldsource=journal, match={Journal of Statistical Mechanics: Theory and Experiment}, replace={J. Stat. Mech. Theory Exp.}]
      \step[fieldsource=journal, match={Journal of Physics A: Mathematical and Theoretical}, replace={J. Phys. A Math. Theor.}]
      \step[fieldsource=journal, match={SciPost Physics Lecture Notes}, replace={SciPost Phys. Lect.}]
      \step[fieldsource=journal, match={SciPost Physics}, replace={SciPost Phys.}]
      \step[fieldsource=journal, match={SciPost Physics Codebases}, replace={SciPost Phys. Codebases}]
      \step[fieldsource=journal, match={SIAM Journal on Scientific Computing}, replace={SIAM J. Sci. Comput.}]
      \step[fieldsource=journal, match={Annals of Physics}, replace={Ann. Phys.}]
      \step[fieldsource=journal, match={Linear Algebra and Its Applications}, replace={Linear Algebra Appl.}]
      \step[fieldsource=journal, match={Reports on Mathematical Physics}, replace={Rep. Math. Phys.}]
      \step[fieldsource=journal, match={Nature Reviews Physics}, replace={Nat. Rev. Phys.}]
      \step[fieldsource=journal, match={Communications in Mathematical Physics}, replace={Commun. Math. Phys.}]
      \step[fieldsource=journal, match={New Journal of Physics}, replace={New J. Phys.}]
      \step[fieldsource=journal, match={Journal of Physics A: Mathematical and General}, replace={J. Phys. A Math. Gen.}]
      \step[fieldsource=journal, match={Journal of the Physical Society of Japan}, replace={J. Phys. Soc. Jpn.}]
      \step[fieldsource=journal, match={Journal of Open Source Software}, replace={J. Open Source Softw.}]
      \step[fieldsource=journal, match={Annual Review of Physical Chemistry}, replace={Annu. Rev. Phys. Chem.}]
      \step[fieldsource=journal, match={International Journal of Quantum Chemistry}, replace={Int. J. Quantum Chem.}]
      \step[fieldsource=journal, match={Materials Research Bulletin}, replace={Mater. Res. Bull.}]
      \step[fieldsource=journal, match={Progress of Theoretical Physics}, replace={Prog. Theor. Phys.}]
      \step[fieldsource=journal, match={Nature Communications}, replace={Nat. Commun.}]
    }
  }
}

\hypersetup{
  pdftitle={Introduction to matrix-product states and tensor networks},
  pdfauthor={Grégoire Misguich},
}

\usepackage[frozencache,cachedir=_minted-notes]{minted}

\usepackage[most]{tcolorbox}
\usepackage[compatibility=false]{caption}
\usepackage{xstring}
\usepackage{environ}
\tcbset{
  exercisebox/.style={
    enhanced,
    breakable,
    colback=black!1,
    colframe=black!35,
    boxrule=0.5pt,
    arc=1mm,
    left=6pt,
    right=6pt,
    top=6pt,
    bottom=6pt,
    colbacktitle=black!5,
    coltitle=black,
    fonttitle=\bfseries,
  },
  codebox/.style={
    enhanced,
    breakable,
    colback=black!2,
    colframe=black!35,
    boxrule=0.4pt,
    arc=1mm,
    left=4pt,
    right=4pt,
    top=4pt,
    bottom=4pt,
  },
  codecaptionbox/.style={
    enhanced,
    breakable,
    boxsep=0pt,
    colback=black!1,
    colframe=black!35,
    boxrule=0.4pt,
    arc=1.5mm,
    left=6pt,
    right=6pt,
    top=6pt,
    bottom=6pt,
    colbacktitle=black!5,
    coltitle=black,
    fonttitle=\bfseries,
    toptitle=2pt,
    bottomtitle=2pt,
  },
  remarkbox/.style={
    enhanced,
    breakable,
    colback=black!1,
    colframe=black!35,
    boxrule=0.5pt,
    arc=1.5mm,
    left=6pt,
    right=6pt,
    top=6pt,
    bottom=6pt,
    colbacktitle=black!5,
    coltitle=black,
    fonttitle=\bfseries,
    title=Remark,
  },
}
\newenvironment{remarkblock}[1][Remark]{%
  \begin{tcolorbox}[remarkbox,title={#1}]
}{%
  \end{tcolorbox}
}
\newsavebox{\longlistingcaptionbox}
\newif\iflonglistingcaptionboxopen
\newcommand{\closelonglistingcaptionbox}{%
  \iflonglistingcaptionboxopen
    \end{minipage}%
    \end{lrbox}%
    \endgroup
    \par\medskip
    \global\longlistingcaptionboxopenfalse
  \fi
}
\newcommand{\longlistingsourceonlinemarker}{\ifjcodeincludesource\else\href{https://github.com/gregoire-misguich/Introduction-to-matrix-product-states-and-tensor-networks/blob/main/code_examples/\jcodefile}{\emph{(source online)}} \fi}
\newcommand{\jcodeboxtitle}{}
\newenvironment{longlisting}{%
  \captionsetup{type=listing,labelfont=bf,name=Code example}%
  \let\longlistingorigcaption\caption
  \RenewDocumentCommand{\caption}{o m}{%
    \ifjcodeincludesource
      \closelonglistingcaptionbox
    \fi
    \IfNoValueTF{##1}{%
      \ifjcodeincludesource
        \longlistingorigcaption{\longlistingsourceonlinemarker ##2}%
      \else
        {\captionsetup{labelformat=empty}\longlistingorigcaption{\longlistingsourceonlinemarker ##2}}%
      \fi
      \gdef\jcodeboxtitle{Code example~\thelisting: ##2}%
    }{%
      \ifjcodeincludesource
        \longlistingorigcaption[##1]{\longlistingsourceonlinemarker ##2}%
      \else
        {\captionsetup{labelformat=empty}\longlistingorigcaption[##1]{\longlistingsourceonlinemarker ##2}}%
      \fi
      \gdef\jcodeboxtitle{Code example~\thelisting: ##1}%
    }%
    \ifjcodeincludesource
      \jcodeinputsource{\jcodefile}%
    \fi
  }%
  \global\longlistingcaptionboxopenfalse
  \par\medskip\noindent\begingroup
  \setlength{\fboxsep}{6pt}%
  \begin{lrbox}{\longlistingcaptionbox}%
  \begin{minipage}{\dimexpr\linewidth-2\fboxsep-2\fboxrule\relax}%
  \global\longlistingcaptionboxopentrue
}{%
  \iflonglistingcaptionboxopen
    \end{minipage}%
    \end{lrbox}%
    \begin{tcolorbox}[codecaptionbox,title={\jcodeboxtitle}]
    \usebox{\longlistingcaptionbox}%
    \end{tcolorbox}%
    \endgroup
    \par\medskip
    \global\longlistingcaptionboxopenfalse
  \fi
}

\AtEveryBibitem{
  \clearfield{issn}
  \clearfield{language}
  \ifentrytype{article}{
    \clearfield{publisher}
    \clearfield{note}
  }{}
}

\DeclareFieldFormat[article,inbook,incollection,inproceedings,book,thesis,report,misc]
  {title}{\mkbibemph{#1}}

\DefineBibliographyStrings{english}{
    in    = {},
    page  = {},
    pages = {}
 }

\usepackage[sort&compress, english]{cleveref}
\crefname{figure}{Fig.}{Figs.}     
\Crefname{figure}{Figure}{Figures} 
\crefname{equation}{Eq.}{Eqs.}     
\Crefname{equation}{Equation}{Equations} 
\crefname{section}{Sec.}{Secs.}     
\Crefname{section}{Section}{Sections} 
\Crefname{appsec}{Appendix}{Appendices}
\crefname{appsec}{App.}{App.}   
\crefname{listing}{code example}{code examples}
\Crefname{listing}{Code example}{Code examples}
\creflabelformat{listing}{#2\##1#3}
\newif\ifincludeexercises
\includeexercisesfalse

\DeclareCaptionType{exercise}[Exercise][List of exercises]
\crefname{exercise}{exercise}{exercises}
\Crefname{exercise}{Exercise}{Exercises}

\NewEnviron{exerciseblock}[1]{%
  \ifincludeexercises
    \refstepcounter{exercise}%
    \addcontentsline{loexercise}{exercise}{\protect\numberline{\theexercise}{#1}}%
    \begin{tcolorbox}[exercisebox,title={Exercise~\theexercise: #1}]
    \BODY
    \end{tcolorbox}
  \fi
}

\newcommand{\ket}[1]{\left|#1\right\rangle}
\newcommand{\bra}[1]{\left\langle#1\right|}
\newcommand{\braket}[2]{\left\langle#1\middle|#2\right\rangle}

\newcommand{\jcodefile}{}
\newcommand{\jcodefiletext}{}
\newcommand{\codeexamplespath}{}
\IfFileExists{../code_examples/mps.jl}{%
  \renewcommand{\codeexamplespath}{../code_examples/}%
}{%
  \renewcommand{\codeexamplespath}{code_examples/}%
}
\newread\jcodelinereader
\newcount\jcodelinecount
\newif\ifjcodeincludesource
\newif\ifjcodefilefound
\newcommand{\jcodesetinclusion}[1]{%
  \global\jcodeincludesourcefalse
  \global\jcodefilefoundfalse
  \begingroup
    \catcode`\\=12
    \catcode`\{=12
    \catcode`\}=12
    \catcode`\#=12
    \catcode`\%=12
    \catcode`\_=12
    \catcode`\$=12
    \catcode`\&=12
    \catcode`\^=12
    \catcode`\~=12
    \global\jcodelinecount=0
    \openin\jcodelinereader=\codeexamplespath#1\relax
    \ifeof\jcodelinereader
    \else
      \global\jcodefilefoundtrue
      \loop\unless\ifeof\jcodelinereader
        \read\jcodelinereader to \jcodeline
        \global\advance\jcodelinecount by 1
      \repeat
    \fi
    \closein\jcodelinereader
  \endgroup
  \ifjcodefilefound
    \ifnum\jcodelinecount<40
      \global\jcodeincludesourcetrue
    \fi
  \fi
}
\newcommand{\jcodeinputsource}[2][]{%
  \begin{tcolorbox}[codebox,title={\jcodeboxtitle}]
  \IfEndWith{\jcodefile}{.py}{%
    \inputminted[linenos,breaklines,fontsize=\scriptsize,#1]{python}{\codeexamplespath#2}%
  }{%
    \inputminted[linenos,breaklines,fontsize=\scriptsize,#1]{julia}{\codeexamplespath#2}%
  }%
  \end{tcolorbox}%
}
\newcommand{\jcode}[1]{%
  \gdef\jcodefile{#1}%
  \gdef\jcodefiletext{\detokenize{#1}}%
  \jcodesetinclusion{#1}%
}
\DeclareRobustCommand{\jcodegithublink}{%
  \par\smallskip\noindent{\footnotesize\hfill \href{https://github.com/gregoire-misguich/Introduction-to-matrix-product-states-and-tensor-networks/blob/main/code_examples/\jcodefile}{\texttt{\jcodefiletext}} on GitHub.}}

\begin{document}

\title{Introduction to matrix-product states and tensor networks}

\author[1]{Grégoire Misguich\orcidlink{0000-0002-7012-3204}}

\affil[1]{%
  Université Paris-Saclay, CEA, CNRS, Institut de Physique Théorique
  \protect\\
  91191 Gif-sur-Yvette, France
  \protect\\
  \href{mailto:gregoire.misguich@ipht.fr}{\texttt{\small gregoire.misguich@ipht.fr}}
}

\maketitle

\begin{abstract}
These notes provide an introduction to tensor-network methods in quantum many-body physics, with an emphasis on matrix-product states (MPS). They develop the basic tensor-network language, including graphical notation, virtual indices, bond dimensions, gauge freedom, canonical forms, QR and singular-value decompositions, and the role of entanglement in controlling the efficiency of the representation. The main MPS algorithms are then introduced, including contractions, correlation functions, matrix-product operators, DMRG, and time-evolution methods. The notes also briefly discuss projected entangled-pair states (PEPS) as a higher-dimensional generalization of MPS, together with the basic ideas behind approximate PEPS contraction. Finally, tensor-network representations of mixed states, quantum channels, and Lindblad dynamics are presented, with applications to thermal states and open quantum systems. The presentation is accompanied by short Julia code examples based on \texttt{ITensor}, \texttt{ITensorMPS}, and \texttt{TensorMixedStates}. These notes were written for the \textit{9$^{\rm th}$ Les Houches Summer School on Computational Physics: Open Quantum Systems}, held in June 2026.
\end{abstract}

\tableofcontents
\listoflistings

\section{Introduction}
\label{sec:introduction}


These notes provide a beginner-friendly introduction to tensor-network (TN) methods in quantum many-body physics, with an emphasis on matrix-product states (MPS) and their applications to lattice systems, including the simulation of open quantum systems.

A substantial part of the notes is devoted to MPS. The reason is that, among TNs, MPS are the best understood, the easiest to manipulate numerically, and the most widely used, with a history spanning roughly three decades. They also provide the simplest setting in which to introduce central notions of the TN language: virtual indices and bond dimensions, tensor contractions, reshaping and factorization, gauge freedom, truncations, matrix-product operators, and variational algorithms such as the density-matrix renormalization group (DMRG) and time-evolution methods. Many of these ideas have natural generalizations to more complex TNs, such as projected entangled-pair states (PEPS), which are briefly discussed.

The presentation is intended to be concrete and practical, and is accompanied by several code examples~\cite{gm_github} illustrating the main ideas. These examples use the Julia libraries \texttt{ITensor} and \texttt{ITensorMPS}~\cite{FISHMAN_ITensorSoftwareLibrary_2022}. Some examples involving dissipative or mixed-state systems use the \texttt{TensorMixedStates} library~\cite{HOUDAYER_TensorMixedStatesJuliaLibrary_2026}, which is built on top of \texttt{ITensor}.

To make full use of the code examples, the reader is expected to have some familiarity with models such as spin chains or simple free fermion Hamiltonians. Very basic notions from quantum information theory---in particular bipartite entanglement, reduced density matrices, Schmidt decomposition, and entanglement entropy---will be used throughout the notes, with only brief reminders.

The field of TN has become quite interdisciplinary and the associated literature is vast.
The applications of TNs extend well beyond lattice quantum many-body problems. These include
numerous problems in classical statistical mechanics, quantum computing, the simulation of quantum circuits, and the use of tensor cross interpolation in a variety of non-quantum settings. Although these notes focus on lattice quantum many-body physics, we have included a few references that may serve as entry points into other areas of application of TN.

\subsection{What are TNs useful for?}
\label{sec:tensor_networks_usefulness}

TNs have profoundly changed the way we think about many-body quantum physics, from condensed matter physics to quantum computing. Instead of viewing a many-body state as a vector in an exponentially large Hilbert space, they offer structured objects whose essential content is the structure of the entanglement between spatial regions. They provide a language to describe and manipulate such states, and a framework for designing algorithms.

While an (exponentially large) vector state representation can by definition encode any state, TN representations are well suited to encoding states with particular patterns of entanglement and some underlying low-rank structure.
They brought a major computational breakthrough: by encoding only the relevant entanglement, they make it possible to represent and manipulate several types of states that are relevant to physics (in particular low-energy states of local Hamiltonians and those obeying area-law entanglement) with resources that grow in a controlled way, rather than exponentially with system size.

At the same time, tensor-network methods are not a universal remedy for the exponential complexity of quantum many-body physics. Highly entangled states (those obeying volume-law entanglement in particular), long-time dynamics, critical systems, and generic contractions in dimensions higher than one are examples of situations in which the computational cost can still grow prohibitively large.

TN applications extend beyond quantum many-body physics and quantum information, and are growing rapidly across physics and related fields. Some of these applications will be mentioned below, although these notes focus on TN applications to (lattice) quantum many-body physics.

\subsubsection{Compressed representation of quantum many-body states}
\label{sec:compressed_many_body_states}

\paragraph{Compression, entanglement, and bond dimension}

A central application of TNs is to provide ``compressed'' representations of quantum many-body states.
Here, compression means that TNs can be used to represent (exactly or approximately) certain states using a number of parameters that is much smaller than the dimension of the full Hilbert space. This turns out to be particularly useful in the context of numerical simulations of quantum many-body systems.
A crucial quantity in this context is the bipartite entanglement entropy of the state, which quantifies the amount of quantum correlations between two subsystems.\footnote{A common way to quantify such bipartite entanglement is to compute the von Neumann entropy of one of the subsystems.
  The von Neumann entropy of a subsystem $A$ is defined as $S_{\rm vN} = -\mathrm{Tr}(\rho_A \log \rho_A)$ where $\rho_A$ is the reduced density matrix of the subsystem $A$.
  $\rho_A = \mathrm{Tr}_B |\psi\rangle\langle\psi|$ is obtained by tracing out the degrees of freedom of the complementary subsystem $B$ from the pure state $|\psi\rangle$ of the full system.
  The von Neumann entropy is a measure of the entanglement between the subsystem $A$ and its complement. It is zero for product states (no entanglement) and can be as large as $\log d_A$ for maximally entangled states, where $d_A$ is the dimension of the Hilbert space of subsystem $A$.
} Generally, the more entangled a state is, the more parameters (larger tensors and more memory) are needed to represent it as a TN.
The number of parameters in a TN can be varied by changing the so-called bond dimensions of the TN. The associated indices are called virtual or auxiliary indices. TNs can be viewed as wave-function ansatzes with tunable entanglement properties. By increasing the bond dimension of the TN, one can increase the amount of entanglement that can be represented and thus improve the accuracy of the approximation of a given state.
There exist many approaches to tackle quantum many-body problems, such as exact diagonalization, quantum Monte Carlo, series expansions, mean-field decoupling, variational wave functions, etc. They all have their own advantages and limitations, and they are often complementary to each other. From this perspective, TNs are quite different from the approaches mentioned above since they are well suited to representing low-entangled states, and in particular states with an area law of entanglement.

In practice, such approaches are efficient for ground states or low-energy states of local Hamiltonians in low dimensions (typically 1d and 2d). States obeying the area law of entanglement~\cite{eisertColloquiumAreaLaws2010} can be efficiently represented as TNs, while states with volume-law entanglement (such as generic highly excited states) can generally not be well represented by TNs unless the bond dimension is allowed to grow exponentially with the system size (in which case the TN representation is much less useful).

\paragraph{Physical systems beyond spin chains}

TNs can be used to encode states of lattice spin models, but fermionic systems can also be treated.
In 1d, the Jordan-Wigner transformation allows one to treat fermionic models essentially as spin models. In higher dimensions
it is also possible to study interacting fermions using TNs, but a different technique (not discussed in these notes) is needed to maintain the locality of the Hamiltonian~\cite{corbozSimulationStronglyCorrelated2010,mortierFermionicTensorNetwork2025}.

DMRG and MPS methods have also become useful tools in quantum chemistry in order to approximate electronic wave functions of molecules. In this setting the sites of the MPS are not lattice sites in real space, but molecular orbitals. The bond dimension then controls the amount of correlation between different groups of orbitals~\cite{chanDensityMatrixRenormalization2011,szalayTensorProductMethods2015}.

\paragraph{Infinite and continuous systems}

The system often has a finite size, but it is also possible to work directly in the thermodynamic limit (infinite system size) using infinite TNs~\cite{vidalClassicalSimulationInfiniteSize2007,mccullochInfiniteSizeDensity2008,jordanClassicalSimulationInfiniteSize2008a}.
Many algorithms have been developed to work directly on an infinite system,
by considering an infinite TN with a repeated unit cell.
For simplicity, however, these notes will focus on finite TNs and on algorithms formulated for finite systems.

MPS can also be extended to continuous models and quantum field theories~\cite{verstraeteContinuousMatrixProduct2010}.
The fractional quantum Hall effect is an example of a domain where problems are defined in the continuum and where MPS have been used successfully~\cite{feiguinDensityMatrixRenormalization2008,zaletelExactMatrixProduct2012,zaletelTopologicalCharacterizationFractional2013}.
There, Landau level orbitals provide a natural way to formulate the Hamiltonians in a discrete basis well suited to an MPS formulation, in spherical or cylindrical geometries.

\paragraph{Main tensor-network ansatzes}

MPS~\cite{OSTLUND_ThermodynamicLimitDensity_1995}\footnote{In the mathematical literature MPS are also known as tensor-train decompositions~\cite{OseledetsTensorTrainDecomposition2011}.} are a particular class of TNs that are well suited to represent low-entangled states of 1d quantum systems. They can sometimes also be used in 2d~\cite{stoudenmireStudyingTwoDimensionalSystems2012}. Among all TNs, MPS are the best understood, the easiest to manipulate numerically, and the most developed both from the theoretical and numerical points of view.
MPS are now the standard framework for performing numerical simulations of 1d quantum systems.

Another class of TNs is PEPS~\cite{NISHINO_TwoDimensionalTensorProduct_2001,VERSTRAETE_RenormalizationAlgorithmsQuantumMany_2004}, which are a natural generalization of MPS to higher dimensions. The associated states can support area-law entanglement in $d>1$ (unlike MPS, which can only support area-law entanglement in 1d), and they have mostly been used in $d=2$.

\begin{figure}
  \begin{center}
    \includegraphics[width=0.88\textwidth]{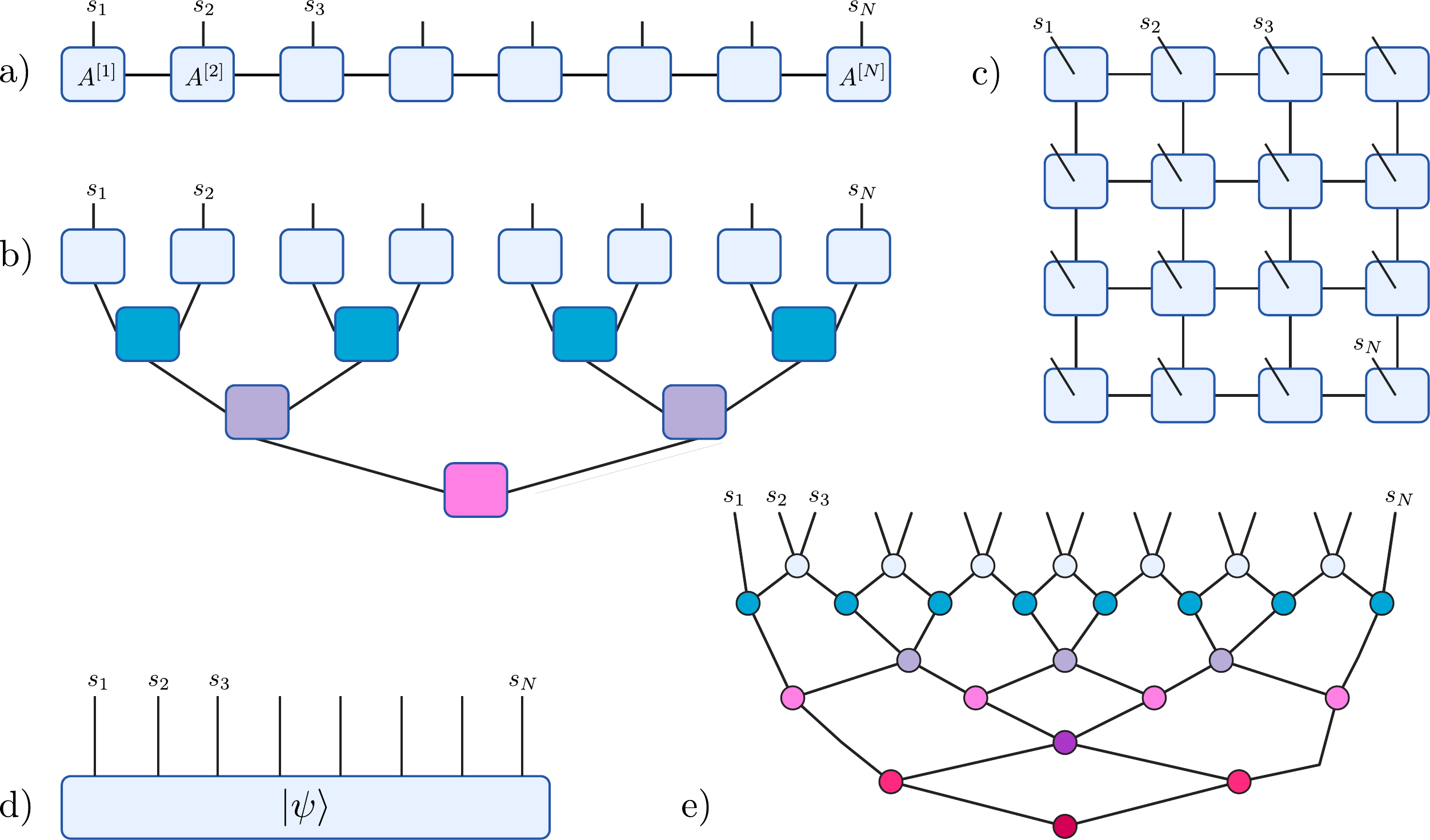}
  \end{center}
  \caption{Examples of TNs: a) MPS for a chain of length 8, b) binary tree TN for a system of 8 sites, c) projected entangled pair state for a system with $4\times4$ sites in a square lattice geometry, d) many-body wave function represented as a single $N$-leg tensor and e) multi-scale entanglement renormalization ansatz (MERA) TN.
  Each box (or circle) represents a tensor, and each leg represents an index of the tensor. In this representation the open legs (labeled $s_1,\dots,s_N$) that are not connected to any other tensor represent the physical indices. They correspond to the degrees of freedom of the physical system (e.g., spin states, where for a spin-$S$ system such indices can take $2S+1$ values). The legs that connect two tensors represent the virtual indices, which encode the entanglement between different parts of the system. The dimensions of the virtual indices, called the bond dimensions, determine the amount of entanglement that can be represented. Larger bond dimensions allow for a more accurate representation of entangled states.
  }
  \label{fig:MPS_TTN_PEPS_MERA}
\end{figure}

Another important TN ansatz is the multi-scale entanglement renormalization ansatz (MERA)~\cite{VIDAL_EntanglementRenormalization_2007,VIDAL_ClassQuantumManyBody_2008,EVENBLY_AlgorithmsEntanglementRenormalization_2009}. It
has a hierarchical structure and is built from layers of isometries
(tensors with three legs in the example of \cref{fig:MPS_TTN_PEPS_MERA}e) and (unitary) disentanglers
(tensors with four legs in the example of \cref{fig:MPS_TTN_PEPS_MERA}e). Such networks, which can be used in 1d but also in $d>1$, are particularly well suited to describing scale-invariant or critical systems.
In a MERA, some tensors have physical indices (top layer of \cref{fig:MPS_TTN_PEPS_MERA}e),
while others only have virtual indices. 
The further away a tensor is from the physical layer, the larger the length scale of the correlations it encodes. 
This structure gives MERA a causal-cone property: local observables and correlation functions can be efficiently evaluated by contracting only the tensors that influence the corresponding physical sites.
Thanks to this scale-by-scale organization, a MERA with finite bond dimension can represent states with logarithmic violations of the area law of entanglement, which is the case for critical systems in 1d~\cite{holzheyGeometricRenormalizedEntropy1994a,vidalEntanglementQuantumCritical2003a,calabreseEntanglementEntropyQuantum2004}.
This structure also has connections with holography and the AdS/CFT correspondence~\cite{swingleEntanglementRenormalization2012}. In a MERA, the additional network direction can be interpreted as a renormalization scale, which is reminiscent of the radial direction in holographic descriptions of conformal field theories.
Concerning entanglement entropy, counting the bonds cut by a minimal curve in a MERA resembles the Ryu--Takayanagi formula~\cite{ryuHolographicDerivation2006}, where the entanglement entropy of a boundary region is related to the area of a minimal surface in the bulk.

\subsubsection{Some algorithms}
\paragraph{Ground-state search}
\label{sec:ground_state_search}

Several algorithms exist to find the ground state (or low-energy excited states) of Hamiltonians as TNs. This is, for instance, the idea of the Density Matrix Renormalization Group (DMRG) algorithm~\cite{whiteDensityMatrixFormulation1992,WHITE_DensitymatrixAlgorithmsQuantum_1993,SCHOLLWOCK_DensitymatrixRenormalizationGroup_2011}, which constructs (approximate) ground states in the space of MPS.
This algorithm will be described in more detail in \cref{sec:dmrg}.

Once a state is expressed as an MPS, it is possible (and often easy) to {\em efficiently} compute expectation values of observables, correlation functions, entanglement properties, and more. Here, efficiently means that the computational cost scales polynomially with the system size and with the TN bond dimensions, avoiding operations that would be {\em exponential} in $N$.\footnote{Obtaining expectation values of observables in a state represented by a TN such as a PEPS is in general more complex and can usually not be carried out exactly in polynomial time. Approximate methods are then used to compute such expectation values.
See \cref{sec:peps_contraction} for more details.}

\paragraph{Time evolution}
\label{sec:time_evolution_applications}

Computing time evolution is also an important domain where TNs are very useful.
A typical application is to compute the time evolution $\ket{\psi(t)} = e^{-iHt} \ket{\psi(0)}$ of an initial state $\ket{\psi(0)}$ under the action of a Hamiltonian $H$. It is also possible to study problems where the Hamiltonian is time-dependent.
Examples include the
tDMRG algorithm~\cite{WHITE_RealTimeEvolutionUsing_2004,DALEY_TimedependentDensitymatrixRenormalizationgroup_2004,GOBERT_RealtimeDynamicsSpin12_2005},
TEBD~\cite{vidalEfficientClassicalSimulation2003,VIDAL_EfficientSimulationOneDimensional_2004},
the time-dependent variational principle (TDVP) for MPS~\cite{haegemanTimeDependentVariationalPrinciple2011,haegemanUnifyingTimeEvolution2016}, and the W$^{\rm I}$ and W$^{\rm II}$ algorithms~\cite{zaletelTimeevolvingMatrixProduct2015}. Such dynamics often leads to entanglement-entropy growth. The bond dimension of the TN representation of the state will then need to grow as well (to maintain the precision of the approximation). This can lead to a difficulty known as the {\em entropy barrier}, which limits the time up to which the time evolution can be accurately computed with TN methods.
A very similar difficulty arises when simulating quantum circuits with TNs, where the circuit depth plays a role similar to time in the Hamiltonian evolution case~\cite{zhouWhatLimitsSimulation2020,ayral_density-matrix_2023}.
Finding methods that are able to overcome this entropy barrier is currently an active area of research in the field of TNs and quantum many-body dynamics~\cite{CEREZO-ROQUEBRUN_SpatiotemporalTensornetworkApproaches_2025,CARIGNANO_OvercomingEntanglementBarrier_2025}.

\paragraph{Thermal states, mixed states and open quantum systems}
\label{sec:thermal_mixed_open_systems}
TNs can also be used to represent thermal states of quantum many-body systems, such as the density matrix $\rho = e^{-\beta H}/Z$, where $H$ is the Hamiltonian of the system and $Z = \mathrm{Tr}(e^{-\beta H})$ is the partition function. It is also possible to simulate the Markovian dynamics of open quantum systems
(\cref{sec:lindblad_MPS_form}). In both cases, this can be done by extending the TN descriptions to {\em mixed states}~\cite{VERSTRAETE_MatrixProductDensity_2004,ZWOLAK_MixedStateDynamicsOneDimensional_2004}. In the case of MPS, this will be discussed in more detail in \cref{sec:mixed_states_open_systems}.
We also mention that TNs can be used to study the {\em non-Markovian} dynamics of open quantum systems, using so-called {\em process tensors}~\cite{keelingProcessTensorApproaches2025}.

\paragraph{Simulating quantum circuits}
\label{sec:simulating_quantum_circuits}

The simulation of quantum circuits is another important field of application of TNs (and of MPS in particular).

\begin{figure}
  \begin{center}
    \includegraphics[width=0.645\textwidth]{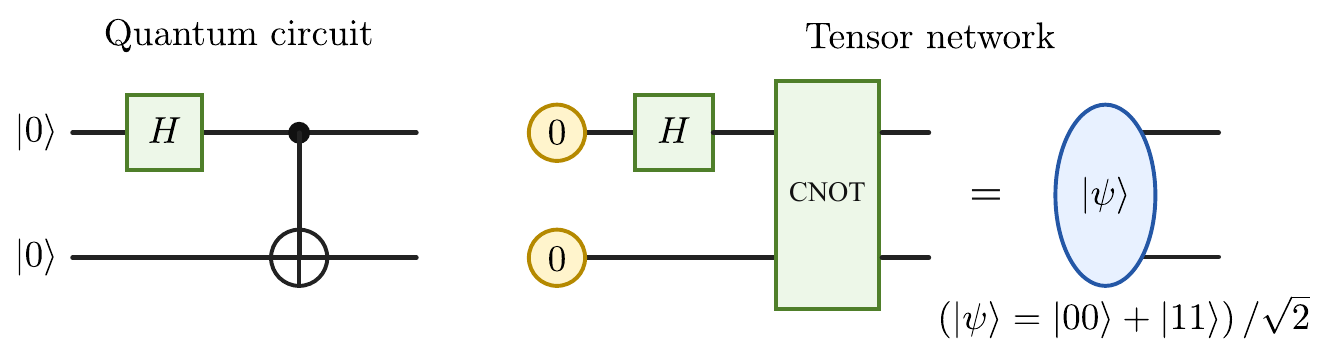}
    \caption[Bell state as a TN]{The Bell state $|\psi\rangle=\frac{1}{\sqrt{2}}(|00\rangle+|11\rangle)$ viewed as a two-index tensor, as the output of a quantum circuit, and as the corresponding TN. $H$ is the Hadamard gate (matrix) $H = \frac{1}{\sqrt{2}}\begin{pmatrix}1 & 1 \\ 1 & -1\end{pmatrix}$ and 
    ${\rm CNOT} = \ket{0}\bra{0}\otimes I + \ket{1}\bra{1}\otimes X$, where $X$ is the Pauli $X$ matrix.}
    \label{fig:bell_circuit_tensor_network}
  \end{center}
\end{figure}

First, a quantum circuit is nothing but a special type of TN.
In such a representation, each qubit wire carries an index. The initial state is often a product state, so that the state of each qubit can be represented by a one-index tensor.
For instance, if all qubits are initialized in the state \( |0\rangle \), one attaches a copy of the vector \( |0\rangle \) to each input leg.
The unitary gates appearing in the circuit are then represented as tensors: a one-qubit gate is a rank-2 tensor, a two-qubit gate is a rank-4 tensor, and so on.
After all internal indices have been contracted, the TN associated with the circuit evaluates a transition amplitude.
More precisely, for an output bitstring \(x_1x_2\dots x_N\), with \(x_i\in\{0,1\}\), one attaches the corresponding dual vectors \(\langle x_i|\) to the output legs.
The full contraction then gives $\langle x_1x_2\dots x_N|\,U_{\rm circuit}\,|0\dots 0\rangle$. An example is shown in \cref{fig:bell_circuit_tensor_network}.
This basic observation also makes clear why TN contraction is closely related to the classical simulation of quantum circuits and quantum computers.

The effect of noise and decoherence can be modeled by non-unitary quantum channels, and the output of such a circuit is no longer a pure state, but a mixed state described by a density matrix $\rho$. The simulation of such noisy quantum circuits can be done using TNs and MPS in particular (\cref{sec:mixed_states_open_systems}). Since noise generally limits the growth of entanglement and correlations, it is possible to simulate larger systems than in the ideal (unitary) case~\cite{zhouWhatLimitsSimulation2020,ayral_density-matrix_2023}. More broadly, in the field of quantum computing, TN calculations have raised the performance bar that quantum processors must clear in order to offer a quantum advantage~\cite{panSimulationQuantumCircuits2022,tindallEfficientTensorNetwork2024a}.

\subsubsection{Classifying quantum phases of matter}
\label{sec:classifying_quantum_phases}

Beyond the computational aspects mentioned above, TNs and MPS in particular have also been used as mathematical tools to classify phases of matter. MPS have in particular been used, in 1d, to prove the absence of intrinsic topological order and to classify gapped symmetry-protected phases~\cite{CHEN_ClassificationGappedSymmetric_2011,CHEN_CompleteClassificationOnedimensional_2011,schuchClassifyingQuantumPhases2011}.
A central concept here is the fact that, under some injectivity condition (see the definition in \cref{sec:injective_MPS}), an MPS that is invariant under a global onsite symmetry transformation carries a projective representation of the symmetry group on its virtual indices. More concretely, let us denote by \(u_g\) the onsite action of a global symmetry element \(g\)
in the group $G$.\footnote{For instance, in the case $G=U(1)$ of spin rotation symmetry about the $z$ axis,
and if $g$ is a rotation of angle $\phi$, then $u_g=e^{i\phi S^z}$.} 
The action of $g$ is then $A^s \mapsto \sum_{s'} (u_g)_{s s'} A^{s'}$.
For injective MPS the fundamental theorem of MPS states
that two MPS describe the same state if and only if their tensors are related by a gauge transformation on the virtual indices.
So, if the MPS is injective and if it is invariant under $g$,
the transformed tensor must be related to the original one by a gauge transformation.
Then for any $g\in G$ there exists an angle $\theta_g$ and
an invertible matrix \(V_g\) acting on the virtual indices such that
\[
  \sum_{s'} (u_g)_{s s'} A^{s'} =
  e^{i\theta_g} V_g^{-1} A^s V_g.
\]
The equation above is interpreted as the fact that the
action of $g$ on the physical space can be ``pushed'' to the virtual space. 
Identifying the successive actions of two symmetries $g$ and $h$ on the MPS with the action of $gh$, one finds that there exist phase factors \(\omega(g,h)\) such that
the matrices \(V_g\) satisfy \(V_g V_h = \omega(g,h) V_{gh}\). 
The phase factors \(\omega(g,h)\) depend on the phase convention chosen for the matrices \(V_g\), and therefore do not have a physical meaning by themselves. What is meaningful, however, is whether these phases can be removed by a redefinition \(V_g\to \beta(g)V_g\). If this is possible, the symmetry action is equivalent to an ordinary representation; if not, the MPS carries a genuinely {\em projective} representation. This distinction is what underlies the classification of one-dimensional symmetry-protected topological phases in the MPS formalism.

In the absence of symmetry, it has been proven that all injective MPS (\cref{sec:injective_MPS}) are smoothly connected\footnote{Two states are smoothly connected if there exists a continuous path of Hamiltonians such that i) the initial and final states are the ground states of the initial and final Hamiltonians, respectively and ii) the gap of the Hamiltonian does not close along the path.
An equivalent definition is that two states are smoothly connected if one can obtain one from the other by applying a finite-depth local unitary circuit, i.e., a quantum circuit with a finite number of layers of local unitary gates.}
to a product state, and they therefore belong to the topologically trivial phase. However, when global symmetries are imposed, gapped states without spontaneous symmetry breaking form distinct phases. These are symmetry-protected topological phases, classified in the MPS formalism by the projective representation carried by the virtual indices of the MPS tensors. 



\subsubsection{Classical statistical mechanics}
\label{sec:classical_statistical_mechanics}

In the field of nonequilibrium classical statistical physics, a notable achievement is the formulation of the steady state of a totally asymmetric exclusion process (TASEP) as an MPS~\cite{DERRIDA_ExactSolution1D_1993}. This seminal work opened the way to the use of MPS in the study of the dynamics of 1d classical stochastic models~\cite{MALLICK_ExactResultsExclusion_2011}.

When computing the partition function of a 2d classical model, it can be useful to consider the associated transfer matrix, which is a linear operator that describes how the system evolves from one row to the next. The DMRG algorithm has been used as early as 1995~\cite{NISHINO_DensityMatrixRenormalization_1995} to compute the dominant eigenvector of such transfer matrices, which gives access to the thermodynamic properties of the system.
Along this line, another important development is the use of iTEBD to study the partition function of infinite classical models~\cite{ORUS_InfiniteTimeevolvingBlock_2008}.
As for the 2d TN known as PEPS in the context of quantum many-body physics, they were originally introduced in the context of 3d classical models~\cite{NISHINO_TwoDimensionalTensorProduct_2001}.

\subsubsection{Tensor cross interpolation}
\label{sec:tensor_cross_interpolation}

Tensor Cross Interpolation (TCI)~\cite{oseledetsTTcrossApproximationMultidimensional2010,NUNEZFERNANDEZ_LearningTensorNetworks_2025} is a learning and compression algorithm that allows one to represent approximately a tensor with $N$ indices (or a function of $N$ variables on a discrete grid) as an MPS. Assuming that each index can take $d$ values, the full tensor has $d^N$ components. The TCI algorithm allows one to find an interpolation of the tensor in the form of an MPS with bond dimension $\chi$, using only $\mathcal{O}(N d \chi^2)$ evaluations of the tensor or the function.
Once the function has been put in MPS form, the TN toolbox can be exploited to perform various operations in
a time that scales polynomially with $N$ and $\chi$.
TCI has a growing number of applications, in particular when combined with the representation of multivariate functions in quantics form\footnote{As an illustration, consider a function $f$ of a single real variable $x$ defined on the interval $[0,1]$. The interval is discretized into $2^N$ points, so that each point $x_{\bf s}$ can be labeled by a binary string ${\bf s}=(s_1,s_2,\dots,s_N)$, where $s_i\in\{0,1\}$ and $x_{\bf s}=\sum_{i=1}^N s_i 2^{-i}$ in the interval $[0,1[$.
On the discretized set, $f$ becomes a function of $N$ binary variables (like the wave function of a spin chain), which can be represented as an MPS with $N$ physical indices. This is the so-called quantics representation of $f$.
Increasing $N$ amounts to increasing (exponentially) the number of discretization points, but the MPS representation of $f$ can be kept compact if $f$ has some underlying low-rank structure.}~\cite{RITTER_QuanticsTensorCross_2024}.
We can for instance mention fast numerical differentiation, integration, convolution, Fourier transforms, and solving PDEs. Ref.~\cite{waintalWhoCanCompete2026} provides a recent introduction to TCI and its applications. 

\subsection{Reviews and lecture notes}
\label{sec:reviews_and_lecture_notes}

There are many reviews and lecture notes on TNs and MPS in particular. We mention here a few of them, but this list is certainly far from being exhaustive.
General beginner-friendly reviews and lecture notes on TNs include for instance (chronological order):
Orus~\cite{ORUS_PracticalIntroductionTensor_2014},
Bridgeman {\it et al.}~\cite{BRIDGEMAN_HandwavingInterpretiveDance_2017},
Silvi {\it et al.}~\cite{SILVI_TensorNetworksAnthology_2019},
Orus~\cite{ORUS_TensorNetworksComplex_2019},
Cirac {\it et al.}~\cite{CIRAC_MatrixProductStates_2021},
Ba\~nuls~\cite{BANULS_TensorNetworkAlgorithms_2023},
Vancraeynest-De Cuiper {\it et al.}~\cite{CUIPER_HouchesLectureNotes_2025}.
For beginners, the reviews by Orus~\cite{ORUS_PracticalIntroductionTensor_2014}
and Ba\~nuls~\cite{BANULS_TensorNetworkAlgorithms_2023} are particularly useful starting points.
Concerning MPS in particular, the review from Schollw{\"o}ck~\cite{SCHOLLWOCK_DensitymatrixRenormalizationGroup_2011} is one of the main references on MPS and DMRG. 
We also wish to mention the review on MPS methods for time evolution by Paeckel {\it et al.}~\cite{PAECKEL_TimeevolutionMethodsMatrixproduct_2019}.
Finally, we recommend the very comprehensive lecture notes by J. von Delft \cite{Delft_TN_course_2025}.

\subsection{Software and libraries}
\label{sec:software_libraries}

Many open-source libraries for doing simulations with TNs and MPS are available, providing low-level routines for basic tensor operations to high-level functions that implement algorithms such as DMRG, time evolution, and more. Here we just mention three of the most widely used ones:
\begin{itemize}
  \item \texttt{ITensor} is in Julia and is one of the most widely used libraries for working with TNs and MPS in particular. See \cite{FISHMAN_ITensorSoftwareLibrary_2022}. Most of the examples in these notes are based on this library and its MPS extension \texttt{ITensorMPS}.
  \item \texttt{Quimb} is a Python library for quantum information calculation that is focused on TNs~\cite{gray2018quimb}.
  \item \texttt{TenPy} is in Python and is another popular library~\cite{HAUSCHILD_EfficientNumericalSimulations_2018,tenpy}.
\end{itemize}

We also point to \texttt{tensornetwork.org}, which is a website with many TN resources, applications, and a longer list of software libraries \cite{TensorNetwork}. \texttt{tensors.net} is another website with resources on TNs, including tutorials~\cite{tensor.net}.

Finally, \texttt{TensorMixedStates} is a recent Julia library built on top of \texttt{ITensor} for working with MPS representations of mixed states \cite{HOUDAYER_TensorMixedStatesJuliaLibrary_2026} and is particularly well suited to simulating open quantum systems. A few code examples using it are included in these notes; see \cref{lst:WII_onsite_field_exact,lst:lindblad_XX_spin_chain,lst:noisy_random_circuit_xeb}.

\section{MPS and associated algorithms}
\label{sec:MPS}

We now turn to MPS, which are
one of the simplest yet most powerful classes of TNs. They are particularly well suited to represent low-entangled states of 1d quantum systems, and they have been widely used for three decades in the field of quantum many-body physics. They also provide the simplest setting in which to introduce several central notions of the TN language. 

We will first define MPS and present simple examples, before turning to the gauge freedom of the representation and to canonical forms. These canonical forms make contractions, expectation values, truncations, and entanglement calculations efficient. We then introduce MPOs, which are the operator analogue of MPS and can be used to represent Hamiltonians, density matrices, and other operators
in a way that is particularly well suited to the MPS representation of states.
Finally, we describe DMRG and time-evolution algorithms.

\subsection{Definition}
\label{sec:MPS_definition}

Consider $N$ sites, each site having a local Hilbert space of dimension $d$. An MPS of bond dimension $\chi$ is a state of the form
\begin{equation}
  |\psi\rangle = \sum_{s_1,s_2,\dots,s_N} A^{[1]\,s_1} A^{[2]\,s_2} \cdots A^{[N]\,s_N} |s_1 s_2 \dots s_N\rangle
  \label{eq:mps}
\end{equation}
where $A^{[i]\,s_i}$ are $\chi\times \chi$ matrices, except for the first and last sites, where they are $1\times \chi$ and $\chi\times 1$, respectively (this ensures that the product of the $A^{[i]\,s_i}$ matrices evaluates to a scalar). The parameter $\chi$ is called the bond dimension of the MPS.

It is sometimes also useful to consider {\em periodic} MPS, where the first and last tensors
share a common index and are also $\chi\times \chi$ matrices. The state is thus obtained by taking the trace of the product of the $N$ matrices:
\begin{equation}
  |\psi\rangle = \sum_{s_1,s_2,\dots,s_N} \mathrm{Tr}(A^{[1]\,s_1} A^{[2]\,s_2} \cdots A^{[N]\,s_N}) |s_1 s_2 \dots s_N\rangle.
  \label{eq:periodic_mps}
\end{equation}
A periodic MPS has the topology of a ring, and the corresponding tensor network therefore contains a {\em loop}. This difference from open-boundary MPS has an important consequence: canonical forms are more subtle to construct and use for periodic MPS (see \cref{sec:mixed_canonical_form}). As a result, periodic MPS are generally more difficult and more expensive to manipulate than open-boundary MPS, although efficient algorithms exist~\cite{pippanEfficientMatrixproductState2010}. For this reason, most numerical studies use open-boundary MPS, even when the physical problem is translationally invariant.

Another notation amounts to absorbing the physical index into the entries of a matrix. This defines a matrix $\mathcal{A}^{[i]}$ whose entries are single-site basis states
of the type $\ket{s_i}$, rather than complex numbers. More precisely:
\begin{equation}
  \mathcal{A}^{[i]}_{\alpha_{i-1},\alpha_i}
  =
  \sum_{s_i} A^{[i]\,s_i}_{\alpha_{i-1},\alpha_i}\ket{s_i}.
  \label{eq:matrix_of_kets}
\end{equation}
For open boundary conditions, the MPS is recovered as
\begin{equation}
  \ket{\psi}
  =
  \sum_{\alpha_1,\ldots,\alpha_{N-1}}
  \mathcal{A}^{[1]}_{1,\alpha_1}
  \otimes
  \mathcal{A}^{[2]}_{\alpha_1,\alpha_2}
  \otimes \cdots \otimes
  \mathcal{A}^{[N]}_{\alpha_{N-1},1}.
  \label{eq:mps_matrix_of_kets}
\end{equation}

If $\chi$ is large enough, any state can be represented as an MPS.
In particular, an arbitrary state can be put in the form of an MPS with $\chi$ at most equal to $d^{\lfloor N/2 \rfloor}$ (see \cref{sec:MPS_construction_QR}). In practice, for many physically relevant states, a good approximation can be obtained with a much smaller $\chi$ (even $\chi=\mathcal{O}(1)$ for some states).

The MPS representation is particularly well suited to represent low-entangled states of 1d quantum systems. For instance, the ground state of a gapped local Hamiltonian in 1d can be efficiently approximated by an MPS with a bond dimension that does not grow with the system size.

TNs and MPS in particular can be represented graphically, as shown in \cref{fig:MPS_TTN_PEPS_MERA}.
For an MPS, each tensor $A^{[k]}$ is represented as a box, with three legs corresponding to its three indices. The vertical legs correspond to the physical indices $s_k=1,\dots,d$ of the local Hilbert space, while the horizontal legs represent the virtual indices of dimension $\chi$ that are contracted between neighboring tensors. For an open MPS the first and last tensors have only one horizontal leg.

\begin{figure}
  \begin{center}
    \includegraphics[width=0.7\textwidth]{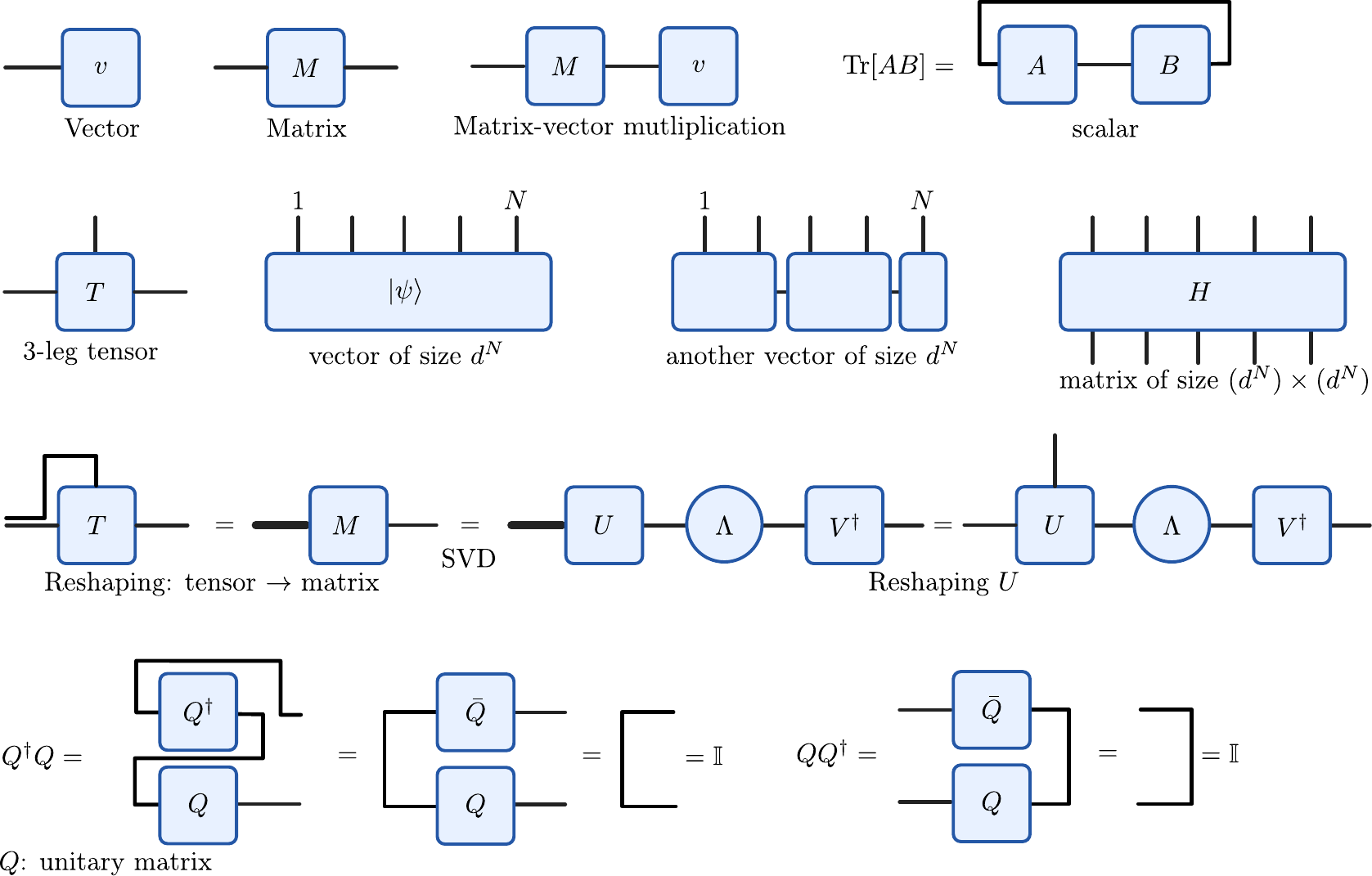}
  \end{center}
  \caption{Graphical representation of a few tensor examples and associated operations.}
  \label{fig:tensor_notation_basics}
\end{figure}

The graphical representation of MPS and TNs in general is a very useful tool to visualize the structure of the states, of the operators, and also to manipulate them. It allows us to represent operations such as contractions, reshaping, SVD, etc. in a simple and intuitive way. Graphical representations of a few elementary tensor operations are shown in \cref{fig:tensor_notation_basics}.
A more complete description of the graphical rules associated with the most common operations can be found in Sec.~2 of~\cite{BRIDGEMAN_HandwavingInterpretiveDance_2017}. The most common ones will appear in the figures of these notes.

\subsection{Gauge redundancy}
\label{sec:MPS_gauge_redundancy}

The number of parameters in an MPS described by \cref{eq:periodic_mps} is $N d \chi^2$ (for $N$ sites, each with $d$ local states, and each matrix $A^{[i]\,s_i}$ having $\chi^2$ parameters).
However, there is a {\em gauge} redundancy in the MPS representation:
one can multiply the matrices $A^{[i]\,s_i}$ by invertible matrices $X_i$ without changing the state $|\psi\rangle$. More precisely,
the MPS defined by the matrices $A^{[i]\,s_i} \to \widetilde A^{[i]\,s_i} = X_{i-1} A^{[i]\,s_i} X_i^{-1}$ describes the same state as the original MPS (each $X_i$ will simplify with its inverse $X_i^{-1}$ in the product of matrices in \cref{eq:periodic_mps}).
This gauge redundancy plays an important role in the theory and algorithms of MPS. For instance, it is often useful to choose a particular gauge (for instance the left or right canonical gauge, see \cref{sec:MPS_construction_QR}), which simplifies calculations.

The converse statement is known as the {\em fundamental theorem of MPS}.
Consider two MPS tensors $A^{s}_{\alpha\beta}$ and $B^{s}_{\alpha\beta}$, which define two translationally invariant MPS $|\psi_A\rangle=\sum_{s_1,\dots,s_N} \mathrm{Tr}(A^{s_1}\cdots A^{s_N})|s_1\dots s_N\rangle$ and $|\psi_B\rangle=\sum_{s_1,\dots,s_N} \mathrm{Tr}(B^{s_1}\cdots B^{s_N})|s_1\dots s_N\rangle$. 
If the two MPS
are injective (see \cref{sec:injective_MPS}) and represent the same state (up to a global phase) for all sufficiently large system sizes, then their tensors are related by an invertible gauge transformation, up to a phase. More explicitly, there exists a complex phase \(\omega\), with \(|\omega|=1\), and an invertible matrix \(X\) such that \(A^{s} = \omega X B^{s} X^{-1}\) for all \(s\).
This result is important for the classification of phases of matter using MPS, and for understanding symmetries and topological properties of MPS states~\cite{PEREZ-GARCIA_MatrixProductState_2007,CHEN_ClassificationGappedSymmetric_2011,CHEN_CompleteClassificationOnedimensional_2011,CIRAC_MatrixProductStates_2021,schuchClassifyingQuantumPhases2011}.

\subsection{A few examples}
\label{sec:MPS_simple_examples}
We start with a trivial case: product states. They correspond to MPS with bond dimension $\chi=1$.
$$|\psi\rangle = |\phi_1\rangle \otimes |\phi_2\rangle \otimes \cdots \otimes |\phi_N\rangle$$ can be represented as an MPS with bond dimension $\chi=1$. The matrix $A^{[i]\,s}$ is just a scalar that gives the amplitude of the local state $|\phi_i\rangle$ in the local basis:
$  A^{[i]\,s} = \langle s|\phi_i\rangle$. Equivalently,
\begin{equation}
  \mathcal{A}^{[i]} = |\phi_i\rangle.
\end{equation}

\subsubsection{GHZ state}
\label{sec:GHZ_state}

The GHZ state is defined as the equal superposition of all sites in state $|0\rangle$ and all sites in state $|1\rangle$:
\begin{equation}
  |\mathrm{GHZ}\rangle = \frac{1}{\sqrt{2}}(|00\dots 0\rangle + |11\dots 1\rangle).
\end{equation}
It can be represented as an MPS with bond dimension $\chi=2$ by choosing the following tensors (in \cref{eq:periodic_mps}):
\begin{equation}
  A^{[k]\,0} = \begin{pmatrix}1 & 0 \\ 0 & 0 \end{pmatrix}, \quad
  A^{[k]\,1} = \begin{pmatrix}0 & 0 \\ 0 & 1 \end{pmatrix}
  \label{eq:ghz_mps}
\end{equation}
for $k=1,\dots,N$. The product of two consecutive matrices $A^{[k]\,s} A^{[k+1]\,s'}$ gives zero if $s\neq s'$. The only non-zero contributions to the sum in \cref{eq:periodic_mps} are then for $s_1=s_2=\cdots = s_N=0$ and for $s_1=s_2=\cdots = s_N=1$, which gives the GHZ state (up to a normalization factor $\sqrt{2}$).
One can also represent the GHZ state as an open MPS. In this case, the first and last tensors are $1\times 2$ and $2\times 1$ matrices, respectively, while the tensors in the bulk are $2\times 2$ matrices. For instance, one can choose
\begin{equation}
  A^{[1]\,0} = \begin{pmatrix}1/\sqrt{2} & 0\end{pmatrix}, \quad
  A^{[1]\,1} = \begin{pmatrix}0 & 1/\sqrt{2}\end{pmatrix}
\end{equation}
for the first site, and
\begin{equation}
  A^{[N]\,0} = \begin{pmatrix}1 \\ 0\end{pmatrix}, \quad
  A^{[N]\,1} = \begin{pmatrix}0 \\ 1\end{pmatrix}
\end{equation}
for the last site. The matrices in the bulk (sites $k=2,\dots,N-1$) are the same as for the periodic representation \cref{eq:ghz_mps}.

Using the state-valued matrix notation of \cref{eq:matrix_of_kets}, this open-boundary representation can equivalently be written as
\begin{equation}
  \mathcal{A}^{[1]}
  =
  \frac{1}{\sqrt{2}}
  \begin{pmatrix}
    \ket{0} & \ket{1}
  \end{pmatrix},
  \qquad
  \mathcal{A}^{[N]}
  =
  \begin{pmatrix}
    \ket{0}\\
    \ket{1}
  \end{pmatrix},
\end{equation}
and, for $k=2,\dots,N-1$,
\begin{equation}
  \mathcal{A}^{[k]}
  =
  \begin{pmatrix}
    \ket{0} & 0\\
    0       & \ket{1}
  \end{pmatrix}.
\end{equation}
Again, the zero entries denote the zero vector in the one-site Hilbert space.

\subsubsection{W state}
\label{sec:W_state}

The W state is defined as the equal superposition of all states with exactly one site in state $|1\rangle$ and the remaining sites in state $|0\rangle$:
\begin{equation}
  |\mathrm{W}\rangle = \frac{1}{\sqrt{N}}(|100\dots 0\rangle + |010\dots 0\rangle + \cdots + |000\dots 1\rangle).
\end{equation}
It can also be represented as an MPS with bond dimension $\chi=2$. 
On the first and last sites one can choose (state-valued matrix notation)
\begin{equation}
  \mathcal{A}^{[1]}
  =
  \frac{1}{\sqrt{N}}
  \begin{pmatrix}
    \ket{0} & \ket{1}
  \end{pmatrix},
  \qquad
  \mathcal{A}^{[N]}
  =
  \begin{pmatrix}
    \ket{1}\\
    \ket{0}
  \end{pmatrix},
\end{equation}
and, for $k=2,\dots,N-1$,
\begin{equation}
  \mathcal{A}^{[k]}
  =
  \begin{pmatrix}
    \ket{0} & \ket{1}\\
    0       & \ket{0}
  \end{pmatrix}.
\end{equation}
Going from left to right in the product of \cref{eq:mps_matrix_of_kets}, the auxiliary index $\alpha \in\{1,2\}$ of the MPS tensors keeps track of whether we have already encountered a site in state $|1\rangle$ ($\alpha=2$) or not ($\alpha=1$). The matrix $A^{[k]\,0}$, which corresponds to a $|0\rangle$, does not change this information.
On the other hand, the matrix $A^{[k]\,1}$ changes the auxiliary index from $\alpha=1$ to $\alpha=2$. It corresponds to the fact that one has encountered a site in state $|1\rangle$. Once we have encountered such a $|1\rangle$ state, the auxiliary index remains $\alpha=2$ until the last site. This ensures that there is a single site in state $|1\rangle$ in the superposition.

\subsubsection{Valence-bond crystal}
\label{sec:valence_bond_crystal}
A valence-bond crystal (VBC) is a state of a spin-$\frac{1}{2}$ system where some pairs of neighboring spins form singlets.
In 1d and for an even number of sites \(N\), the state reads
    \[|\Psi_{\mathrm{VBC}}\rangle=\bigotimes_{k=1}^{N/2}
      \frac{1}{\sqrt2}
      \Bigl(|\uparrow_{2k-1}\downarrow_{2k}\rangle - |\downarrow_{2k-1}\uparrow_{2k}\rangle \Bigr).
    \]
Using the notation of \cref{eq:matrix_of_kets}, this can be written as an open-boundary MPS with alternating tensors
\begin{equation}
  \mathcal{A}^{[2k-1]}
  =
  \begin{pmatrix}
    \ket{\uparrow} & \ket{\downarrow}
  \end{pmatrix},
  \qquad
  \mathcal{A}^{[2k]}
  =
  \frac{1}{\sqrt{2}}
  \begin{pmatrix}
    \ket{\downarrow}\\
    -\ket{\uparrow}
  \end{pmatrix},
  \qquad k=1,\dots,N/2.
\end{equation}
The contraction of the virtual index inside one pair gives
$
  \mathcal{A}^{[2k-1]}\mathcal{A}^{[2k]}
  =
  \frac{1}{\sqrt{2}}
  \left(
    \ket{\uparrow_{2k-1}\downarrow_{2k}}
    -
    \ket{\downarrow_{2k-1}\uparrow_{2k}}
  \right)
$,
so the product of the alternating MPS tensors gives the VBC state above.
The bond dimension is 1 (no entanglement) between sites $2k$ and $2k+1$, while the bond dimension is $\chi=2$ between sites $2k-1$ and $2k$ (singlet state).

\subsubsection{AKLT state}
\label{sec:AKLT}

In 1987 Affleck, Kennedy, Lieb and Tasaki (AKLT) introduced a spin-1 chain Hamiltonian with a ground state that can be exactly represented as an MPS with bond dimension $\chi=2$~\cite{AFFLECK_RigorousResultsValencebond_1987}. The idea is to start from a state of $2N$ spin-$\frac{1}{2}$ degrees of freedom, where each pair of neighboring spin-$\frac{1}{2}$ degrees of freedom is in a singlet state. This state can be represented as an MPS with bond dimension $\chi=2$.
This state is a paradigmatic example of a state with short-range correlations and non-trivial (symmetry-protected) topological properties. Its construction (detailed in \cref{sec:AKLT_MPS}) from virtual spin-$\frac{1}{2}$ degrees of freedom is also at the heart of the construction of PEPS in higher dimensions.

\begin{figure}
  \begin{center}
    \includegraphics[width=0.7\textwidth]{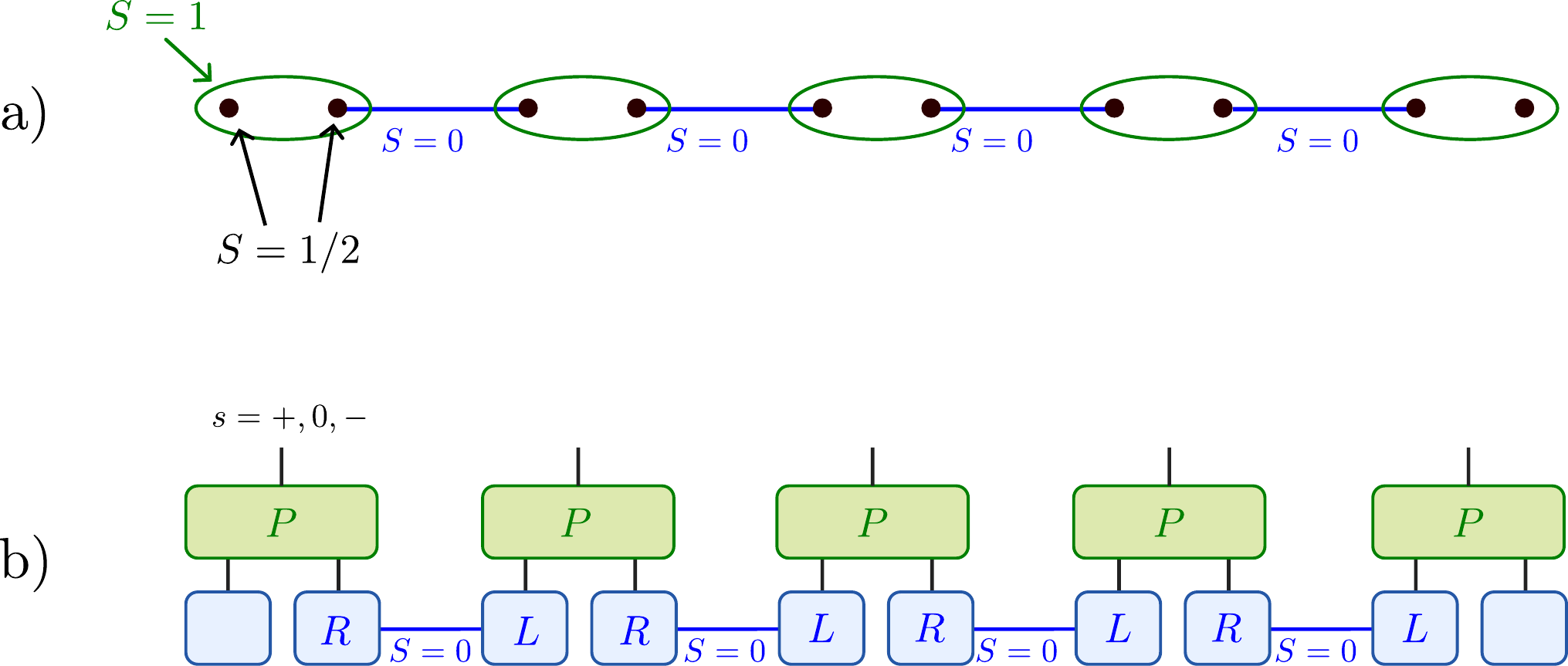}
  \end{center}
  \caption{AKLT state for a spin-1 chain.
    a) The Hilbert space is enlarged by replacing each spin-1 (green ovals) by two fictitious spin-$\frac{1}{2}$ degrees of freedom (black dots).
    The state of the $2N$ spin-$\frac{1}{2}$ degrees of freedom is a product of spin singlet states between neighboring spin-$\frac{1}{2}$ degrees of freedom (blue bonds). b) This product of singlets is an MPS with bond dimension $\chi=2$ for the fictitious spins.
    By projecting the two spin-$\frac{1}{2}$ degrees of freedom at each site onto the $S=1$ triplet subspace (operator $P$), one obtains a state for the original spin-1 chain that can also be represented as an MPS with bond dimension $\chi=2$. Each tensor of this MPS is obtained by contracting three tensors: $P$, $L$ and $R$.}
    \label{fig:AKLT}
\end{figure}

By projecting the two spin-$\frac{1}{2}$ degrees of freedom at each site onto the $S=1$ triplet subspace, one obtains a state of $N$ spin-1 degrees of freedom that can also be represented as an MPS with bond dimension $\chi=2$ (\cref{fig:AKLT}). Now consider the following spin-1 Hamiltonian:
    \begin{equation}
      H = \sum_{i=1}^{N-1} \Pi^{S=2}_{i,i+1}
    \end{equation}
where $\Pi^{S=2}_{i,i+1}$ is the projector onto the total spin $S_{\rm tot}=2$ subspace of the two neighboring spin-1 degrees of freedom at sites $i$ and $i+1$. This projector can be constructed explicitly by noting that the square of the total spin $S^2_{\rm tot}=(\vec S_i + \vec S_j)^2=2\vec S_i\cdot\vec S_{j}+4$ has three eigenvalues $S_{\rm tot}(S_{\rm tot}+1)$ with $S_{\rm tot}=0,1,2$. The projector onto the $S_{\rm tot}=2$ subspace is thus $\Pi^{S=2}_{i,j} = \frac{1}{24} S^2_{\rm tot}(S^2_{\rm tot}-2)$. After a few lines of elementary algebra one finds
    \begin{equation}
      H = \frac{1}{6}\sum_{i=1}^{N-1} (
      (\vec S_i \cdot \vec S_{i+1})^2
      + 3 \vec S_i \cdot \vec S_{i+1}
      + 2
      )
    \end{equation}
By construction, any state $\ket{\psi}$ that has no component in the $S_{\rm tot}=2$ subspace of any pair of neighboring sites is annihilated by the above Hamiltonian since it gives $ \Pi^{S=2}_{i,i+1} \ket{\psi} = 0$. Since $H$ is a sum of projectors, its eigenvalues are necessarily $\geq 0$. A state that is annihilated by $H$ is thus a ground state.
By construction, two neighboring spin-1 sites have a total spin at most equal to 1 since two of the four spin-$\frac{1}{2}$ degrees of freedom on these two spin-1 sites are in a singlet state. 
So, the state of \cref{fig:AKLT} satisfies $\Pi^{S=2}_{i,i+1} \ket{\psi} = 0$.
This state is thus a ground state of the above Hamiltonian.
The construction shown schematically in \cref{fig:AKLT} also shows that this state can be represented as an MPS with bond dimension $\chi=2$. This state is usually called the AKLT state or valence bond solid (VBS) state. The derivation is detailed in \cref{sec:AKLT_MPS}, together with the explicit expression of the resulting MPS tensors.

Readers interested in additional simple MPS examples can for instance refer to \cite{ORUS_PracticalIntroductionTensor_2014}, which discusses the Majumdar-Ghosh state and the 1d cluster state.


%
\begin{longlisting}
  \jcode{mps.jl}
  \caption[Simple Néel-state MPS]{Creation of a simple (Néel) MPS state for a system of spin-$\frac{1}{2}$ using the \texttt{ITensorMPS} library in Julia.
  The code builds the site indices, chooses the alternating product-state configuration \texttt{Up,Dn,Up,Dn,\dots}, converts it directly into an MPS, and prints the maximal bond dimension. Since this is a product state, the bond dimension is trivially equal to 1.
  Before executing the code above, install the \texttt{ITensor} and \texttt{ITensorMPS} libraries by running the following commands in the Julia REPL:
  {\tt  using Pkg ; Pkg.add("ITensors") ; Pkg.add("ITensorMPS")}. \jcodegithublink}
\end{longlisting}

\subsection{MPS construction by successive QR and canonical forms}
\label{sec:MPS_construction_QR}

A generic many-body wavefunction can be viewed as a large (many indices) tensor with one index per site. By repeatedly reshaping this tensor into a matrix and factorizing it, one can ``peel off'' one site at a time and obtain an MPS. If QR decompositions are used, the resulting tensors satisfy so-called left- or right-orthonormality conditions, leading to canonical forms.

\subsubsection{Construction of a left-canonical MPS by successive QR decompositions}
\label{sec:left_canonical_QR_construction}

\begin{figure}
  \begin{center}
    \includegraphics[width=0.6\textwidth]{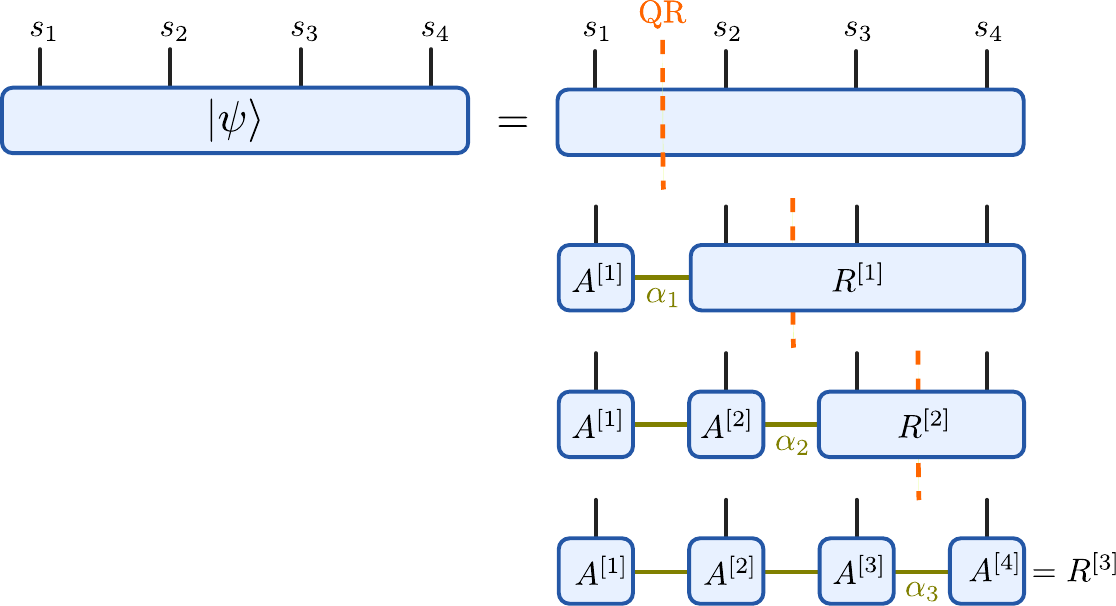}
  \end{center}
  \caption{Construction of an MPS from a vector representation by successive QR decompositions. At each step, the $Q$ matrix gives an MPS tensor $A^{[i]}$, while the $R$ matrix is reshaped and factorized again at the next site. Such a construction leads to an MPS in a left-canonical gauge (see \cref{fig:MPS_mixed_canonical}).
  }
  \label{fig:successive_QR}
\end{figure}

Starting from a generic state $|\psi\rangle$ expressed in the local basis $\ket{s_1 s_2 \dots s_N}$, one can construct an MPS representation of this state by performing successive QR decompositions (\cref{sec:QR}). 
Such a construction is illustrated in \cref{fig:successive_QR} for an $N=4$-site state with coefficients $\psi_{s_1s_2s_3s_4}$. First, one {\em reshapes} this tensor into a matrix with row index $s_1$ and combined column index $(s_2,s_3,s_4)$. This matrix, of size $d \times d^{N-1}$, is then factorized using a QR decomposition. The $Q$ matrix is interpreted as the first MPS tensor $A^{[1]}_{s_1,\alpha_1}$, while the $R$ matrix is reshaped as a new tensor $R^{[1]}_{\alpha_1,s_2,s_3,s_4}$. One then groups $(\alpha_1,s_2)$ as the row index and $(s_3,s_4)$ as the column index of $R^{[1]}$ and repeats the QR decomposition, producing $A^{[2]}_{\alpha_1s_2\alpha_2}$ and $R^{[2]}_{\alpha_2s_3s_4}$. Repeating this step gives $A^{[3]}_{\alpha_2s_3\alpha_3}$ and a final tensor $A^{[4]}_{\alpha_3s_4}$, yielding an open-boundary MPS.
A code example based on the \texttt{ITensor} library transforms a general state (here random) into an MPS by successive QR decompositions; it is given in \cref{lst:random_state_to_MPS}.

\begin{remarkblock}
  The decomposition depends on the way the sites, or local degrees of freedom, have been ordered. This is an important point to keep in mind when working with MPS, as the choice of the ordering can have a significant impact on the efficiency of the representation. If the original physical system is 1d, there is usually a natural ordering of the sites that follows the geometry of the system. However, for higher-dimensional systems or systems with more complex geometries, the choice of ordering can be more subtle. \Cref{fig:snake_ordering_2d_mps} shows an example of a 2d system where the sites are ordered in a snake-like pattern to create a 1d MPS representation. Other choices, however, are possible and may lead to more efficient representations depending on the structure of the state. 
\end{remarkblock}

\begin{figure}
  \begin{center}
    \includegraphics[width=\textwidth]{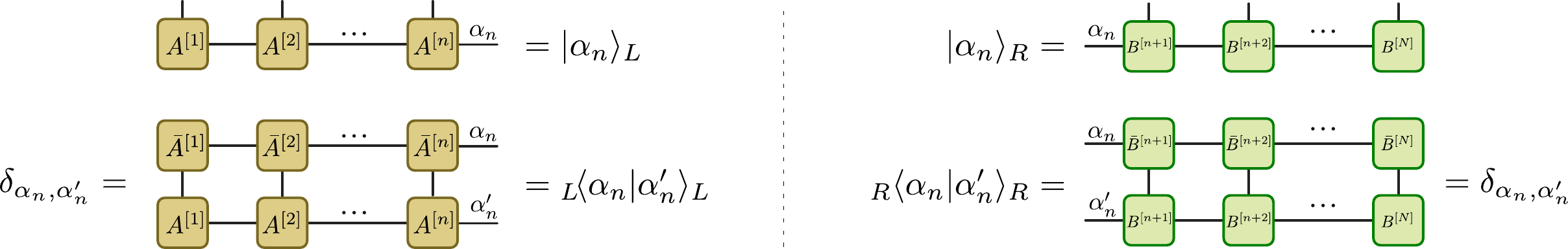}
  \end{center}
  \caption{Partial contractions of a left-canonical MPS and a right-canonical MPS. For each value of the open virtual index $\alpha_n$, the left contraction defines a state $\ket{\alpha_n}_L$ in the Hilbert space of the first $n$ sites, while the right contraction defines a state $\ket{\alpha_n}_R$ in the Hilbert space of sites $n+1,\dots,N$. The corresponding canonical conditions imply that these states form orthonormal families. See also \cref{fig:MPS_mixed_canonical}.}
  \label{fig:MPS_partial_left_right_states}
\end{figure}

The MPS produced by this construction is said to be in the {\em left-canonical gauge}, which means that the tensors $A^{[i]}$ satisfy the left-orthonormalization condition $\sum_{s_i} (A^{[i]\,s_i})^\dagger A^{[i]\,s_i} = I$ for all $i$. This is a consequence of the fact that the $Q$ matrices obtained from the QR decomposition are unitary (or isometric in the case of reduced QR). This property of the $A$ tensors is illustrated in \cref{fig:MPS_mixed_canonical} and is often exploited to simplify certain calculations.
In particular, consider a cut between site $n$ and site $n+1$ and fix the value of the associated bond index $\alpha_n$.
The partial contraction of the first $n$ tensors defines a state in the Hilbert space of the first $n$ sites:
\begin{equation}
  \ket{\alpha_n}_L
  =
  \sum_{s_1,\dots,s_n}
  \left(A^{[1]\,s_1}A^{[2]\,s_2}\cdots A^{[n]\,s_n}\right)_{\alpha_n}
  \ket{s_1\dots s_n}.
\end{equation}
If the MPS is left-canonical, these states form an orthonormal family in the Hilbert space of the first $n$ sites
\begin{equation}
  {}_L\!\braket{\alpha_n}{\beta_n}_L = \delta_{\alpha_n,\beta_n},
\end{equation}
as illustrated in \cref{fig:MPS_partial_left_right_states}.

\begin{longlisting}
  \jcode{random_state_to_MPS.jl}
  \caption[Random state to MPS]{\label{lst:random_state_to_MPS}Julia code using the \texttt{ITensor} and \texttt{ITensorMPS} libraries to generate a random state of $N$ spin-$\frac{1}{2}$ degrees of freedom as a dense vector of dimension $2^N$, convert it to an MPS, and print the bond dimensions of the resulting MPS.
  As expected for a random state, the bond dimension between sites $i$ and $i+1$ takes its maximum possible value  $\chi_i = 2^{\min(i,N-i)}$. Output of the code: \texttt{bond dimensions = [2, 4, 8, 16, 32, 64, 128, 64, 32, 16, 8, 4, 2]}.
  \jcodegithublink
  }
\end{longlisting}

\begin{remarkblock}
What are the bond dimensions obtained by rewriting a generic state $\ket{\psi}$ with local physical dimension $d$ as an MPS? The first QR decomposition involves a matrix of size $d \times d^{N-1}$. The first bond index, $\alpha_1$, thus takes $d$ values. At the next step, the tensor $R^{[1]}_{\alpha_1,s_2,s_3,\dots, s_N}$ is reshaped into a matrix with row index $(\alpha_1,s_2)$ and column index $(s_3,\dots,s_N)$, of size $d^2 \times d^{N-2}$, so that the second bond dimension is $d^2$. More generally, at step $i\leq N/2$ the QR decomposition is applied to a ``wide'' matrix of size $d^i\times d^{N-i}$, leading to a unitary $Q$ matrix of size $d^i\times d^i$ and a wide upper triangular $R$ matrix of size $d^i\times d^{N-i}$.
Beyond the middle of the chain, the matrices to be QR-decomposed are ``tall'' matrices of size $d^i\times d^{N-i}$ with $i>N/2$.
In such a case one instead uses a {\em reduced} QR decomposition (\cref{sec:QR}) where the matrix $Q$ is now rectangular of size $d^i\times d^{N-i}$, and the $R$ matrix is square of size $d^{N-i}\times d^{N-i}$. The associated bond dimension is thus $d^{N-i}$ for $i>N/2$.
Thus, as expected, for a generic state the QR decompositions give bond dimensions
\begin{equation}
  \chi_i =  d^{\min(i,N-i)}.
\end{equation}
The code in \cref{lst:random_state_to_MPS} confirms this for a random state of spin-$\frac{1}{2}$ ($d=2$), where the bond dimensions are found to be $\chi_i=(2,4,8,16,32,64,128,64,32,16,8,4,2)$.
\end{remarkblock}

\subsubsection{Making an MPS left or right canonical}
\label{sec:MPS_left_right_canonical}

\begin{figure}
  \begin{center}
    \includegraphics[width=0.422\textwidth]{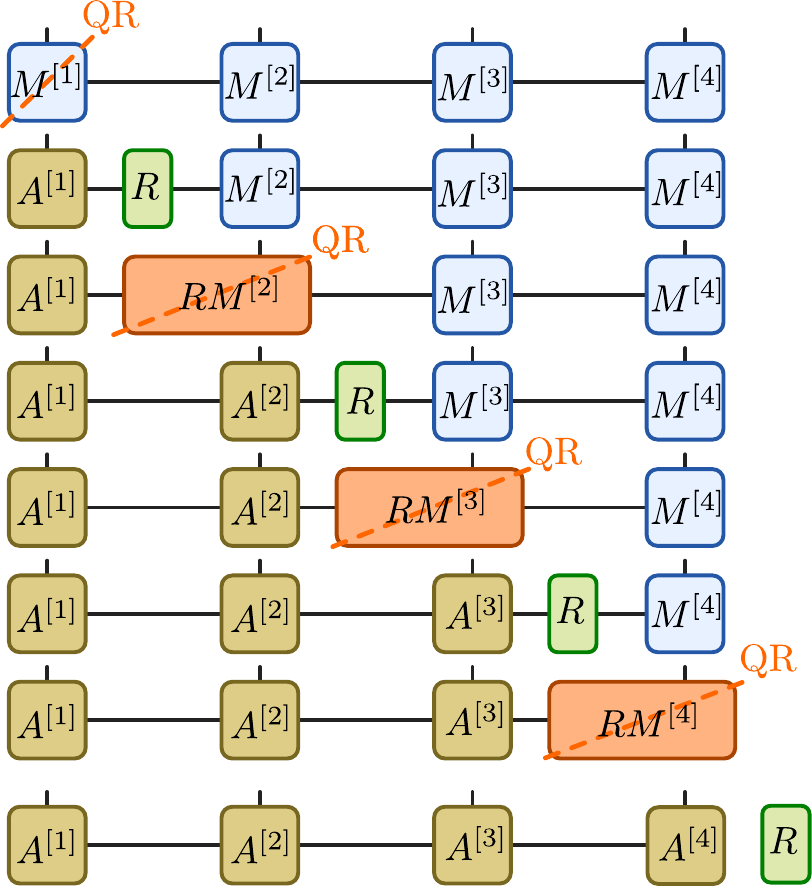}
  \end{center}
  \caption{Transformation of an MPS into left-canonical form by successive QR decompositions. At each step, the active tensor is reshaped into a matrix, QR-decomposed, and the resulting $R$ tensor is contracted into the next tensor on the right. The $Q$ tensors produced in this way become the left-canonical tensors $A^{[i]}$.}
  \label{fig:MPS_left_canonical_QR}
\end{figure}

The construction of an MPS from a vector representation is conceptually important, but in most cases MPS are used for large systems where the full vector representation (size $d^N$) is too large to be tractable. On the other hand, it is frequent to perform a similar series of QR decompositions starting from an MPS, in order to bring it to a canonical gauge. The construction of a left-canonical MPS from a generic MPS by successive QR decompositions is illustrated in \cref{fig:MPS_left_canonical_QR}. One starts from the leftmost tensor of the MPS, reshapes it into a matrix, and performs a QR decomposition. The $Q$ matrix gives the first left-canonical tensor $A^{[1]}$, while the $R$ matrix is contracted into the next tensor on the right. One then repeats this step for the second site, and so on until the end of the chain. The last QR decomposition involves a matrix with a single column, so that the $Q$ matrix is a vector and the $R$ matrix is a scalar. This scalar is the normalization factor of the MPS ($\braket{\psi}{\psi}=|R|^2$), which can be set to 1 by rescaling the last tensor $A^{[N]}$.
Performing a similar series of QR decompositions in the other direction, from right to left, gives an MPS in the right-canonical gauge. The resulting tensors $B^{[i]}$ satisfy the right-orthonormalization condition $\sum_{s_i} B^{[i]\,s_i} (B^{[i]\,s_i})^\dagger = I$ for all $i$ (see \cref{fig:MPS_partial_left_right_states} and \cref{fig:MPS_mixed_canonical}).

The existence of such canonical forms is not a peculiarity of MPS, but rather a consequence of the fact that the underlying TN has no loops. For any loop-free TN, and in particular for tree TNs (TTN), one can choose a root tensor and successively QR-decompose the tensors along the branches so that all tensors away from the root become isometries, defining an orthogonality center analogous to the mixed-canonical form of an MPS~\cite{TAGLIACOZZO_SimulationTwodimensionalQuantum_2009,SILVI_TensorNetworksAnthology_2019}. By contrast, the presence of loops, as in PEPS, prevents one from gauging all environments to identities at the same time; exact contraction is generically hard, and tensor truncation or variational optimization then requires approximate environments, making the corresponding algorithms substantially more involved~\cite{ORUS_PracticalIntroductionTensor_2014,SCHUCH_ComputationalComplexityProjected_2007}. See also \cite{pippanEfficientMatrixproductState2010} concerning {\em periodic} MPS. 

\subsubsection{Mixed-canonical form and orthogonality center}
\label{sec:mixed_canonical_form}

\begin{figure}
  \begin{center}
    \includegraphics[width=0.494\textwidth]{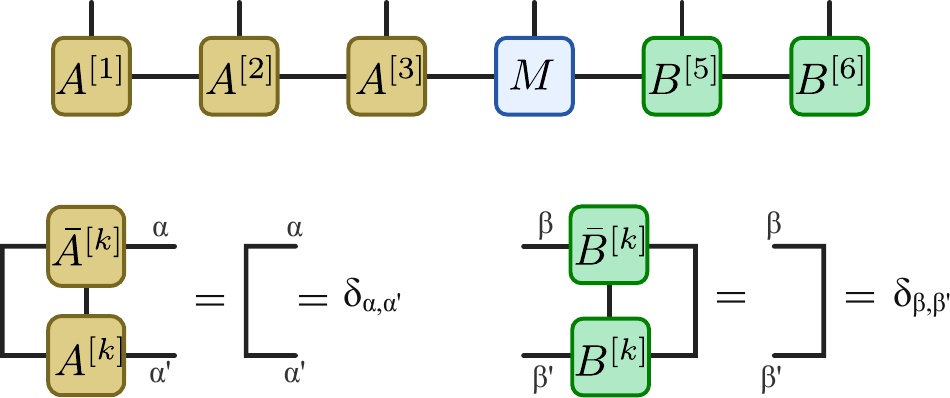}
  \end{center}
  \caption{Top: mixed-canonical form of a six-site MPS with the orthogonality center at site $k=4$. The tensors to its left are left-canonical, and those to its right are right-canonical. Bottom: left- and right-canonical conditions for the corresponding tensors.
  The isometric properties of $A^{[i]}$ (left-canonical) and $B^{[j]}$ (right-canonical) imply that $A^\dagger A = 1$ and $B B^\dagger= 1$.
  }
  \label{fig:MPS_mixed_canonical}
\end{figure}

It is also useful to mix the two sweeps above to obtain a {\em mixed-canonical form} of the MPS, where the tensors to the left of a given site $k$ are left-canonical, the tensors to the right of site $k$ are right-canonical, and the tensor at site $k$ is unconstrained, as illustrated in \cref{fig:MPS_mixed_canonical}.
Starting from the left boundary, one performs QR decompositions only up to site $k-1$, absorbing each remaining factor into the tensor on its right; the tensors
at sites $1,\dots,k-1$ are then left-canonical. Starting from the right boundary, one performs the analogous right-to-left decompositions down to site $k+1$, absorbing the remaining factors into the tensor on their left; the tensors at sites $k+1,\dots,N$
are then right-canonical. All factors left over from these two sweeps are finally gathered into the tensor at site $k$. This tensor is generally not constrained to be left- or right-canonical: it is called the orthogonality center.
In the \texttt{ITensorMPS} library, the function \texttt{orthogonalize!} can be used to bring an MPS into such a mixed-canonical form with an orthogonality center at a given site $k$. An example is given in \cref{lst:MPS_orthogonality_center}.

\begin{longlisting}
  \jcode{MPS_orthogonality_center.jl}
  \caption[MPS orthogonality center]{\label{lst:MPS_orthogonality_center}Use of the \texttt{orthogonalize!} function of \texttt{ITensorMPS} to bring an MPS into mixed-canonical form with an orthogonality center at site $k$. Once this is done, the expectation value of the single-site operator $\sigma_k^z$ can be computed from the center tensor alone (tensor $M$ in \cref{fig:MPS_mixed_canonical}). The first contraction illustrates that the same local calculation does not give the correct result before the orthogonality center has been moved to site $k$. \jcodegithublink}
\end{longlisting}

\subsubsection{Entanglement entropy and MPS compression from mixed-canonical form}
\label{sec:entanglement_entropy_center_matrix}

The mixed canonical form is also useful for computing entanglement entropies associated with a bipartition of the system into two parts $(1,\dots,n)$ and $(n+1,\dots,N)$.
In terms of the states $\ket{\alpha_{n-1}}_L$ and $\ket{\alpha_n}_R$ defined by the left and right canonical parts of the MPS,
the full state can be written as
\begin{equation}
  \ket{\psi}
  =
  \sum_{\alpha_{n-1},s_n,\alpha_n}
  M^{[n]\,s_n}_{\alpha_{n-1},\alpha_n}
  \ket{\alpha_{n-1}}_L \ket{s_n} \otimes  \ket{\alpha_n}_R .
\end{equation}
Since the states $\ket{\alpha_{n-1}}_L\ket{s_n}$
are orthonormal (in subsystem $(1,\dots,n)$), and since the states $\ket{\alpha_n}_R$ are also orthonormal (in subsystem $(n+1,\dots,N)$)
the SVD of the center matrix $M^{[n]}$ (obtained by grouping the indices $\alpha_{n-1}$ and $s_n$ as the row index, and $\alpha_n$ as the column index)
gives the Schmidt decomposition (\cref{sec:SVD_Schmidt}) of the state $|\psi\rangle$ across the bond $(n,n+1)$. 
The von Neumann entropy then follows directly from these singular values.
It is interesting that, with a well-defined orthogonality center, the Schmidt decomposition across a given bond can be obtained from a single tensor.

A code example (based on ITensor/ITensorMPS) that computes the von Neumann entanglement entropy across a given bond from the SVD of the center tensor is given in \cref{lst:dmrg_spin_half_heisenberg}.

When an MPS has a well-defined orthogonality center, the SVD of the center matrix can also be used to compress the MPS by truncating the small singular values. This is a key step in many algorithms. To do this,
as in the entanglement-entropy calculation,
one reshapes the center tensor $M^{[n]}$ into a matrix with combined row index $(\alpha_{n-1},s_n)$ and column index $\alpha_n$. One then performs an SVD of this matrix, and truncates the small singular values to obtain a new center matrix with smaller bond dimension on bond $(n,n+1)$.

\subsubsection{Expectation value of a local observable}
\label{sec:MPS_local_observable}

\begin{figure}
  \begin{center}
    \includegraphics[width=0.6\textwidth]{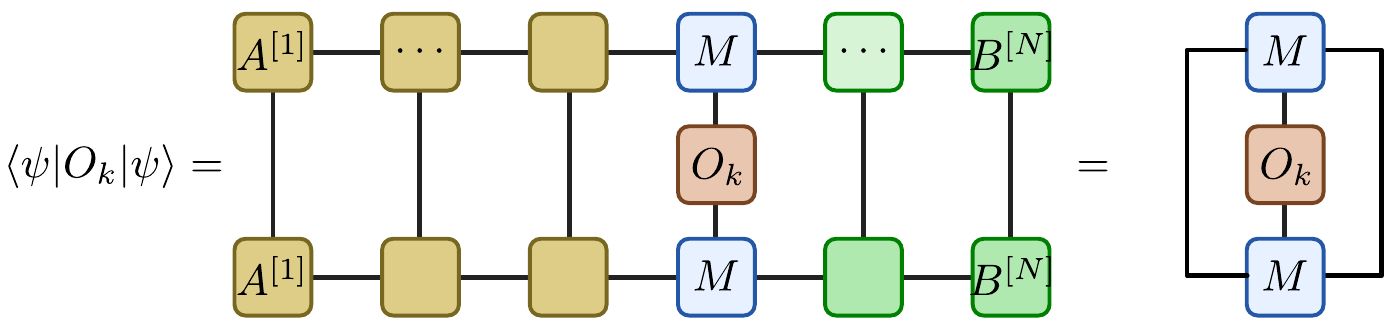}
    \caption{Expectation value of a single-site observable $O$ at site $k$ in an MPS written in mixed-canonical form with orthogonality center at site $k$. The contraction simplifies and only the tensor $M$ at the orthogonality center remains.}
    \label{fig:MPS_local_observable}
  \end{center}
\end{figure}

Consider a local observable $O_k$ acting on site $k$. The expectation $\langle \psi | O_k | \psi \rangle$ can be computed by contracting the double-layer TN shown in the left of \cref{fig:MPS_local_observable}. If the MPS is written in mixed-canonical form with the orthogonality center at site $k$,
one can use the rules illustrated in \cref{fig:MPS_mixed_canonical} to simplify the calculation. As a result, the contraction simplifies considerably, as shown in the right part of \cref{fig:MPS_local_observable}. The calculation finally reduces to a contraction of $M$, its conjugate and the observable $O_k$. The computational cost of this contraction is $\mathcal{O}(d^2\chi^2)$, where $d$ is the physical dimension of the local Hilbert space and $\chi$ is the bond dimension of the MPS.
An elementary code example where such a calculation is performed is given in \cref{lst:MPS_orthogonality_center}.

\subsection{Elementary manipulations}
\label{sec:MPS_elementary_manipulations}

\subsubsection{Addition of two MPS}
\label{sec:MPS_addition}
Consider two MPS $|\psi_A\rangle$ and $|\psi_B\rangle$ with the same local Hilbert space and the same number of sites. We can write them as
\begin{equation}
  |\psi_A\rangle = \sum_{s_1,s_2,\dots,s_N} A^{[1]\,s_1} A^{[2]\,s_2} \cdots A^{[N]\,s_N} |s_1 s_2 \dots s_N\rangle
\end{equation}
and
\begin{equation}
  |\psi_B\rangle = \sum_{s_1,s_2,\dots,s_N} B^{[1]\,s_1} B^{[2]\,s_2} \cdots B^{[N]\,s_N} |s_1 s_2 \dots s_N\rangle.
\end{equation}
The sum of the two MPS can be written as
\begin{equation}
  |\psi_A\rangle + |\psi_B\rangle = \sum_{s_1,s_2,\dots,s_N} C^{[1]\,s_1} C^{[2]\,s_2} \cdots C^{[N]\,s_N} |s_1 s_2 \dots s_N\rangle.
\end{equation}
For open boundary conditions, the new tensors are defined as
\begin{equation}
  C^{[1]\,s_1} =
  \begin{pmatrix} A^{[1]\,s_1} & B^{[1]\,s_1} \end{pmatrix},
  \qquad
  C^{[N]\,s_N} =
  \begin{pmatrix} A^{[N]\,s_N} \\ B^{[N]\,s_N} \end{pmatrix},
\end{equation}
and, for $k=2,\dots,N-1$,
\begin{equation}
  C^{[k]\,s_k} = \begin{pmatrix} A^{[k]\,s_k} & 0 \\ 0 & B^{[k]\,s_k} \end{pmatrix}.
\end{equation}
For periodic boundary conditions, the block-diagonal construction can instead be used on every site.
The new MPS has a bond dimension that is the sum of the bond dimensions of the original MPS. This is a simple way to construct an MPS representation of a linear combination of two states, but it is in general not optimal. For instance, if one tries to write the W state as an MPS using this method, one would get an MPS with bond dimension $\chi=N$, whereas the W state admits an MPS representation with bond dimension $\chi=2$.
\subsubsection{Scalar product}
\label{sec:MPS_scalar_product}

\begin{figure}
  \begin{center}
    \includegraphics[width=0.8\textwidth]{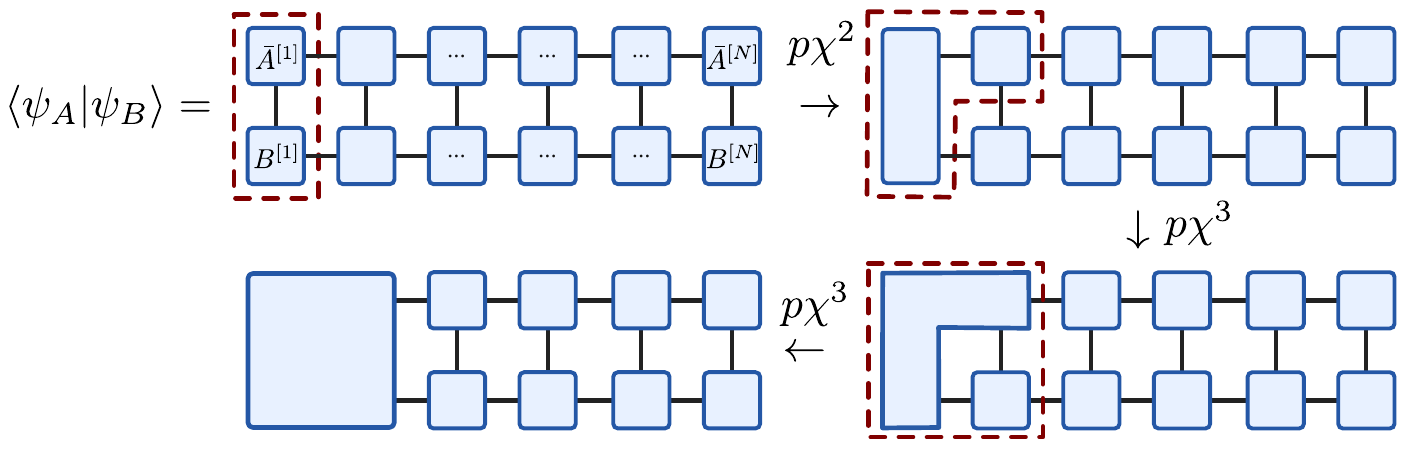}
    \caption{MPS representation of the scalar product $\langle\psi_A|\psi_B\rangle$, together with the left-to-right sequence of partial contractions.}
    \label{fig:mps_scalar_prod}
  \end{center}
\end{figure}

The scalar product $\langle\psi_A|\psi_B\rangle$
between two MPS $|\psi_A\rangle$ and $|\psi_B\rangle$ can be computed by contracting the double-layer TN displayed in \cref{fig:mps_scalar_prod}. The contraction of this TN can be performed efficiently, from left to right (or from right to left), by contracting one tensor ($\bar A^{[i]}$ or $B^{[i]}$) at a time (\cref{fig:mps_scalar_prod}). The computational cost of this contraction is $\mathcal{O}(Nd\chi^3)$, where $N$ is the number of sites, $d$ is the physical dimension of the local Hilbert space, and $\chi$ is the bond dimension of the MPS.

\subsubsection{Correlation function}
\label{sec:MPS_correlation_function}

\begin{figure}
  \begin{center}
    \includegraphics[width=0.75\textwidth]{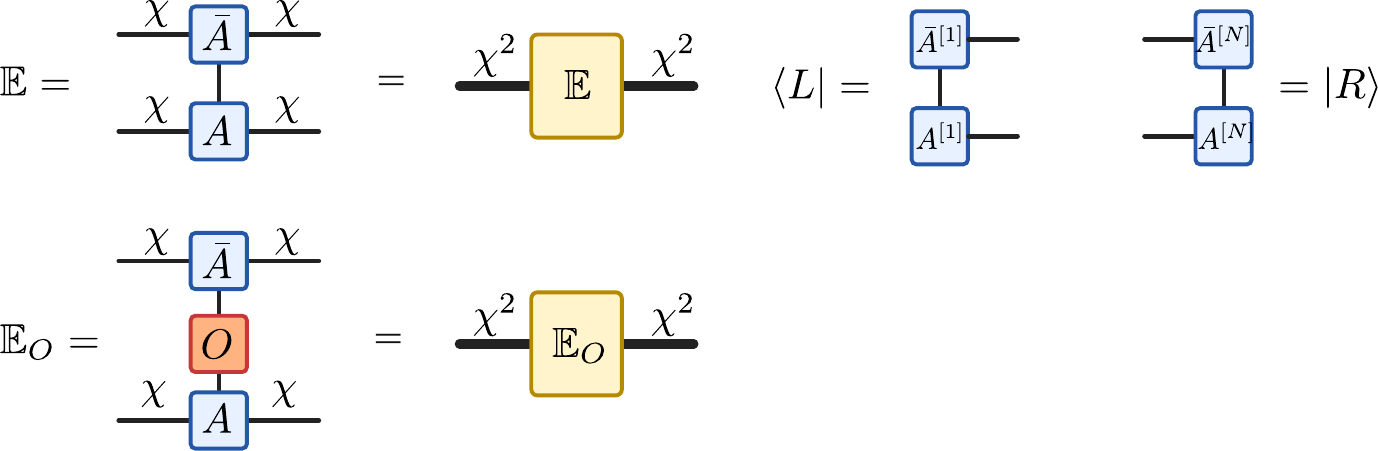}
    \caption{Graphical definition of the transfer matrix for a translationally invariant MPS. The physical index of $A^s$ and $\overline{A^s}$ is contracted, and the pairs of left and right virtual indices are grouped to form the matrix indices of $\mathbb{E}$. The second line shows the corresponding transfer matrix $\mathbb{E}_O$ with a single-site observable inserted on the physical leg. The bottom line represents the boundary vectors $\langle L\vert$ and $\vert R\rangle$ used in \cref{eq:correlation_function_transfer_matrix}.}
    \label{fig:transfer_matrix}
  \end{center}
\end{figure}

For a translationally invariant MPS $A^{[i]}=A$ for all $i$, the correlation function $\langle O_k O'_{k+r}\rangle$ can be computed using the transfer matrix
\begin{equation}
  \mathbb{E} = \sum_{s=1}^d A^{s} \otimes \overline{A^{s}},
  \label{eq:transfer_matrix}
\end{equation}
where $A^s$ are the tensors of the MPS and $\overline{A^{s}}$ are their complex conjugates. 
This matrix, of dimension $\chi^2 \times \chi^2$, is represented graphically in \cref{fig:transfer_matrix}. The physical indices of $A^s$ and $\overline{A^{s}}$ are contracted together, and the pairs of left and right virtual indices are grouped to form the matrix indices of $\mathbb{E}$.\footnote{Note that $\mathbb{E}$ is unchanged if one performs a local basis change on the physical index of $A^s$ (i.e., if one replaces $A^s$ by $\sum_{s'} U_{ss'} A^{s'}$ with $U$ a unitary matrix).}
It is also useful to define the transfer matrix associated with a single-site observable $O$:
\begin{equation}
  \mathbb{E}_O = \sum_{s,s'=1}^d \langle s|O|s'\rangle A^{s} \otimes \overline{A^{s'}},
\end{equation}
as shown in the second line of \cref{fig:transfer_matrix}.

\begin{figure}
  \begin{center}
    \includegraphics[width=0.65\textwidth]{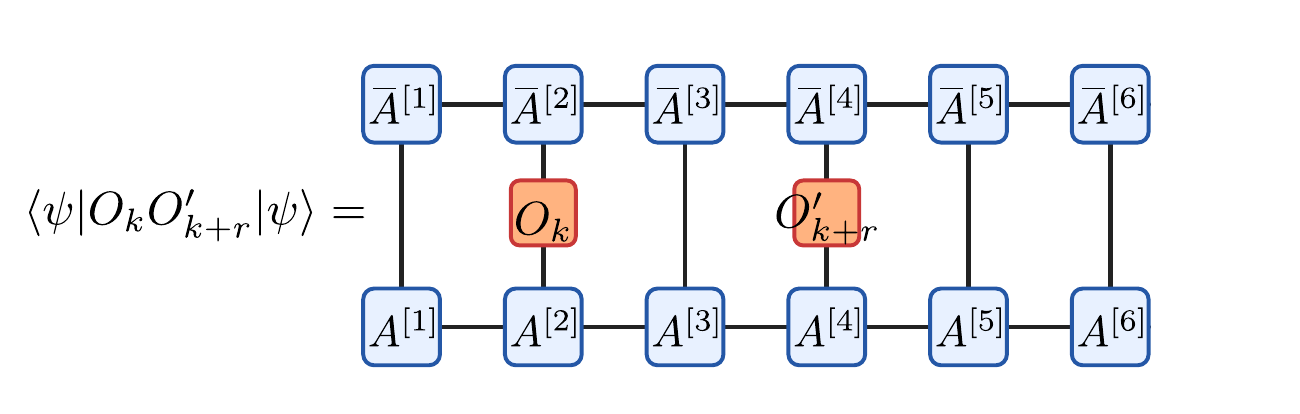}
    \caption{TN representation of the two-point correlation function of the type $\langle\psi|O_k O'_{k+r}|\psi\rangle$.
    No canonical form is assumed here.}
    \label{fig:MPS_two_point_correlation}
  \end{center}
\end{figure}

For a finite MPS, the two-point function can first be represented directly as a double-layer TN, with the operators inserted on the corresponding physical legs, as illustrated in \cref{fig:MPS_two_point_correlation}.

The two-point correlation function can be expressed as
\begin{equation}
\langle O_k O'_{k+r}\rangle
=
\frac{
\langle L|\mathbb{E}^{k-2}\mathbb{E}_O \mathbb{E}^{r-1}\mathbb{E}_{O'}\mathbb{E}^{N-k-r-1}|R\rangle
}{
\langle L|\mathbb{E}^{N-2}|R\rangle
}.
\label{eq:correlation_function_transfer_matrix}
\end{equation}
Using the spectral decomposition
\[
\mathbb{E}=\sum_{\alpha=1}^{\chi^2}
\lambda_\alpha |r_\alpha\rangle\langle l_\alpha|,
\qquad
\langle l_\alpha|r_\beta\rangle=\delta_{\alpha\beta},
\]
and assuming that the eigenvalue \(\lambda_1\) with largest modulus is non-degenerate, we have in the bulk limit ($k$ and $k+r$ are very far from the boundaries)
\[
\mathbb{E}^{k-2}\simeq \lambda_1^{k-2}|r_1\rangle\langle l_1|,
\qquad
\mathbb{E}^{N-k-r-1}\simeq
\lambda_1^{N-k-r-1}|r_1\rangle\langle l_1|.
\]
As for the denominator, it becomes
\[
\langle L|\mathbb{E}^{N-2}|R\rangle
\simeq
\lambda_1^{N-2}
\langle L|r_1\rangle
\langle l_1|R\rangle .
\]
Therefore,
\[
\langle O_k O'_{k+r}\rangle
=
\sum_{\alpha=1}^{\chi^2}
\left(\frac{\lambda_\alpha}{\lambda_1}\right)^{r-1}
\frac{
\langle l_1|\mathbb{E}_O|r_\alpha\rangle
\langle l_\alpha|\mathbb{E}_{O'}|r_1\rangle
}{\lambda_1^2}.
\]
Similarly,
\[
\langle O_k\rangle
=
\frac{\langle l_1|\mathbb{E}_O|r_1\rangle}{\lambda_1},
\qquad
\langle O'_{k+r}\rangle
=
\frac{\langle l_1|\mathbb{E}_{O'}|r_1\rangle}{\lambda_1}.
\]
The \(\alpha=1\) contribution to the two-point function is
\[
\frac{
\langle l_1|\mathbb{E}_O|r_1\rangle
\langle l_1|\mathbb{E}_{O'}|r_1\rangle
}{\lambda_1^2}
=
\langle O_k\rangle \langle O'_{k+r}\rangle .
\]
Hence, the connected correlation function is
\[
\langle O_k O'_{k+r}\rangle_c
=
\langle O_k O'_{k+r}\rangle
-
\langle O_k\rangle \langle O'_{k+r}\rangle
=
\sum_{\alpha=2}^{\chi^2}
\left(\frac{\lambda_\alpha}{\lambda_1}\right)^{r-1}
\frac{
\langle l_1|\mathbb{E}_O|r_\alpha\rangle
\langle l_\alpha|\mathbb{E}_{O'}|r_1\rangle
}{\lambda_1^2}.
\]
 If the second eigenvalue has degeneracy \(n\), $\lambda_2=\lambda_3=\cdots=\lambda_{n+1}$ and at large distance \(r\) we find
\[
\langle O_k O'_{k+r}\rangle_c
\sim
\left(\frac{\lambda_2}{\lambda_1}\right)^{r-1}
\sum_{\alpha=2}^{n+1}
\frac{
\langle l_1|\mathbb{E}_O|r_\alpha\rangle
\langle l_\alpha|\mathbb{E}_{O'}|r_1\rangle
}{\lambda_1^2}
\]
with corrections of order $\mathcal{O}\!\left[\left(\frac{\lambda_{n+2}}{\lambda_1}\right)^{r-1}\right]$.
The connected correlation decays exponentially with the distance $r$, with a correlation length given by
\begin{equation}
  \xi^{-1}
=
-\log\left|
\frac{\lambda_2}{\lambda_1}
\right|.
\label{eq:correlation_length_MPS}
\end{equation}
This shows that, {\em generically},\footnote{The GHZ state described in \cref{sec:GHZ_state} is an example of (non-injective) MPS where the second eigenvalue of the transfer matrix is degenerate with the first one, leading to long-range correlations.}
an MPS (with finite bond dimension) has exponentially decaying correlations. As will be discussed in \cref{sec:PEPS}, this is not the case for higher-dimensional TNs such as PEPS, which can have algebraically decaying correlations even with finite bond dimension.

\begin{longlisting}
  \jcode{AKLT_transfer_matrix.jl}
  \caption[MPS transfer matrix]{\label{lst:AKLT_transfer_matrix}Code using the \texttt{ITensor} library that constructs the local tensor $A$ of the AKLT MPS (with bond dimension $\chi=2$, see \cref{sec:AKLT} and \cref{sec:AKLT_MPS}), contracts it with its complex conjugate to form the transfer matrix $\mathbb{E}$, reshapes $\mathbb{E}$ as a $\chi^2\times\chi^2$ matrix, and computes the two largest eigenvalues. The modulus of their ratio gives the correlation length of the MPS (\cref{eq:correlation_length_MPS}).
  \jcodegithublink}
\end{longlisting}

\subsection{Matrix product operators (MPO) and local Hamiltonians}
\label{sec:MPO}

\begin{figure}[htbp]
    \centering
    \includegraphics[width=0.942\textwidth]{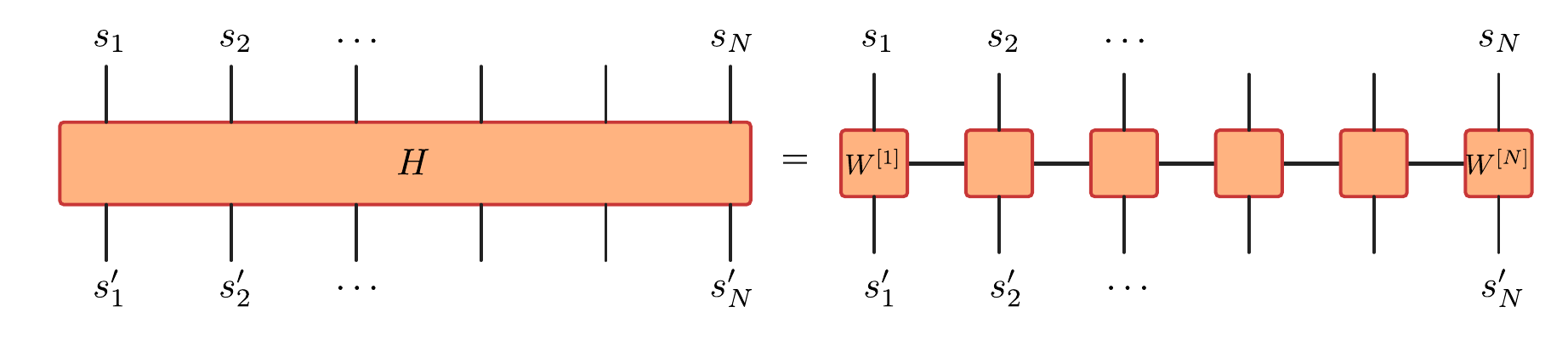}
    \caption{A Hamiltonian (or some generic operator) viewed either as a single tensor or factorized as a matrix product operator (MPO).}
    \label{fig:Hamiltonian_MPO}
\end{figure}

In the same way that a pure state can be represented as an MPS, an operator can be represented as a matrix product operator (MPO)~\cite{mccullochInfiniteSizeDensity2008,crosswhiteFiniteAutomataCaching2008,SCHOLLWOCK_DensitymatrixRenormalizationGroup_2011}.
An MPO is a TN of the same structure as an MPS, but with two physical indices per site, one for the input state and one for the output state. This is illustrated in \cref{fig:Hamiltonian_MPO}. The tensor on the left is the generic representation for a matrix of size $d^N \times d^N$, while the tensor on the right is the factorized MPO representation.
In the same way that any vector of dimension $d^N$ (state $\ket{\psi}$) can be exactly represented by an MPS provided the bond dimension is large enough, any operator acting in the Hilbert space can be represented as an MPO with a large enough bond dimension. The MPO representation can be used for many types of operators: Hamiltonians, observables, unitary evolution operators, quantum gates, density matrices, quantum channels, Lindbladians, etc. In this section we focus on the representation of Hamiltonians as MPOs, which is particularly useful for performing DMRG calculations (\cref{sec:dmrg}) or to simulate time evolution.

For a state $\ket{\psi}$, one generally needs to perform some truncation to get an MPS representation with a reasonable bond dimension (say polynomial in $N$). In contrast, for a Hamiltonian $H$ containing only local interactions, one can find an {\em exact} MPO representation with a bond dimension that is independent of the system size $N$. 

\subsubsection{Example 1: onsite Hamiltonian}
\label{sec:MPO_example_onsite}
The simplest example is a Hamiltonian that is a sum of strictly onsite terms,
\[
H=\sum_{i=1}^N h_i,
\]
where $h_i$ acts only on site $i$.
It can be represented as an MPO with bond dimension $\chi=2$.
In the bulk, one possible local tensor is the following operator-valued matrix,
\[
W^{[1<i<N]} =
\begin{pmatrix}
\mathbbm{1}_i & h_i \\
0 & \mathbbm{1}_i
\end{pmatrix}.
\]
For an open chain, the first and last tensors can be chosen as
\begin{equation}
W^{[1]} =\begin{pmatrix} \mathbbm{1}_1 & h_1 \end{pmatrix}
\qquad , \quad
W^{[N]} =
\begin{pmatrix}
  h_N \\
  \mathbbm{1}_N
  \end{pmatrix}.
\end{equation}
Expanding the product $W^{[1]}\cdots W^{[N]}$ gives paths that stay in the first virtual state, make a single transition to the second virtual state by inserting one of the operators $h_i$, and then stay in the second virtual state. The sum over all such paths is precisely $\sum_i h_i$. Note that this pattern is very similar to that of the MPS representation of the W state in \cref{sec:W_state}.

\subsubsection{Example 2: transverse-field Ising chain}
\label{sec:MPO_example_ising}
As a simple example, consider the transverse-field Ising chain with open boundary conditions,
\[H = J \sum_{i=1}^{N-1} \sigma^z_i \sigma^z_{i+1} +
h \sum_i \sigma^x_i . \]
An MPO for this Hamiltonian can be constructed by letting the virtual index keep track of whether a nearest-neighbor term has already started. In the bulk, one possible local tensor is the following operator-valued matrix,
\[ W^{[2\leq i \leq N-1]} =
\begin{pmatrix}
\mathbbm{1} & J\sigma^z & h \sigma^x \\
0 & 0 & \sigma^z \\
0 & 0 & \mathbbm{1}
\end{pmatrix} = \mathbb{W}.
\]
In the expression above, each entry is an operator acting on the local Hilbert space at site $i$ ($\sigma^z=\sigma^z_i$ etc.)
For an open chain, the first and last tensors can be chosen as
\begin{align}
L^t &= \begin{pmatrix} 1 & 0 & 0 \end{pmatrix},
\qquad
W^{[1]} = L^t \mathbb{W}
=\begin{pmatrix} \mathbbm{1}_1 & J\sigma^z_1 & h \sigma^x_1 \end{pmatrix},
\label{eq:W1}
\\
R &= \begin{pmatrix} 0 \\ 0 \\ 1 \end{pmatrix},
\qquad
W^{[N]} =  \mathbb{W} R
=
\begin{pmatrix}
  h \sigma^x_N \\
  \sigma^z_N \\
  \mathbbm{1}_N
  \end{pmatrix}.
\label{eq:WN}
\end{align}
To understand how the product of the tensors reconstructs $H$, one should think of the virtual index $\alpha\in\{1,2,3\}$ as a virtual state that keeps track of the terms in the Hamiltonian.
This is the so-called finite-state automaton picture~\cite{crosswhiteFiniteAutomataCaching2008}.
Expanding the product $L^t \mathbb{W} \cdots \mathbb{W} R $ amounts to summing over all possible paths
$(\alpha_0, \alpha_1, \dots, \alpha_N)$ in the virtual space, where each transition $\alpha_{i-1} \to \alpha_i$ adds the factor $\mathbb{W}^{[i]}_{\alpha_{i-1},\alpha_i}$ to the operator product.

The choice of the left vector $L$ in \cref{eq:W1} means that the path starts from the first virtual state ($\alpha_0=1$) at site $1$ and \cref{eq:WN} means that the path ends in the third virtual state ($\alpha_N=3$) at site $N$. The matrix $\mathbb{W}$ is designed so that only the two following types of paths contribute to the sum: 
\begin{itemize}
  \item path $\alpha=(1,\dots,1_{i-1},3_i,3,\dots,3)$ $\longrightarrow$ operator $ \mathbbm{1}_1 \otimes \cdots \otimes \mathbbm{1} \otimes h \sigma^x_i \otimes \mathbbm{1} \otimes \cdots \otimes \mathbbm{1}_N$
  \item path $\alpha=(1,\dots,1_{i-1},2_i,3_{i+1},3,\dots,3)$ $\longrightarrow$ operator $\mathbbm{1}_1 \otimes \cdots \otimes \mathbbm{1} \otimes J\sigma^z_i \otimes \sigma^z_{i+1} \otimes \mathbbm{1} \otimes \cdots \otimes \mathbbm{1}_N$. 
\end{itemize}

\subsubsection{Example 3: Heisenberg model}
\label{sec:MPO_example_heisenberg}
A straightforward generalization of the MPO construction above can be used to find an exact MPO representation of the Heisenberg model,
\[ H = J \sum_{i=1}^{N-1} \left( \sigma^x_i \sigma^x_{i+1} + \sigma^y_i \sigma^y_{i+1} + \sigma^z_i \sigma^z_{i+1} \right). \]
The local tensor in the bulk can be chosen as
\[
W^{[2\leq i \leq N-1]} =
\begin{pmatrix}
\mathbbm{1} & J \sigma^x & J \sigma^y & J \sigma^z & 0\\
0 & 0 & 0 & 0 & \sigma^x \\
0 & 0 & 0 & 0 & \sigma^y \\
0 & 0 & 0 & 0 & \sigma^z \\
0 & 0 & 0 & 0 & \mathbbm{1}
\end{pmatrix} = \mathbb{W}.
\]
The first and last tensors can be chosen as
\begin{eqnarray}
L^t &=& \begin{pmatrix} 1 & 0 & 0 & 0 & 0 \end{pmatrix}
\,\,  ,   \,\,
W^{[1]} = L^t \mathbb{W}
= \begin{pmatrix} \mathbbm{1} & J\sigma^x & J\sigma^y & J\sigma^z & 0 \end{pmatrix}
\\
R &=& \begin{pmatrix} 0 \\ 0 \\ 0 \\ 0 \\ 1 \end{pmatrix}
\,\,  ,   \,\,
W^{[N]} = \mathbb{W} R
= \begin{pmatrix} 0 \\ \sigma^x \\ \sigma^y \\ \sigma^z \\ \mathbbm{1} \end{pmatrix}.
\end{eqnarray}
One can check that the above tensors reconstruct the Heisenberg Hamiltonian by realizing that the possible $\alpha_i$-paths are of the following types:
\begin{itemize}
  \item path $\alpha=(1,\dots,1_{i-1},2_i,5_{i+1},5,\dots,5)$ $\longrightarrow$ operator $\mathbbm{1}_1 \otimes \cdots \otimes \mathbbm{1} \otimes J\sigma^x_i \otimes \sigma^x_{i+1} \otimes \mathbbm{1} \otimes \cdots \otimes \mathbbm{1}_N$
  \item path $\alpha=(1,\dots,1_{i-1},3_i,5_{i+1},5,\dots,5)$ $\longrightarrow$ operator $\mathbbm{1}_1 \otimes \cdots \otimes \mathbbm{1} \otimes J\sigma^y_i \otimes \sigma^y_{i+1} \otimes \mathbbm{1} \otimes \cdots \otimes \mathbbm{1}_N$
  \item path $\alpha=(1,\dots,1_{i-1},4_i,5_{i+1},5,\dots,5)$ $\longrightarrow$ operator $\mathbbm{1}_1 \otimes \cdots \otimes \mathbbm{1} \otimes J\sigma^z_i \otimes \sigma^z_{i+1} \otimes \mathbbm{1} \otimes \cdots \otimes \mathbbm{1}_N$.
\end{itemize}

\subsubsection{Example 4: 2-qubit gate}
\label{sec:MPO_two_qubit_gate}

\begin{figure}[h]
  \centering
  \includegraphics[width=0.942\textwidth]{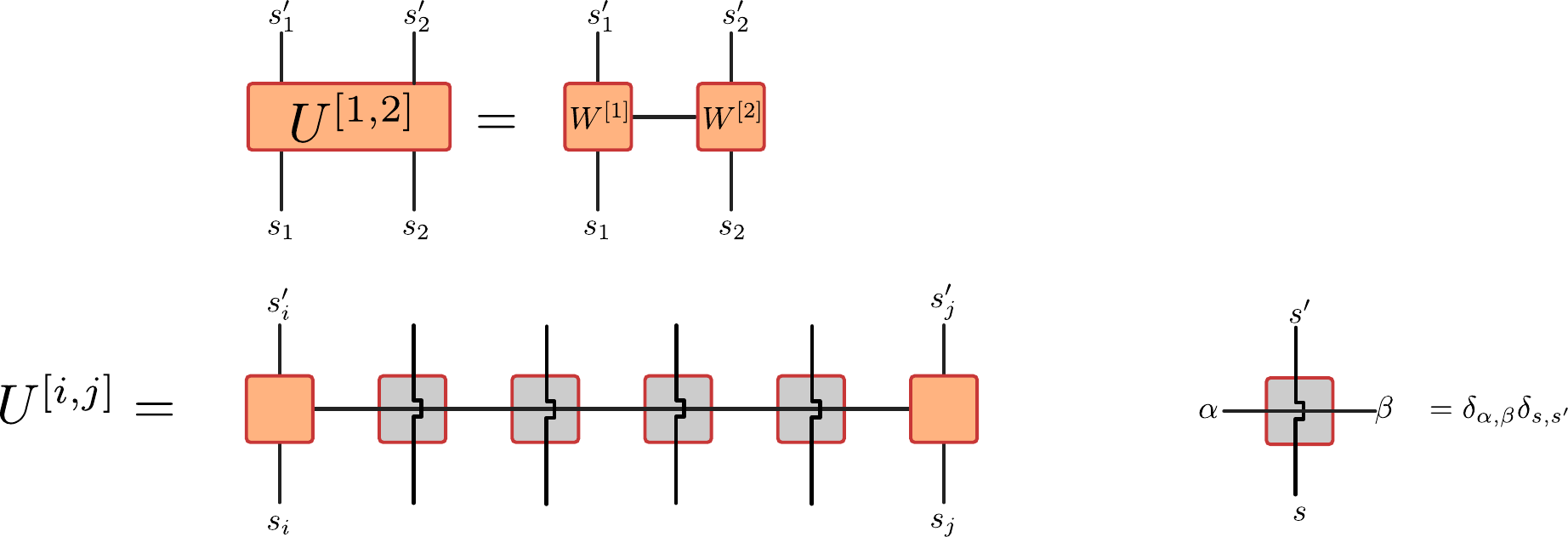}
  \caption{MPO representations of a two-qubit gate $U$ acting on neighboring sites (top) or on non-neighboring sites (bottom).
  The diagonal tensors (gray) simply propagate the virtual index from site $i$ to site $j$ while acting as the identity in the physical space.
  }
  \label{fig:MPO_2_qubit_gate}
\end{figure}

A two-qubit gate $U$ acting on sites $i$ and $j$ is a $d^2 \times d^2$ matrix, and it can be factorized using a QR factorization or an SVD.
If $U$ acts on neighboring sites, such factorization means that $U$ can be represented as an MPO with bond dimension (at most) $d^2$.
This is represented in the upper part of \cref{fig:MPO_2_qubit_gate}.
If $U$ acts on non-neighboring sites, one can still find an MPO representation by inserting diagonal tensors on the sites between $i$ and $j$ to propagate the virtual index from site $i$ to site $j$.
These tensors act as the identity in the physical space and in the virtual space. This is represented in the lower part of \cref{fig:MPO_2_qubit_gate}.

\subsubsection{Finite-state automaton picture}
\label{sec:MPO_finite_state_automata}

We describe a general method to construct an MPO representation of a Hamiltonian based on the finite-state automaton picture~\cite{crosswhiteFiniteAutomataCaching2008}.
We follow the presentation given in Ref.~\cite{zaletelTimeevolvingMatrixProduct2015}.

Consider a bond $(i,i+1)$. The terms in the Hamiltonian can be classified into three types:
\begin{itemize}
  \item terms that act on sites $1,2,\dots,i$ only (type L)
  \item terms that act on sites $i+1,i+2,\dots,N$ only (type R)
  \item the terms of the form $h_{L,\alpha} \otimes h_{R,\alpha}$ that act on both sides of the bond (type C)
\end{itemize}
This corresponds to the following expression for the Hamiltonian:
\begin{equation}
H = H_{L}(i) \otimes \mathbbm{1}_{[i+1,N]} + \mathbbm{1}_{[1,i]} \otimes H_{R}(i) + \sum_{\alpha=1}^k h_{L,\alpha}(i) \otimes h_{R,\alpha}(i).
\end{equation}
$H_{L}(i)$ is the sum of all terms of type L, $H_{R}(i)$ is the sum of all terms of type R, and $h_{L,\alpha}(i) \otimes h_{R,\alpha}(i)$ are the $k$ terms of type C.
For instance, a purely onsite Hamiltonian corresponds to $k=0$.
For a nearest-neighbor Ising Hamiltonian, $k=1$ (see \cref{sec:MPO_example_ising}).
An Ising model in transverse field with both nearest-neighbor and next-nearest neighbor interactions has $k=2$.
For a Heisenberg Hamiltonian, $k=3$ (see \cref{sec:MPO_example_heisenberg}). 
In principle $k$ can be different for each site (at least close to the chain boundaries); here, to simplify the notation, we will ignore
this dependence and simply write $k$ instead of $k(i)$.

The next step is to realize that the decomposition across the bond ($i-1,i$) is related to that across the next bond ($i,i+1$) through a recursion relation.
This recursion relation takes the form
\begin{equation}
\begin{pmatrix} H_R(i-1) \\ h_{R}(i-1) \\ \mathbbm{1}_{[i,N]} \end{pmatrix}
=
\begin{pmatrix}
\mathbbm{1} & C^{[i]} & D^{[i]} \\
0 & A^{[i]} & B^{[i]} \\
0 & 0 & \mathbbm{1}
\end{pmatrix} \begin{pmatrix} H_R(i) \\ h_{R}(i) \\ \mathbbm{1}_{[i+1,N]} \end{pmatrix}.
\label{eq:recursion_relation}
\end{equation}
The matrix $A^{[i]}$ is a $k\times k$ matrix,
$B^{[i]}$ is a $k\times 1$ matrix,
$C^{[i]}$ has dimension $1\times k$, $D^{[i]}$ is $1 \times 1$.
In total, the local MPO tensor has bond dimension $\chi=k+2$, and the block matrix above has size $(k+2)\times(k+2)$ in the virtual indices.
$D^{[i]}$ encodes the terms in the Hamiltonian that act on site $i$ only, $C^{[i]}$ encodes the terms that start at site $i$, $B^{[i]}$ encodes the terms that end at site $i$, and $A^{[i]}$ encodes the terms that are propagating through site $i$.
It turns out that this matrix is precisely the local tensor of the MPO representation of $H$ at site $i$: 
\begin{equation}
  W^{[i]} =
\begin{pmatrix}
\mathbbm{1} & C^{[i]} & D^{[i]}\\
0 & A^{[i]} & B^{[i]}\\
0 & 0 & \mathbbm{1}
\end{pmatrix}.
\label{eq:general_H_MPO}
\end{equation}
As in the examples of \cref{sec:MPO},  the first and last values of the virtual index $\alpha$ have a special meaning
in the finite-state automaton picture: $\alpha=1$ means that no term of the Hamiltonian has started yet, while $\alpha=\chi$ means that all terms have already ended. The $k$ intermediate values of $\alpha$ keep track of the terms that are currently being added to the operator product. 

A short example is given in \cref{lst:MPO_transverse_field_ising_finite_range}. The Hamiltonian is specified as an \texttt{OpSum}, and the call \texttt{MPO(os, sites)} constructs the corresponding MPO automatically.
In the context of mixed states, the same finite-state automaton picture can be used to construct MPO representations of Lindbladians, as discussed in \cref{sec:lindblad_MPS_form}.

\begin{longlisting}
  \jcode{MPO_transverse_field_ising_finite_range.jl}
  \caption[Finite-range transverse-field Ising Hamiltonian as an MPO]{\label{lst:MPO_transverse_field_ising_finite_range}Transverse-field Ising Hamiltonian on an open chain using \texttt{ITensors}. The Ising couplings $J_1,J_2,\dots,J_k$ act at distances $1,2,\dots,k$, and a local transverse field is included. The Hamiltonian is first defined as an \texttt{OpSum}, then converted to an MPO. The code prints the MPO bond dimensions on all bonds.
  In the example shown, the couplings are chosen as a truncated power-law profile \(J_r\) up to range \(k\). The printed bond dimensions illustrate how the finite interaction range is encoded in the MPO auxiliary space.
  \jcodegithublink}
\end{longlisting}

\subsubsection{Multiplication of an MPO by an MPS and zip-up algorithm}
\label{sec:zipup}

\begin{figure}[h]
  \centering
  \includegraphics[width=1.\textwidth]{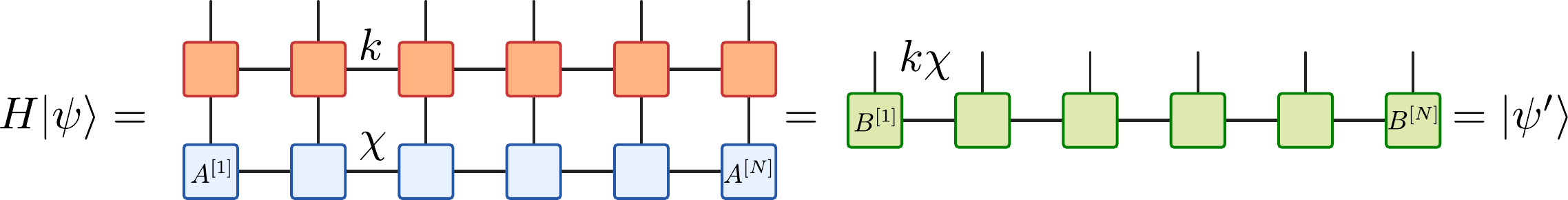}
  \caption{Exact multiplication of an MPO (bond dimension $\chi_{\rm MPO}$) by an MPS (bond dimension $\chi$). The resulting MPS for $\ket{\psi'}=H\ket{\psi}$ has bond dimension $\chi'=\chi\,\chi_{\rm MPO}$.}
  \label{fig:MPO_MPS_multiplication}
\end{figure}

Given an operator $H$ (not necessarily a Hamiltonian) represented as an MPO and a pure state $\ket{\psi}$ represented as an MPS, one can compute the state $\ket{\psi'} = H \ket{\psi}$ by contracting the corresponding TN, as shown in \cref{fig:MPO_MPS_multiplication}. If such multiplication is performed exactly, the resulting MPS for $\ket{\psi'}$ has a bond dimension that is the product of the bond dimensions of the MPO and the MPS. Since one often has many such MPS-MPO multiplications to perform, and the bond dimension of the resulting MPS grows exponentially with the number of multiplications, one usually wants to find an approximate MPS representation of $\ket{\psi'}$ with a smaller bond dimension $\chi' < \chi_{\rm MPO}\chi$, where $\chi_{\rm MPO}$ is the MPO bond dimension.

\begin{figure}[h]
  \centering
  \includegraphics[width=0.5\textwidth]{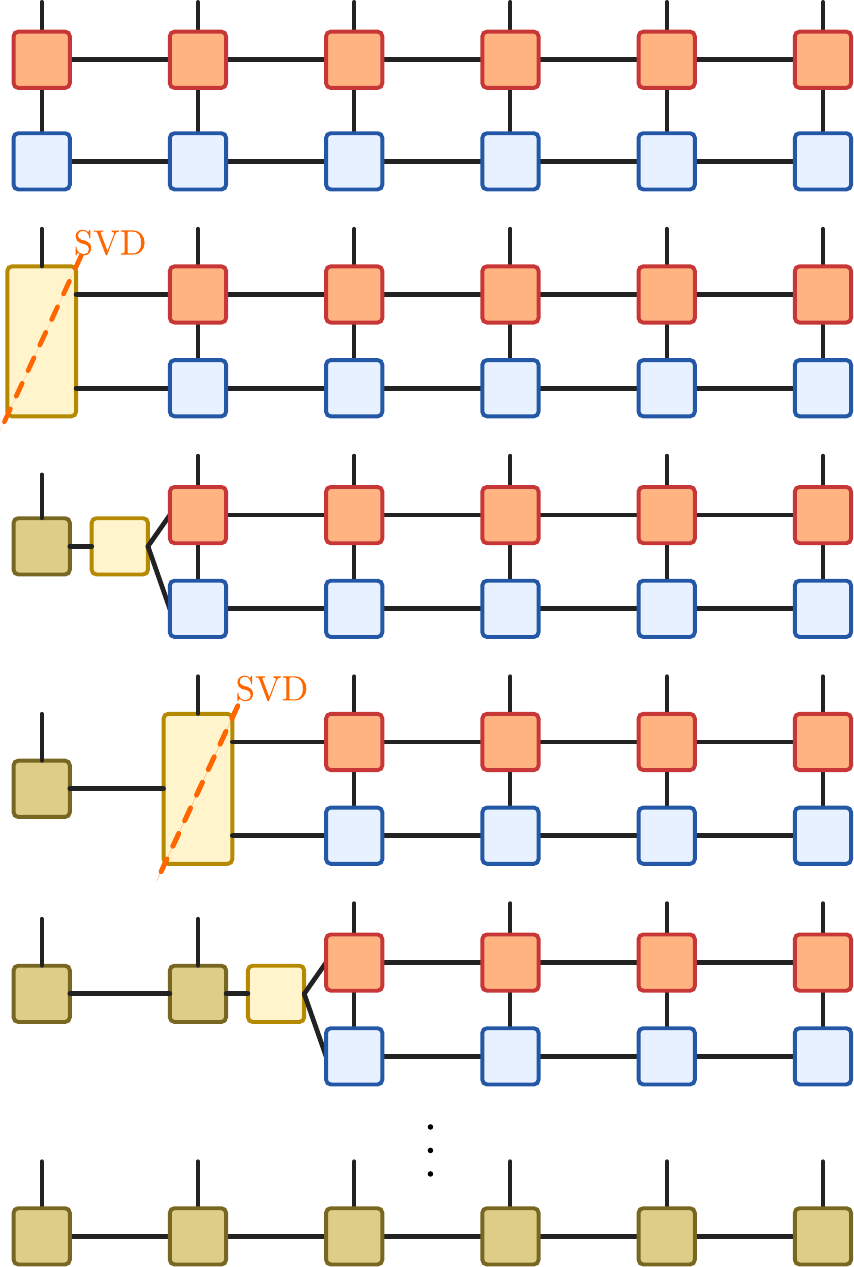}
  \caption{Schematic representation of the zip-up algorithm. The MPO and MPS tensors are contracted and factorized (SVD plus truncation) successively from left to right.}
  \label{fig:zipup_algorithm}
\end{figure}

One possible way to do this is to perform the multiplication and the truncation at the same time, thus computing directly the new truncated tensors of $\ket{\psi'}$. One algorithm to do this is the zip-up algorithm~\cite{STOUDENMIRE_MinimallyEntangledTypical_2010}, illustrated in \cref{fig:zipup_algorithm}.
In the zip-up algorithm one starts from one end of the chain, contracts the first MPO tensor with the first MPS tensor, and factorizes the result by an SVD. The singular values are truncated, and the remaining factor is absorbed into the next site. Repeating this procedure site by site produces an approximate MPS for \(H|\psi\rangle\) without ever forming the MPS that would result from the exact multiplication. 

It is also possible to perform some variational optimization to find the best MPS approximation to $H\ket{\psi}$ with a given bond dimension $\chi'$.
We are thus looking for the MPS $\ket{\psi'}$ with given bond dimension that minimizes the distance $\| H\ket{\psi} - \ket{\psi'} \|^2 = \bra{\psi} H^\dagger H \ket{\psi} - \langle \psi'|H|\psi\rangle - \langle\psi|H^\dagger|\psi'\rangle + \langle \psi'|\psi'\rangle$.
This can be done by optimizing one tensor at a time while keeping the others fixed, and sweeping through the lattice until convergence. 
Both methods can also be combined: one can first perform a zip-up to get an initial guess for the new MPS tensors, and then perform a variational optimization to further improve the approximation.

\subsection{DMRG algorithm}
\label{sec:dmrg}

The Density Matrix Renormalization Group (DMRG) algorithm was introduced by S. White in 1992~\cite{whiteDensityMatrixFormulation1992,WHITE_DensitymatrixAlgorithmsQuantum_1993} and is one of the most powerful numerical methods for computing the ground state of 1d (and more generally low-dimensional) quantum systems. Although it was not formulated in this way in the original paper, the DMRG algorithm is essentially a variational method in the space of MPS, where the tensors $A^{[k]\,s_k}$ are optimized to minimize the energy. The optimization is performed iteratively by sweeping through the lattice and updating one or two tensors at a time while keeping the others fixed. The DMRG algorithm can also be used to find low-energy excited states by targeting states that are orthogonal to the ground state. It can also be used to perform certain contractions, such as the multiplication of an MPO.
Some extensions of the DMRG algorithm can also be used on loop-free TNs such as TTN~\cite{TAGLIACOZZO_SimulationTwodimensionalQuantum_2009,SILVI_TensorNetworksAnthology_2019}. 

We give here a brief sketch of the two-site DMRG algorithm in the case where the Hamiltonian takes the form of an MPO. The two-site algorithm is slightly more complicated than the single-site DMRG algorithm, but it is often used in practice because it allows one to increase the bond dimension of the MPS during the optimization.

\begin{figure}
  \begin{center}
    \includegraphics[width=0.9\textwidth]{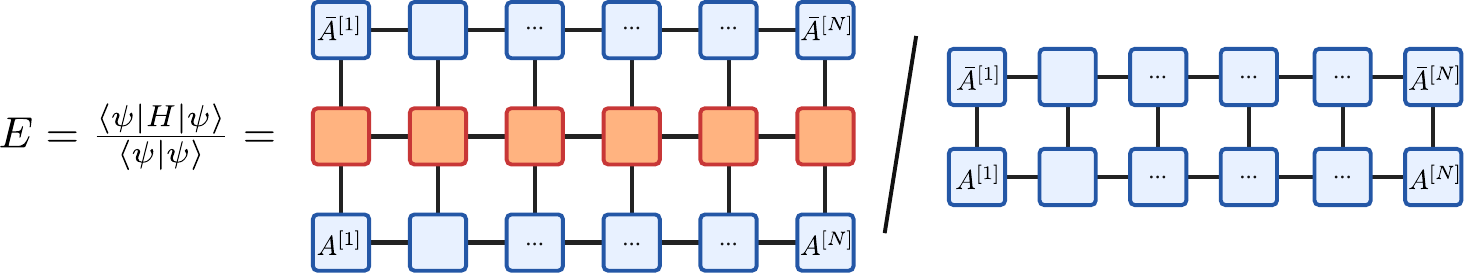}
    \caption{
      Energy expectation value $E=\bra{\psi} H \ket{\psi} / \bra{\psi}\ket{\psi}$ as a TN. The state $\ket{\psi}$ is in an MPS form (tensors $A^{[i]}$) and the Hamiltonian is expressed as an MPO (orange tensors). $E$ is    
      iteratively minimized in the DMRG algorithm by optimizing one or two tensors of the MPS at a time while keeping the others fixed (see \cref{fig:dmrg-2}).
    }
    \label{fig:dmrg-1}
  \end{center}
\end{figure}

\begin{figure}
  \begin{center}
    \includegraphics[width=0.9\textwidth]{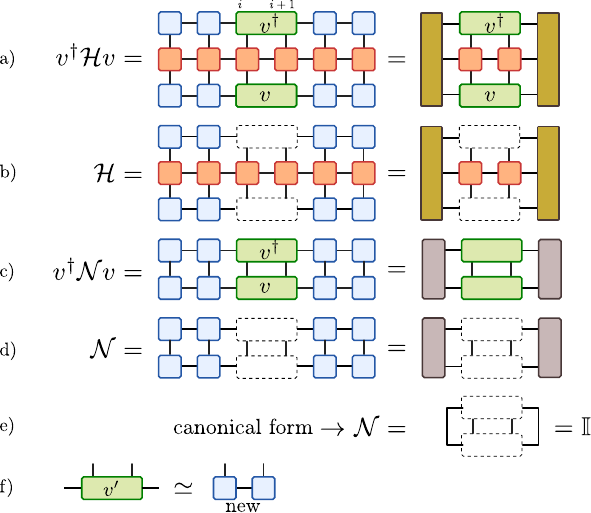}
    \caption{
      a) Expectation value $\bra{\psi} H \ket{\psi}$ as a quadratic function of the tensors $A^{[i]}$ and $A^{[i+1]}$. The vector $v$ contains the coefficients of the product $A^{[i]}A^{[i+1]}$. The vector $v$ has dimension $d^2 \chi^2$, where $d$ is the physical dimension of the local Hilbert space and $\chi$ is the bond dimension of the MPS.
      b) Matrix $\mathcal{H}$ of the effective Hamiltonian for the optimization of the tensors at sites $i$ and $i+1$. $\mathcal{H}$ has dimension $d^2 \chi^2 \times d^2 \chi^2$. 
      c) Norm $\braket{\psi}{\psi}$ as a quadratic function of $v=A^{[i]}A^{[i+1]}$.
      d) Matrix $\mathcal{N}$ of the effective norm for the optimization of the tensors at sites $i$ and $i+1$.
      e) If the MPS is in a mixed-canonical form with the orthogonality center at site $i$ or $i+1$,
      the matrix $\mathcal{N}$ reduces to the identity matrix, and the optimization reduces to a standard eigenvalue problem $\mathcal{H} v = E v$. In this case, the optimal $v'$ is given by the eigenvector associated with the lowest eigenvalue $E$ of $\mathcal{H}$. This lowest eigenvector is usually computed numerically using an iterative eigensolver such as the Lanczos algorithm.
      f) Once the optimal $v'$ is found, it is reshaped back into the two-site tensor and then factorized into two tensors $A^{[i]}$ and $A^{[i+1]}$ using an SVD (and truncating the smallest singular values if needed to get the desired bond dimension). 
    }
    \label{fig:dmrg-2}
  \end{center}
\end{figure}

The starting point is the energy expectation value $E=\bra{\psi} H \ket{\psi} / \bra{\psi}\ket{\psi}$, represented as a TN in \cref{fig:dmrg-1}. In the two-site version, one focuses on two neighboring MPS tensors, say $A^{[i]}$ and $A^{[i+1]}$, while all the other tensors are kept fixed. The two tensors are contracted together and reshaped into a single vector $v$ collecting all variational parameters of this local update, as shown in \cref{fig:dmrg-2}.
This vector $v$ has dimension $d^2 \chi^2$, where $d$ is the physical dimension of the local Hilbert space and $\chi$ is the bond dimension of the MPS. 

Once the rest of the TN has been contracted, both the numerator and the denominator of the Rayleigh quotient become quadratic forms in $v$:
\begin{equation}
  \bra{\psi} H \ket{\psi} = v^\dagger \mathcal{H} v,
  \qquad
  \langle\psi|\psi\rangle = v^\dagger \mathcal{N} v ,
\end{equation}
where $\mathcal{H}$ is the effective Hamiltonian matrix and $\mathcal{N}$ is the effective norm matrix for the two optimized sites.

The minimization can therefore be written as a constrained variational problem. Introducing a Lagrange multiplier $\lambda$ to impose $\langle\psi|\psi\rangle=1$, one has to extremize
\begin{equation}
  v^\dagger \mathcal{H} v
  - \lambda \left( v^\dagger \mathcal{N} v - 1 \right).
\end{equation}
Taking the derivative with respect to $v^\dagger$ gives the generalized eigenvalue problem
\begin{equation}
  \mathcal{H} v = \lambda \mathcal{N} v .
\end{equation}
The optimal local tensor is obtained from the eigenvector associated with the lowest eigenvalue.

If the MPS is brought into the appropriate mixed-canonical form, with the orthogonality center on the optimized bond, the effective norm matrix is simply the identity, $\mathcal{N}=\mathbbm{1}$. The generalized eigenvalue problem then reduces to the standard one,
\begin{equation}
  \mathcal{H} v = \lambda v .
\end{equation}

The dimension of the matrix $\mathcal{H}$ is $(d^2 \chi^2) \times (d^2 \chi^2)$, which can be quite large. For large bond dimensions $\mathcal{H}$ becomes too large to be stored explicitly and fully diagonalized.\footnote{Simulations often involve bond dimension $\chi$ of the order of a few hundreds, or even a few thousands. State-of-the-art DMRG calculations can reach $\chi$ of the order of $10^4$ or more.}
For this reason, and also because one is only interested in the lowest eigenvalue,
the eigenvalue problem is solved with an iterative eigensolver such as the Lanczos algorithm. Such an algorithm does not require the explicit construction and storage of the full matrices but only requires the action of $\mathcal{H}$ and $\mathcal{N}$ on trial vectors, i.e., the ability to compute $v'=\mathcal{H}v$ and $v''=\mathcal{N}v$ for given $v$.  This can be done by contracting the corresponding effective TN.

Once the optimal $v$ is found, it is reshaped back into the two-site tensor. Then, an SVD is performed (with truncation of the smallest singular values if needed) to split the two-site tensor into two tensors $A^{[i]}$ and $A^{[i+1]}$ with the desired new bond dimension. After the update of the two tensors, one moves to the next pair of sites and repeats the process until convergence.

Two minimal DMRG code examples are given in \cref{lst:dmrg_spin_half_heisenberg} and \cref{lst:dmrg_AKLT} for spin chain models, the spin-$\frac{1}{2}$ Heisenberg model and the spin-1 AKLT model~\cite{AFFLECK_RigorousResultsValencebond_1987}. In the second example the exact ground state is known and can be represented as an MPS with bond dimension $\chi=2$ (see \cref{sec:MPS_simple_examples}).
Another example is given in \cref{lst:dmrg_free_fermion} for a noninteracting spinless fermion tight-binding model. There, the exact ground-state energy can be computed analytically and compared to the DMRG result. The energy variance of the DMRG state is also computed, which gives an estimate of how close the DMRG state is to an eigenstate of the Hamiltonian (by construction this energy variance vanishes for an exact Hamiltonian eigenstate).

\begin{longlisting}
  \jcode{dmrg_spin_half_heisenberg.jl}  
  \caption[DMRG for the spin-$\frac{1}{2}$ Heisenberg chain -- energy and entanglement entropy]{\label{lst:dmrg_spin_half_heisenberg}Short DMRG code to find the ground state of a spin-$\frac{1}{2}$ Heisenberg chain. The Hamiltonian is defined as a sum of two-site terms, and the initial state is a random MPS with a small bond dimension. The DMRG algorithm is then used to optimize the MPS representation of the ground state, with increasing bond dimensions and a fixed cutoff for the singular values.
  \texttt{H = MPO(os,sites)} converts the user-friendly \texttt{OpSum} representation of the Hamiltonian into an MPO, which is the format required by the DMRG algorithm.
  \texttt{energy,psi = dmrg(H,psi0;nsweeps,maxdim,cutoff)} performs the DMRG sweeps.
  The maximum bond dimension for each sweep is specified by \texttt{maxdim = [10,20,100,100,200]}; on the last sweep, the maximum bond dimension is $\chi=200$. 
  The last part of the code computes the entanglement entropy for a bipartition in the center of the chain (between sites $i=N/2$ and site $i+1$),
  which grows logarithmically with the system size for critical systems~\cite{holzheyGeometricRenormalizedEntropy1994a,vidalEntanglementQuantumCritical2003a,calabreseEntanglementEntropyQuantum2004}.
  One first moves the orthogonality center to the site $i=N/2$
  (\texttt{orthogonalize!(psi, i)})  
  and then
  (\texttt{U, S, V = svd(psi[i], (linkind(psi, i - 1), siteind(psi, i)))})
  reshapes the MPS tensor at site $i$ into a matrix by grouping the left index
  (\texttt{linkind(psi, i - 1)})
    and physical index
  (\texttt{siteind(psi, i)})
    together and keeping the right index as the second (column) index of the matrix.
  The SVD of $M$ gives the singular values
  (\texttt{S}), which are needed to compute the von Neumann entanglement entropy.
  \jcodegithublink}
\end{longlisting}

\begin{longlisting}
  \jcode{dmrg_AKLT.jl}
  \caption[DMRG for the spin-1 AKLT chain]{\label{lst:dmrg_AKLT}DMRG code for the spin-1 AKLT Hamiltonian~\cite{AFFLECK_RigorousResultsValencebond_1987}, whose exact ground state is an MPS with bond dimension $\chi=2$ (see \cref{sec:AKLT,sec:AKLT_MPS}).
  The Hamiltonian is first defined as an \texttt{OpSum} (lines with \texttt{os +=}) and then converted into an MPO
  (\texttt{H = MPO(os, sites)}).
  The DMRG algorithm is then used to minimize the energy with a given number of sweeps (\texttt{nsweeps}) and maximum bond dimension (\texttt{maxdim}). The Hamiltonian used in the code is \(H_{\rm code}=J\sum_i\left[\vec S_i\cdot\vec S_{i+1}+\frac{1}{3}(\vec S_i\cdot\vec S_{i+1})^2\right]\), which is related to the projector Hamiltonian in \cref{sec:AKLT} by \(H_{\rm code}=2J H_{\rm proj}-\frac{2J}{3}(N-1)\). Thus the same ground state has energy zero for \(H_{\rm proj}\) and energy \(-2J/3\) per bond in the code.
  \jcodegithublink}
\end{longlisting}

\begin{longlisting}
  \jcode{dmrg_FF.jl}
  \caption[DMRG for 1d free fermions]{\label{lst:dmrg_free_fermion}DMRG for a noninteracting spinless fermion 1d tight-binding model (open boundary conditions). The code compares the DMRG estimate of the ground-state energy with the exact analytical result, which can be computed by filling the lowest single-particle energy levels. The code also computes the energy variance of the DMRG state, which gives an estimate of how close the DMRG state is to an eigenstate of the Hamiltonian.
    The code uses fermionic sites with conserved quantum numbers, starts from a half-filled product state, and performs DMRG sweeps with increasing values of \texttt{maxdim}.
    \jcodegithublink
  }
\end{longlisting}

\subsection{Time evolution with MPS}
\label{sec:time_evol_MPS}

The simulation of the time evolution of quantum many-body systems is a central domain of application of MPS and TN methods in general. 
The first MPS-based time-evolution algorithm is usually traced back to Vidal's TEBD approach, introduced in 2003~\cite{vidalEfficientClassicalSimulation2003} and formulated explicitly for 1d quantum many-body systems in 2004~\cite{VIDAL_EfficientSimulationOneDimensional_2004}. Closely related adaptive tDMRG formulations appeared shortly afterwards~\cite{DALEY_TimedependentDensitymatrixRenormalizationgroup_2004,WHITE_RealTimeEvolutionUsing_2004}.
For a review on time-evolution algorithms for MPS, see Paeckel {\it et al.}~\cite{PAECKEL_TimeevolutionMethodsMatrixproduct_2019}.

In the following we give a brief presentation of the simplest time-evolution algorithm, the TEBD algorithm.

\subsubsection{Time-evolving block decimation (TEBD) algorithm}
\label{sec:TEBD}

\begin{figure}
  \begin{center}
    \includegraphics[width=0.411\textwidth]{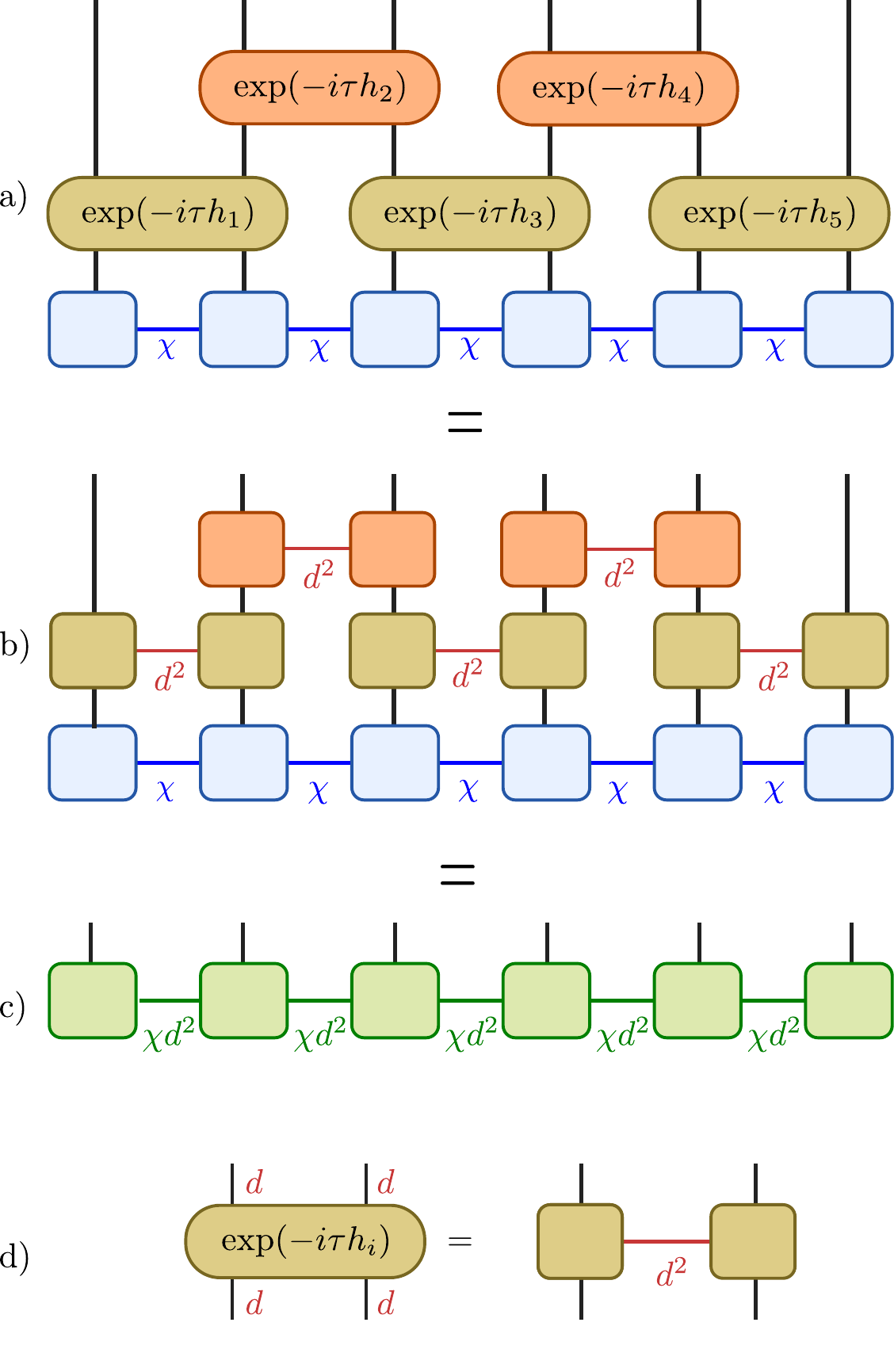}
    \caption{
      Time-evolution of an MPS using the TEBD algorithm.
      a) The time-evolution operator is decomposed into a product of two-site gates $e^{-i h_{i} \tau}$, which can be applied in parallel on the even and odd bonds.
      b) Using the fact that each $e^{-i h_{i} \tau}$ is a two-site operator, the SVD of the associated
      $d^2 \times d^2$ matrix can be used to write it as a product of two tensors with bond dimension at most $d^2$ [as shown in d)].
      c) After one step the MPS bond dimension has increased to (at most) $\chi' = \chi d^2$. After a time step one generally needs to truncate the bond dimension of the MPS back to $\chi$ by keeping only the $\chi$ largest singular values.
    }\label{fig:TEBD}
  \end{center}
\end{figure}

Let us assume for simplicity that the (1d) Hamiltonian of the system is a sum of nearest-neighbor terms:
\begin{equation}
  H = \sum_{i=1}^{N-1} h_{i},
\end{equation}
where $h_i$ acts on sites $i$ and $i+1$ only.
The time evolution operator $U(t) = e^{-iHt}$ can be approximated using a Trotter-Suzuki decomposition. For instance, at lowest order, one has
\begin{equation}
  U(\tau) \approx e^{-i h_{1} \tau} e^{-i h_{2} \tau}\cdots e^{-i h_{N-1} \tau} 
  +\mathcal{O}(\tau^2).
\end{equation}
At this level of approximation the order of the factors does not matter. For notational simplicity, take $N$ even; then we can group the even and the odd terms as
\begin{equation}
  U(\tau) \approx \left( \prod_{i=1}^{N/2} e^{-i h_{2i-1} \tau} \right) \left( \prod_{i=1}^{N/2-1} e^{-i h_{2i} \tau} \right) +\mathcal{O}(\tau^2).
\end{equation}
The $h_{2i}$ (even bonds) commute with each other, and the $h_{2i-1}$ (odd bonds) commute with each other (but $[h_{2i},h_{2i-1}]\neq 0$ in general). Therefore, the $e^{-i h_{2i} \tau}$ can be applied in parallel, and the $e^{-i h_{2i-1} \tau}$ can also be applied in parallel.
In other words:
\begin{equation}
  U(\tau) \approx e^{-i\tau H_{\rm odd}} e^{-i\tau H_{\rm even}} 
  +\mathcal{O}(\tau^2), 
\end{equation}
where $H_{\rm odd} = \sum_{i=1}^{N/2} h_{2i-1}$ and $H_{\rm even} = \sum_{i=1}^{N/2-1} h_{2i}$.
This is illustrated in \cref{fig:TEBD}.

It is also possible to decompose each time step into three sub-steps,
\begin{equation}
   U(\tau) \approx e^{-i\tau H_{\rm odd}/2} e^{-i\tau H_{\rm even}} e^{-i\tau H_{\rm odd}/2} +\mathcal{O}(\tau^3),
\end{equation}
which leads to a smaller finite-step error at no extra computational cost (since the operator layer $e^{-i\tau H_{\rm odd}/2}$ of one step is merged with the operator layer $e^{-i\tau H_{\rm odd}/2}$ of the next step).
The same logic can be used to construct higher-order decompositions. It is also possible to generalize the method to more complicated Hamiltonians, where the decomposition of $H$ as a sum of Hamiltonians $H_i$ in which all terms commute requires more than two terms~\cite{SCHOLLWOCK_DensitymatrixRenormalizationGroup_2011,PAECKEL_TimeevolutionMethodsMatrixproduct_2019}.

The algorithm above also works for imaginary time evolution, which can be used to find the ground state of a Hamiltonian by applying $e^{-\beta H}$ to an initial state and taking the limit $\beta \to +\infty$. However, this is not the most efficient way to find the ground state, and the DMRG algorithm is usually preferred for this purpose.

\subsubsection{Time-dependent variational principle (TDVP) algorithm}
\label{sec:TDVP}

Another important class of time-evolution algorithms for MPS is based on the time-dependent variational principle (TDVP).
TDVP is a general strategy for approximating dynamics inside a given variational manifold. Instead of evolving the state in the full Hilbert space and projecting afterwards, one projects the Schrödinger equation itself onto the {\em tangent space of the variational manifold}, obtaining equations of motion for the variational parameters. Let $P_{T[\ket{\psi}]}$ be the projector onto the tangent space of the variational manifold at the point $\ket{\psi}$. The TDVP equation reads
\begin{equation}
  i \frac{d}{dt} \ket{\psi} = P_{T[\ket{\psi}]} H \ket{\psi}.
\end{equation}
The solution of this equation stays by construction in the variational manifold, and this approach is not specific to TNs. In the context of MPS, however, TDVP provides a powerful time-evolution algorithm~\cite{haegemanTimeDependentVariationalPrinciple2011,haegemanUnifyingTimeEvolution2016}.

The simplest version is the one-site TDVP algorithm, which evolves within the manifold of MPS with a fixed bond dimension.
The method is applicable for arbitrary Hamiltonians expressed as MPOs, and it guarantees energy conservation during the evolution.
A two-site extension can also be used; it allows the bond dimension to grow dynamically during the evolution, which is necessary when the entanglement increases. In that case, one uses a projector onto the linear space of {\em two-site} variations. The evolved two-site tensor is then split into two tensors using an SVD to obtain the new MPS tensors with the desired new bond dimension.
This is similar to the two-site DMRG algorithm.

These algorithms are more complicated than TEBD and will not be presented in detail here.
An example of code based on the TDVP implementation in ITensor is given, however, in \cref{lst:tdvp_XX_domain_wall}.

\begin{longlisting}
  \jcode{XX_spin_chain_domain_wall.jl}
  \caption[TDVP for domain-wall melting in XX spin chain]{\label{lst:tdvp_XX_domain_wall}Time evolution of a domain-wall initial state in the XX spin-$\frac{1}{2}$ chain~\cite{antalTransportXXChain1999} using the two-site TDVP algorithm implemented in the \texttt{ITensor} library. The code compares the TDVP result with the exact solution obtained by mapping the XX chain to free fermions.
    The exact reference curve is obtained by diagonalizing the single-particle free-fermion Hamiltonian. The script evolves the domain-wall state up to the time \texttt{t}, plots \(\langle\sigma_i^z(t)\rangle\), and prints the maximum pointwise deviation between TDVP and the exact result.
    The figure generated by the code is displayed in \cref{fig:tdvp_XX_domain_wall}.
    \jcodegithublink
  }
\end{longlisting}

\begin{figure}
  \begin{center}
    \includegraphics[width=0.6\textwidth]{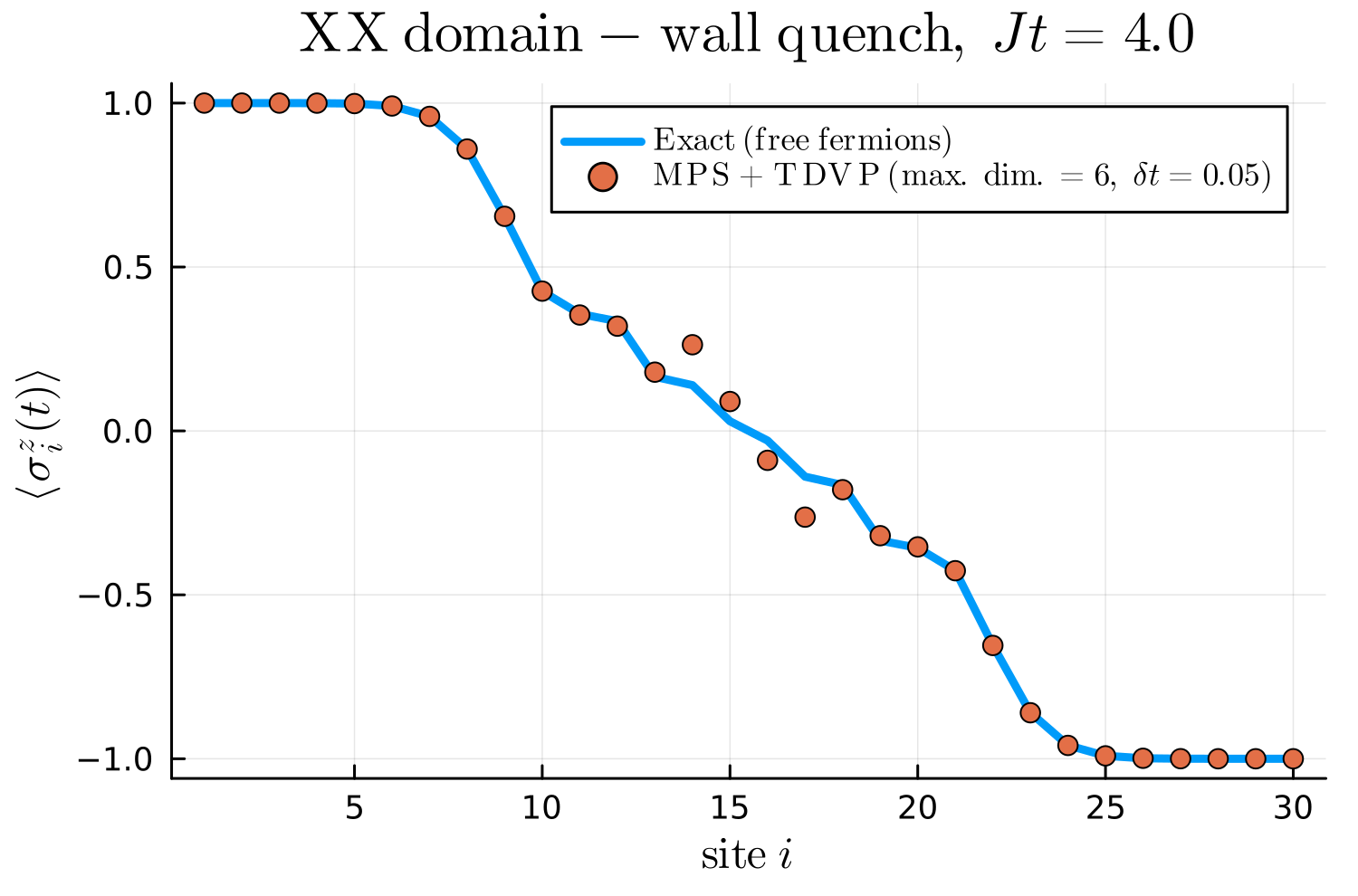}
    \caption{Magnetization profile for the domain wall quench in the XX spin-$\frac{1}{2}$ chain: exact versus TDVP results.
      Output of the code given in \cref{lst:tdvp_XX_domain_wall}.}
    \label{fig:tdvp_XX_domain_wall}
  \end{center}
\end{figure}

\subsubsection[MPO methods (W I and W II)]{MPO methods (${\rm W}^{I}$ and ${\rm W}^{II}$)}
\label{sec:W}

We mention here another family of time-evolution algorithms for MPS. The idea is to construct an MPO $W$ which approximates the time-evolution operator
$U=e^{t H}$
where $t$ is a small (possibly complex) evolution parameter and where $H$ is given as an MPO.
We assume that $H$ is extensive and write it, as in \cite{zaletelTimeevolvingMatrixProduct2015}, as a sum $H=\sum_x H_x$ where each term is labeled by a site $x$. Each term $H_x$ acts only on a finite region around $x$, whose range we denote by $k$.

A simple possibility would be to write $U\simeq \mathbbm{1}+t\sum_x H_x$, since this has a simple MPO representation as a sum of two MPOs.
In terms of the time step $t$, the error would scale as $\mathcal{O}(t^2)$, but the number of second-order terms discarded in this approximation is quadratic in the system size $N$:
\[
U-\left(\mathbbm{1}+t\sum_x H_x\right)
=
\frac{t^2}{2} \sum_{x,y} H_x H_y + \mathcal{O}(t^3).
\]
For this reason, this approximation would lead to an error per site that grows with $N$ in large systems.

A second possibility would be to use $U\simeq\prod_x \left(\mathbbm{1}+tH_x\right)$, which now keeps products of spatially separated terms.
Its leading error (order $t^2$ as before) is now proportional to the system size:
\[
U-\prod_x \left(\mathbbm{1}+tH_x\right)
=
\frac{t^2}{2}\sum_x H_x^2
-\frac{t^2}{2}\sum_{x<y}[H_x,H_y]
+\mathcal{O}(t^3).
\]
The sign of the commutator term depends on the convention chosen for the ordering of the product on the left-hand side. Since terms with non-overlapping supports commute, only a number of commutators proportional to $N$ contribute. The first term on the right-hand side is also a sum of $\mathcal{O}(N)$ contributions.
So, the total error is $\mathcal{O}(N t^2)$, instead of $\mathcal{O}(N^2 t^2)$ for the previous approximation. The problem, however, is that $\prod_x \left(\mathbbm{1}+tH_x\right)$ does not generally have a simple MPO form, even if $H$ itself does.

In Ref.~\cite{zaletelTimeevolvingMatrixProduct2015}, Zaletel {\it et al.} proposed two approximations for $U$, called ${\rm W}^{I}$ and ${\rm W}^{II}$, which have (i) a compact MPO representation and (ii) an error of order $\mathcal{O}(Nt^2)$.
These algorithms can be applied to any Hamiltonian that can be expressed as an MPO. This is an important difference from the simple TEBD algorithm presented in \cref{sec:TEBD}, which is limited to Hamiltonians with nearest-neighbor interactions. 

We briefly describe below the ${\rm W}^{I}$ method.
As in \cite{zaletelTimeevolvingMatrixProduct2015}, we use the notation $x<y$ if all sites in the support of $H_x$ are strictly to the left of all sites in the support of $H_y$. If $x<y$, then the supports of $H_x$ and $H_y$ do not overlap (and the two terms commute).
Since we assumed that each term $H_x$ has a finite range around $x$, there is only a finite number of $y$ such that neither $x<y$ nor $y<x$ holds.
In the ${\rm W}^{I}$ approximation the evolution operator is approximated by 
\begin{equation}
U_{\rm I} (t)= \mathbbm{1} + t \sum_x H_x + t^2 \sum_{x<y} H_x H_y + t^3 \sum_{x<y<z} H_x H_y H_z + \cdots .
\label{eq:U_W_I}
\end{equation} 
The difference between $U$ and $U_{\rm I}$ consists of products such as $H_x H_{x'}$, $H_x H_{x'} H_{x''}$, etc., in which at least two terms have overlapping support. The leading error is therefore of order $t^2$ and involves only pairs of overlapping terms. The error thus scales as $\mathcal{O}(N t^2)$. The main advantage over $U\simeq \prod_x \left(\mathbbm{1}+tH_x\right)$ is that $U_{\rm I}$ has a simple MPO form, with local tensors
\begin{equation}
W_{\mathrm I}^{[i]}
=
\begin{pmatrix}
\mathbbm{1} + tD^{[i]} & \sqrt{t}\,C^{[i]}\\[2mm]
\sqrt{t}\,B^{[i]} & A^{[i]}
\end{pmatrix}
\label{eq:W_I_MPO}
\end{equation}
where the blocks $A$, $B$, $C$ and $D$ have been defined in \cref{eq:general_H_MPO} and describe the Hamiltonian MPO.
If $\chi$ is the bond dimension of the MPO representing $H$, the above MPO for $U_{\rm I}$ has a bond dimension $\chi-1$.
We refer the reader to \cite{zaletelTimeevolvingMatrixProduct2015} for the proof that
the MPO tensor \cref{eq:W_I_MPO} indeed gives the expansion \cref{eq:U_W_I} for the evolution operator.

Using \cref{eq:U_W_I} one can compose two time steps:
\begin{equation}
  U_{\rm I}(t_1) U_{\rm I}(t_2) = \mathbbm{1} + (t_1+t_2) \sum_x H_x + (t_1^2+ t_2^2) \sum_{x<y} H_x H_y
  + t_1 t_2 \left(\sum_{x} H_x\right)^2 + \mathcal{O}(t_1^3) + \mathcal{O}(t_2^3).
\end{equation}
It is possible to impose $U_{\rm I}(t_1) U_{\rm I}(t_2) = U(t) + \mathcal{O}(N t^3)$ by choosing $t_1$ and $t_2$ such that $t_1+t_2=t$, $t_1^2+ t_2^2 = 0$ and $t_1 t_2 = t^2/2$. The solution is $t_1 = t (1+i)/2$ and $t_2 = t (1-i)/2$. Composing two complex time steps allows one to reduce the error from $\mathcal{O}(N t^2)$ to $\mathcal{O}(N t^3)$.
In a similar way, composing four different time steps can reduce the error to $\mathcal{O}(N t^4)$
and with seven steps the error is $\mathcal{O}(N t^5)$~\cite{bidzhiev_out--equilibrium_2017}. These schemes are implemented in the \texttt{TensorMixedStates} library.

The \(W^{\mathrm I}\) method is simple but not yet optimal. For example, for a purely onsite Hamiltonian,
\[
H=\sum_x h_x,
\]
the exact evolution factorizes as
\[
e^{tH}=\prod_x e^{t h_x},
\]
which has a (trivial) MPO representation with bond dimension 1. By contrast, in this case \(W^{\mathrm I}\) only gives
\[
  U_{\rm I} = \prod_x (\mathbbm{1}+t h_x).
\]
This motivates the improved \(W^{\mathrm{II}}\) construction. Compared with \(W^{\mathrm I}\), it keeps additional products $H_x H_{x'} \cdots$, provided that no two terms in the product cross the same bond, while still admitting a compact MPO representation. In particular, onsite terms
(which by definition do not cross any bond)
are exponentiated exactly to all orders in \(t\). The \(W^{\mathrm{II}}\) MPO (not given here) is more complicated to construct than \(W^{\mathrm I}\), but it has the same bond dimension. Exactly as before, different time steps can be combined to reduce the discretization error.

A short code illustrating how to call the \(W^{\mathrm{II}}\) function implemented in \texttt{TensorMixedStates} is given in \cref{lst:WII_onsite_field_exact}. This example illustrates that the method exactly solves the dynamics associated with a purely onsite Hamiltonian.
\Cref{lst:lindblad_XX_spin_chain,lst:lindblad_XX_spin_chain_OSEE} are other code examples using the \(W^{\mathrm{II}}\) method.

\begin{longlisting}
  \jcode{WII_onsite_field_exact.jl}
  \caption[Exact time evolution with W$^{\rm II}$ MPO for onsite Hamiltonian]{\label{lst:WII_onsite_field_exact}
    Short \texttt{TensorMixedStates} code showing that the \(W^{\mathrm{II}}\) method {\em exactly} evolves a spin-$\frac{1}{2}$ chain in a uniform transverse field \(H=B\sum_i S_i^x\), even with a large single time step. The initial state is the product state \(\ket{\uparrow\uparrow\dots\uparrow}\), for which \(\langle Z_i(t)\rangle=\cos(Bt)\). Note that the exactness of the time evolution would persist for an arbitrary initial state with bond dimension larger than 1.
    \jcodegithublink}
\end{longlisting}

\section{Projected entangled pair states (PEPS)}
\label{sec:PEPS}

Projected entangled pair states (PEPS) were introduced in 2004 by F. Verstraete and J. I. Cirac in \cite{VERSTRAETE_RenormalizationAlgorithmsQuantumMany_2004}. They are a class of TN states that can efficiently represent quantum states obeying an area law for entanglement entropy in two or more spatial dimensions, and are the natural generalization of MPS and AKLT states.
Each tensor is associated with a site of the lattice and has one physical index and several virtual indices that connect it to the neighboring tensors
(\cref{fig:MPS_TTN_PEPS_MERA}). As for MPS, the bond dimension $\chi$ of the virtual indices controls the amount of entanglement that can be captured by the PEPS. Similarly, any quantum state can be represented as a PEPS if the associated bond dimension is large enough. However, such a representation is useful in practice only if the state of interest can be well approximated by a PEPS with a bond dimension that is reasonably small. One important difference from MPS is that there is no known efficient algorithm (i.e., polynomial in the system size) to contract exactly a generic PEPS. This implies that one has to rely on approximate methods to optimize PEPS and to compute expectation values of observables.

Before discussing PEPS it is useful to understand the difficulties of treating 2d systems with a 1d MPS, which motivates the introduction of PEPS.

\subsection{Area law and difficulty of treating 2d systems with MPS}
\label{sec:PEPS_area_law_MPS_2d}

\begin{figure}[htbp]
  \centering
  \includegraphics[width=0.95\textwidth]{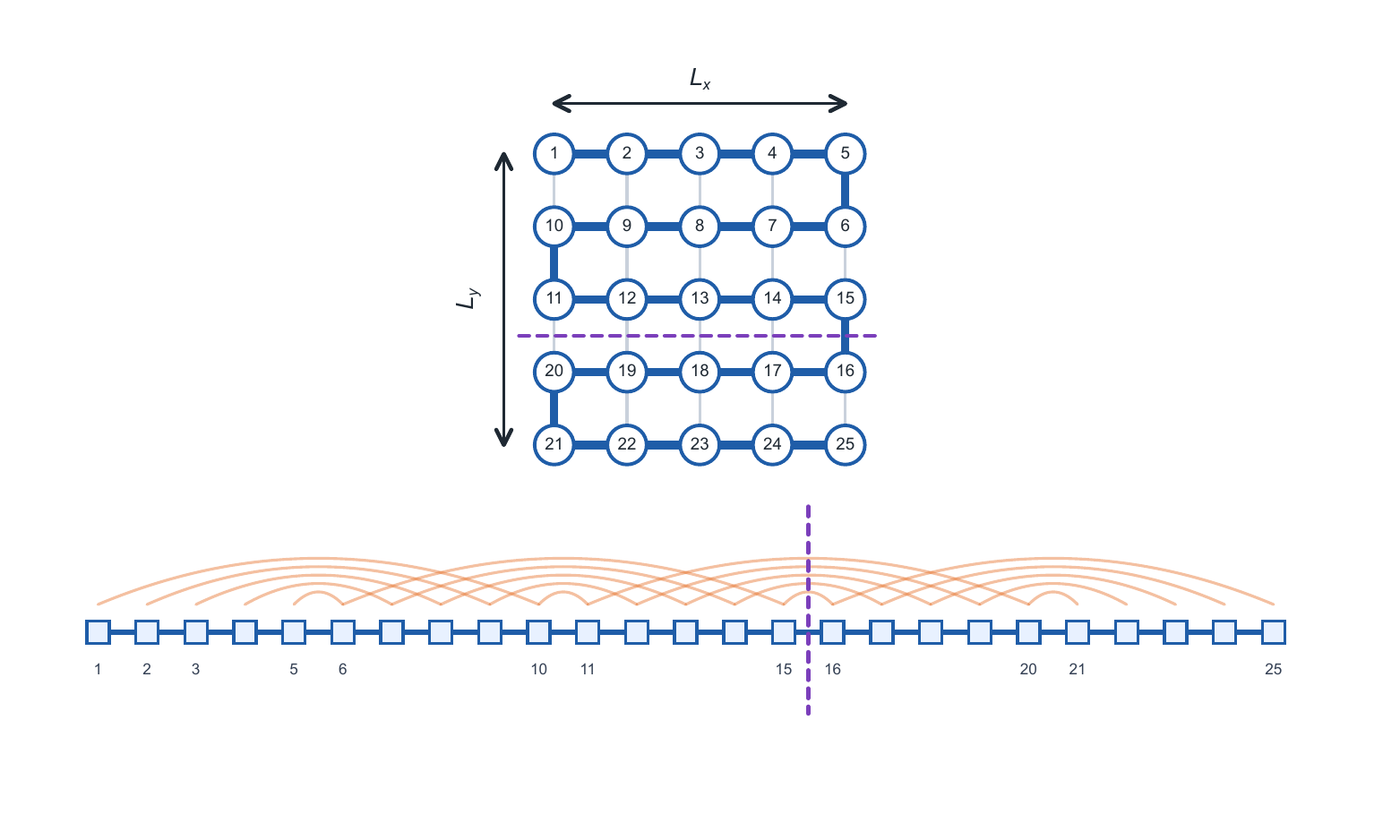}
  \caption{Mapping a Hamiltonian with nearest-neighbor interactions on a $5\times 5$ square lattice to a 1d MPS by choosing a ``snake'' ordering of the sites. The original square lattice is shown in gray, while the blue path gives the ordering of the physical degrees of freedom in the MPS. In the unfolded 1d chain, the nearest-neighbor bonds along the snake path remain local, while vertical bonds of the original square lattice become longer-range couplings (orange arcs). A bipartition after site 15 (dashed purple line) corresponds to a single MPS bond cut, but cuts $L_x$ bonds of the original 2d lattice.}
  \label{fig:snake_ordering_2d_mps}
\end{figure}

One can treat a 2d Hamiltonian with a 1d MPS by choosing an ordering of the lattice sites, for instance the snake ordering shown in \cref{fig:snake_ordering_2d_mps}. This strategy has been successfully used on several 2d models~\cite{stoudenmireStudyingTwoDimensionalSystems2012}. The first drawback, however, is that the locality of the Hamiltonian is partly lost in the 1d representation, since some short-range interactions in the 2d problem translate into long-range interactions along the 1d snake. This is illustrated in \cref{fig:snake_ordering_2d_mps}, where the vertical bonds of the original square lattice become longer-range couplings (orange arcs) in the 1d representation.
We do not choose a specific Hamiltonian here, but we assume that it contains nearest-neighbor interactions on the square lattice.
When expressing this Hamiltonian as an MPO, the bond dimension of the MPO will then grow with the range of the interactions along the 1d path. The bond dimension of the Hamiltonian MPO will thus grow with the dimension $L_x$ of the 2d lattice.

A more serious difficulty is the fact that encoding a state with an area law for the entanglement entropy in 2d requires an MPS bond dimension that grows {\em exponentially} with the width of the system ($L_x$ in \cref{fig:snake_ordering_2d_mps}). This can easily be understood by looking at the bipartition marked with the purple dashed line in \cref{fig:snake_ordering_2d_mps}. Such a bipartition cuts a single MPS bond (between sites 15 and 16). The entanglement entropy across this cut thus satisfies $S_{\rm vN} \leq \log \chi$, where $\chi$ is the MPS bond dimension on bond $(15,16)$. However, this bipartition cuts $L_x$ bonds of the 2d lattice. So, for a state with an area law for the entanglement entropy, the entanglement entropy across this cut scales as $S_{\rm vN}\simeq a L_x$. We thus have $\chi \geq e^{a L_x}$. This means that the bond dimension of the MPS must grow exponentially with the width of the system to capture an area law state in 2d.
This strongly limits the possible values of $L_x$ (typically $L_x \lesssim 12$) but the length $L_y$ can a priori be much larger. 
For an aspect ratio $L_y/L_x$ that is of order one, the bond dimension of the MPS must grow exponentially with the system size in order to capture area law states in 2d.

The \cref{lst:dmrg_2D_fermions} studies noninteracting spinless fermions on a 2d $L_x\times L_y$ square lattice with open boundary conditions. It first builds the single-particle tight-binding Hamiltonian and computes (in polynomial time) the exact half-filled ground-state energy and a connected density-density correlation. It then solves the same problem with DMRG (\texttt{ITensors}) and an MPO formulation of the Hamiltonian. To do so, a snake (or zigzag) site ordering is used to map the 2d lattice to a 1d chain with long-range hoppings. The code reports the DMRG ground-state energy and the connected density-density correlation, and compares them to the exact results.

\begin{figure}
  \begin{center}
    \includegraphics[width=0.8\textwidth]{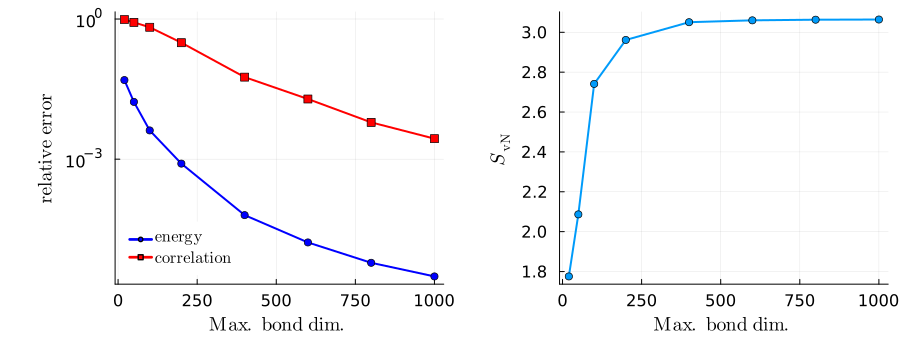}
    \caption{Data from \cref{lst:dmrg_2D_fermions} for a $7\times 6$ lattice.
      The correlation considered is the connected density-density correlation between the two sites that are the farthest apart on the lattice, at opposite corners (see code).
      The left panel shows the relative error on the ground-state energy and on the connected density-density correlation as a function of the MPS bond dimension. The right panel shows the entanglement entropy across the middle bond of the snake ordering as a function of the MPS bond dimension.}
    \label{fig:dmrg_2D_fermions}
  \end{center}
\end{figure}

If one analyzes the {\em relative} error on the ground-state energy and the connected density-density correlation, one finds that the error on the energy is significantly smaller than the error on the correlation. This is shown in \cref{fig:dmrg_2D_fermions} for a $7\times 6$ lattice, where the relative error on the energy is of order $10^{-5}$ while the relative error on the correlation is of order $10^{-2}$ for a bond dimension $\chi=100$.

\begin{longlisting}
  \jcode{dmrg_2D_fermions.jl}
  \caption[DMRG for 2d free fermions]{\label{lst:dmrg_2D_fermions}
    Noninteracting spinless fermions on a 2d $L_x\times L_y$ square lattice: exact free fermion calculation versus DMRG.
    The 2d lattice is mapped to a 1d chain using a snake ordering. The code computes the exact half-filled free-fermion ground-state energy by diagonalizing the single-particle Hamiltonian, then compares it with DMRG results for increasing values of \texttt{maxdim}. It also compares a connected density-density correlation between opposite corners of the lattice and reports the entanglement entropy across the middle bond of the snake-ordered MPS.
    \jcodegithublink
  }
\end{longlisting}

\begin{figure}
  \begin{center}
    \includegraphics[width=0.27\textwidth]{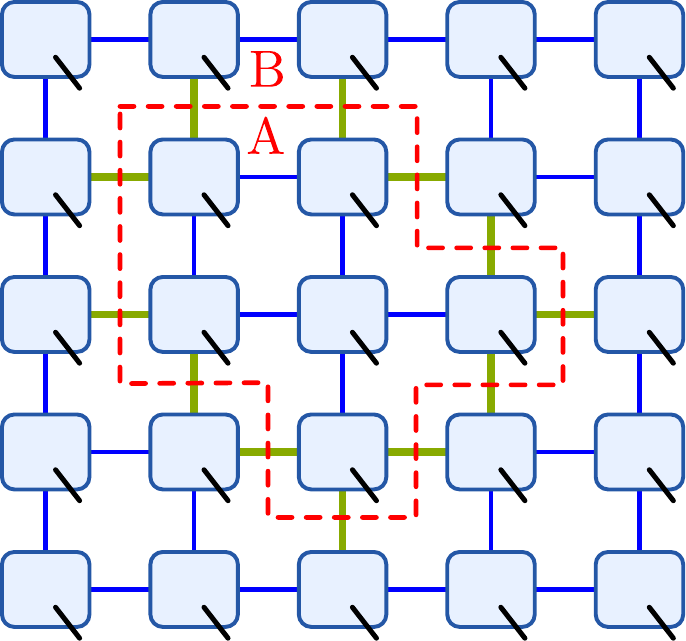}
  \end{center}
  \caption{A PEPS representation of a quantum many-body state on a 2d lattice, with a subsystem $A$ separated from its complement $B$ by a boundary (red dotted line). The virtual indices that cross this boundary are colored in green. The physical indices are the (black) open legs.}
  \label{fig:PEPS_subsystem}
\end{figure}

Contrary to the snake-MPS discussed above, a PEPS can capture an area law for the entanglement entropy with a bond dimension that does not grow with the system size. This is easily understood by noticing that a spatial bipartition of the system cuts a number of virtual bonds that is by construction proportional to the length of the boundary between the two subsystems, as illustrated in \cref{fig:PEPS_subsystem}.
In this way, a PEPS with a fixed bond dimension $\chi$ can capture an area law state in 2d, with an entanglement entropy that scales as $S_{\rm vN} \sim \mathcal{O}(l)$, where $l$ is the length of the boundary between the two subsystems.

\begin{remarkblock}
  Consider a quantum state $|\psi\rangle$ defined on some lattice and represented by a TN, such as the PEPS shown in \cref{fig:PEPS_subsystem}.
The reduced density matrix $\rho_A$ of a subsystem $A$ is obtained by tracing out the degrees of freedom of the complementary subsystem $B$: $\rho_A = \mathrm{Tr}_B |\psi\rangle\langle\psi|$. A certain number of virtual bonds $\alpha_1, \alpha_2, \dots, \alpha_l$ will
cross the boundary between $A$ and $B$. Assume for simplicity that they all have the same bond dimension $\chi$. If we group them
into a single index $\alpha$ which can take $\chi^l$ values, the state $|\psi\rangle$ reads
\begin{equation}
  |\psi\rangle = \sum_{\alpha} |\alpha,A\rangle \otimes |\alpha,B\rangle,
\end{equation}
where $|\alpha,A\rangle$ and $|\alpha,B\rangle$ are states of the subsystems $A$ and $B$ when the virtual indices crossing the boundary are fixed to $\alpha=(\alpha_1, \alpha_2, \dots, \alpha_l)$. 
The reduced density matrix of the subsystem $A$ can then be written as
\begin{equation}
\begin{aligned}
  \rho_A
  &= \sum_{\alpha,\alpha'} {\rm Tr}_B\left[\,|\alpha,A\rangle\langle\alpha',A|\otimes |\alpha,B\rangle\langle \alpha',B|\,\right] \\
  &= \sum_{\alpha,\alpha'} M_{\alpha,\alpha'} |\alpha,A\rangle\langle\alpha',A| ,
\end{aligned}
\label{eq:rho_A_M}
\end{equation}
where the matrix $M$ is given by $M_{\alpha,\alpha'}= {\rm Tr}_B\left[\,|\alpha,B\rangle \langle \alpha',B|\,\right] = \langle \alpha',B |\alpha,B\rangle $.
We see in \cref{eq:rho_A_M} that the image of $\rho_A$ is included in $\mathrm{span}\{|\alpha,A\rangle\}$, whose dimension is upper bounded by $\mathrm{rank}(\rho_A)\leq\chi^l$. The (von Neumann) entropy $S_{\rm vN}$ of $\rho_A$ is therefore upper bounded by $\ln(\chi^l)$ and $S_{\rm vN} \leq l \log \chi$.
\end{remarkblock}

\subsection{Example 1: PEPS derived from a classical thermal state}
\label{sec:PEPS_classical_thermal_state}

Using the transfer matrix formalism we saw in \cref{sec:MPS_correlation_function} that MPS (with finite bond dimension) generically have exponentially decaying correlations.
The situation is different for PEPS in $d>1$. In particular, PEPS (with a finite bond dimension) can have power-law decaying correlations, and thus can represent critical states. This can be shown by constructing an explicit example of critical PEPS. To do so, one starts from a classical model of statistical mechanics defined by local Boltzmann weights. From this model one can construct a PEPS (and thus a quantum state) with the same correlation functions as the classical model.\footnote{Note that such quantum states constructed from a classical model are closely related to so-called Rokhsar-Kivelson states~\cite{rokhsarSuperconductivityQuantumHardCore1988,stephanShannonEntanglementEntropies2009}.} At the critical point, this gives a PEPS with power-law decaying correlations~\cite{verstraeteCriticalityAreaLaw2006}.

Following the presentation given in \cite{ORUS_PracticalIntroductionTensor_2014}, we illustrate this construction in the case of the 2d Ising model with classical spins $\sigma_i=\pm 1$ and energy $E(\{\sigma\}) = -J \sum_{\langle i,j\rangle} \sigma_i \sigma_j$. 
The Boltzmann weight of a configuration $\{\sigma\}$ is given by $e^{-\beta E(\{\sigma\})}$, where $\beta$ is the inverse temperature. The partition function is given by $Z = \sum_{\{\sigma\}} e^{-\beta E(\{\sigma\})}$. From this we define a quantum state $|\psi\rangle$ (parametrized by $\beta J$) as follows:
\begin{equation}
\begin{aligned}
|\psi\rangle
&= \frac{1}{\sqrt{Z}} \sum_{\{\sigma\}} e^{-\frac{\beta}{2} E(\{\sigma\})} |\sigma_1 \sigma_2 \dots \sigma_N\rangle \\
&= \frac{1}{\sqrt{Z}} e^{\frac{\beta J}{2} \sum_{\langle i,j\rangle} \sigma^z_i \sigma^z_j} |++\dots +\rangle ,
\end{aligned}
\label{eq:psi_PEPS_classical}
\end{equation}
where $|+\rangle = |\uparrow\rangle + |\downarrow\rangle$.
By construction, a correlation function of the type $\langle \psi| \sigma^z_i \sigma^z_j |\psi\rangle$ is given by
\begin{equation}
\langle \psi| \sigma^z_i \sigma^z_j |\psi\rangle = \frac{1}{Z} \sum_{\{\sigma\}} e^{-\beta E(\{\sigma\})} \sigma_i \sigma_j,
\end{equation}
which is nothing but the correlation function of the classical model. As we will see, the quantum state $|\psi\rangle$ can be represented as a PEPS with bond dimension $\chi=2$.

The first step is to consider two neighboring sites $i$ and $j$ and the operator $e^{\frac{\beta J}{2} \sigma^z_i \sigma^z_j}$. This is a two-site operator that can be written as an MPO. It turns out that the bond dimension of this MPO is (only) $\chi=2$.
This can be seen by writing $e^{\frac{\beta J}{2} \sigma^z_i \sigma^z_j} = \cosh(\frac{\beta J}{2}) \mathbbm{1}_i \otimes \mathbbm{1}_j + \sinh(\frac{\beta J}{2}) \sigma^z_i \otimes \sigma^z_j$. This can be represented as an MPO with bond dimension $\chi=2$ by introducing a virtual index $\alpha$ that takes two values, $\alpha=0$ and $\alpha=1$, and writing
\begin{equation}
e^{\frac{\beta J}{2} \sigma^z_i \sigma^z_j} = \sum_{\alpha=0,1} W^{[i]}_{\alpha} W^{[j]}_{\alpha},
\end{equation}
where $W^{[i]}_{\alpha}$ and $W^{[j]}_{\alpha}$ are local operators acting on sites $i$ and $j$ respectively, and are given by
\begin{eqnarray}
W^{[i]}_{0} &=& \sqrt{c}\, \mathbbm{1}_i, \qquad W^{[i]}_{1} = \sqrt{s}\, \sigma^z_i, \\
W^{[j]}_{0} &=& \sqrt{c}\, \mathbbm{1}_j, \qquad W^{[j]}_{1} = \sqrt{s}\, \sigma^z_j \\
{\rm with}\;\; c&=&\cosh(\frac{\beta J}{2}) \;{\rm and} \; s=\sinh(\frac{\beta J}{2}).
\end{eqnarray}
The second step is to insert the above MPO representation for $e^{\frac{\beta J}{2} \sigma^z_i \sigma^z_j}$
into \cref{eq:psi_PEPS_classical} for each pair of neighboring sites $\langle i,j\rangle$ of the lattice. One then obtains an expression for $|\psi\rangle$ as a PEPS
\begin{equation}
|\psi\rangle = \frac{1}{\sqrt{Z}}\sum_{\{\alpha\}} \prod_{\langle i,j\rangle} W^{[i]}_{\alpha_{ij}} W^{[j]}_{\alpha_{ij}}|++\dots +\rangle,
\end{equation}
where $\alpha_{ij}=0,1$ is the virtual index associated with the bond between sites $i$ and $j$. By regrouping the operators acting on each site, one can write $|\psi\rangle$ as a PEPS where the local tensors $A^{[i]}$ are:
\begin{eqnarray}
A^{[i]\,s_i}_{\alpha_1,\alpha_2,\alpha_3,\alpha_4} &=& \bra{s_i}W^{[i]}_{\alpha_1} W^{[i]}_{\alpha_2} W^{[i]}_{\alpha_3} W^{[i]}_{\alpha_4} |+_i\rangle \\
&=& \bra{s_i} W^{[i]}_{\alpha_1} W^{[i]}_{\alpha_2} W^{[i]}_{\alpha_3} W^{[i]}_{\alpha_4} \ket{s_i},
\end{eqnarray}
where $\alpha_1,\alpha_2,\alpha_3,\alpha_4$ are the virtual indices corresponding to the four nearest neighbors of site $i$, and $s_i=\uparrow,\downarrow$ is the physical index corresponding to the spin state at site $i$. In the second line we have used the fact that $W^{[i]}_{\alpha}$ is diagonal in the $\sigma^z$ basis. This gives an explicit expression for the 5-index tensor $A^{[i]}$ of the PEPS, which has a bond dimension $\chi=2$ and a physical dimension $d=2$. As explained above, the correlation functions of $|\psi\rangle$ are the same as those of the classical 2d Ising model, and thus decay algebraically at the critical point ($\beta J = \frac{1}{2} \ln(1+\sqrt{2})$).

\subsection{Example 2: Toric code}
\label{sec:PEPS_toric_code}

Kitaev's toric code~\cite{kitaevFaulttolerantQuantumComputation2003,kitaevTopologicalPhasesQuantum2009} provides a simple but paradigmatic example of a topologically ordered state.
It exhibits long-range entanglement, anyonic excitations and topological degeneracy. 
As described below, it is another canonical example of a quantum state that can be exactly represented as a PEPS with small bond dimension ($\chi=2$). This example shows that a PEPS with finite bond dimension can capture topological order.

\begin{figure}
  \begin{center}
    \includegraphics[width=0.63\textwidth]{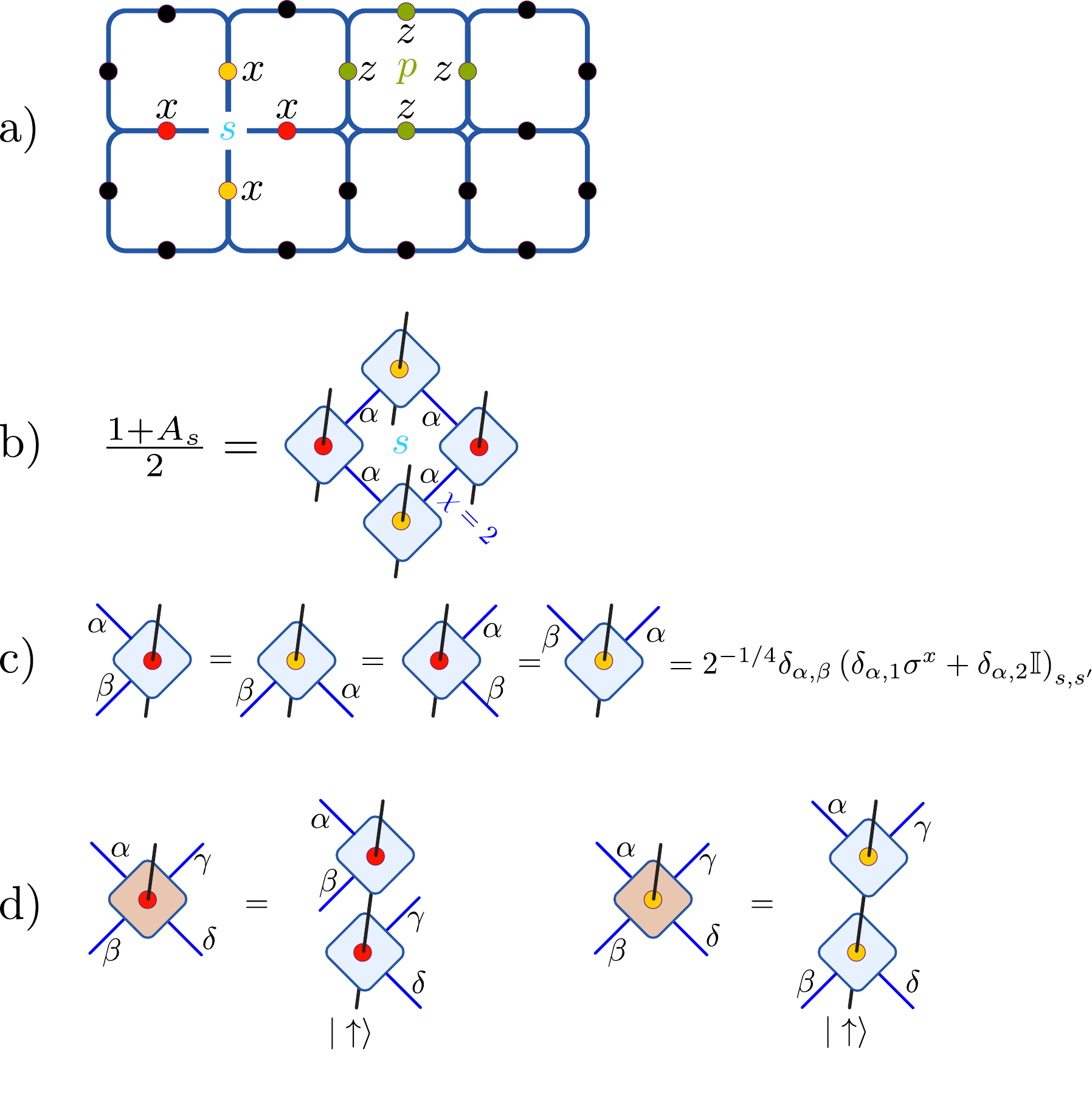}
  \end{center}
  \caption{PEPS representation of the toric-code ground state.
  a) In the toric code each vertex \(s\) is associated with the four spins adjacent to it. 
  The associated operator $A_s$ is the product of $\sigma^x$ operators around the vertex.
  Each plaquette \(p\) is associated with the four spins around it, as well as the operator $B_p$, which is the product of $\sigma^z$ operators around the plaquette.
  b) The projector $\Pi_s=\frac{1+A_s}{2}$ can be written as an MPO using four tensors that are diagonal for the bond index $\alpha=0,1$.
  c) Representation of the tensors used in the MPO. The tensors have virtual indices (blue lines) of dimension $\chi=2$. The MPO that represents $\frac{1+A_s}{2}$ thus also has bond dimension 2.
  d) The local tensors of the PEPS. The form slightly depends on whether the site belongs to one sublattice or the other (yellow or red dot).
  }
  \label{fig:PEPS_toric_code}
\end{figure}

The Hamiltonian of the toric code is defined on a square lattice with spin-$\frac{1}{2}$ degrees of freedom living on the bonds of the lattice
(filled dots in \cref{fig:PEPS_toric_code}a). It is given by
\begin{equation}
H = -\sum_s A_s - \sum_p B_p,
\label{eq:toric_code_Hamiltonian}
\end{equation}
where $A_s = \prod_{i\in s} \sigma^x_i$ is a product of $\sigma^x$ operators around a vertex $s$ and $B_p = \prod_{i\in p} \sigma^z_i$ is a product of $\sigma^z$ operators around a plaquette $p$.
One can easily check that the $A_s$ and $B_p$ have eigenvalues $\pm 1$ and they all commute with each other (even when they share some sites). The (possibly degenerate) ground states of $H$ are the common eigenstates of all $A_s$ and $B_p$ with eigenvalue $+1$.\footnote{Even though it is an important aspect of the toric code, we will not discuss here the (topological) degeneracy of such states, nor their excited states.}

To construct a PEPS representation of this state, one can start from the ferromagnetic product state $|\uparrow\dots\uparrow\rangle$, which is the eigenstate of all the $\sigma^z_i$ with eigenvalue $+1$. This state trivially minimizes the energy of the $B_p$ terms, but not of the $A_s$ terms. To minimize the energy of the $A_s$ terms, one can apply the projector $\prod_s \frac{1+A_s}{2}$ to the state $|\uparrow\dots\uparrow\rangle$:
\begin{equation}
|\psi\rangle = \prod_s \frac{1+A_s}{2} |\uparrow\dots\uparrow\rangle.
\end{equation}
Since the $A_s$ terms commute with the $B_p$ terms, $|\psi\rangle$ is still an eigenstate of all $B_p$ with eigenvalue $+1$. Moreover, it is now also an eigenstate of all $A_s$ with eigenvalue $+1$, and thus minimizes the energy of all terms in the Hamiltonian.

To obtain a PEPS representation of $|\psi\rangle$, we follow \cite{ORUS_PracticalIntroductionTensor_2014} and write the projector $\Pi_s=\frac{1+A_s}{2}$ as a product of four tensors, as illustrated in \cref{fig:PEPS_toric_code}.
The idea is that each star projector can be expanded as a sum over a binary variable \(\alpha_s\in\{0,1\}\):
\[
  \Pi_s=\frac{1+A_s}{2}
  =\frac{1}{2}\sum_{\alpha_s=0,1}
  \prod_{i\in s}(\sigma_i^x)^{\alpha_s}.
\]
The value \(\alpha_s=0\) corresponds to
applying the identity operator on the four spins adjacent to the star, while \(\alpha_s=1\) corresponds to flipping all four of them. This binary variable is the virtual index carried by the MPO representation of $\Pi_s$
(\cref{fig:PEPS_toric_code}b). When all star projectors are multiplied together, each physical spin is touched by the two neighboring stars
(\cref{fig:PEPS_toric_code}d), and its final state depends only on the parity of the two corresponding binary variables. The resulting local tensors therefore have bond dimension \(\chi=2\).

\subsection{Other examples}
\label{sec:PEPS_other_examples}

To conclude this list of examples we mention higher-dimensional generalizations of the AKLT state.
The 1d AKLT construction discussed in \cref{sec:AKLT} can be generalized to higher-dimensional lattices
provided the value $S$ of the spin and the coordination number $z$ of the lattice satisfy $2S \equiv 0 \pmod z$~\cite{Affleck_ValenceBondGroundStates_1988}.
The simplest case, which includes the examples below and gives bond dimension $\chi=2$, has $2S=z$: one places $z$ virtual spin-$\frac{1}{2}$ degrees of freedom on each site and forms one singlet on each bond. The physical state is obtained by projecting the $z$ virtual spin-$\frac{1}{2}$ degrees of freedom onto the spin-$S$ symmetric subspace of dimension $2S+1$.
Notable examples include the spin-$\frac{3}{2}$ AKLT state on the hexagonal lattice ($z=3$) and the spin-2 AKLT state on the square lattice ($z=4$). 

Another family of examples is given by the so-called resonating valence bond (RVB) states, which are superpositions of exponentially many valence-bond (spin singlets) configurations. These states play a role in the theory of high-temperature superconductivity~\cite{andersonResonatingValenceBonds1973,andersonResonatingValenceBond1987a} and quantum spin liquids. Some of these states can be represented as PEPS with a small bond dimension, starting from $\chi=3$~\cite{verstraeteCriticalityAreaLaw2006}. The idea is that the virtual degrees of freedom on each bond encode whether a singlet is present on that bond or not ($\alpha=1$ $\to$ no singlet, $\alpha=2$ $\to$ component $|\uparrow\rangle$ of the singlet, $\alpha=3$ $\to$ component $|\downarrow\rangle$ of the singlet).

\subsection{Contraction of PEPS}
\label{sec:peps_contraction}

The contraction of a PEPS TN (which is required to compute expectation values and correlation functions) is computationally hard if one wants to do it {\em exactly}~\cite{SCHUCH_ComputationalComplexityProjected_2007,haferkampContractingProjectedEntangled2020}. Thus, unlike for MPS, there is no known efficient algorithm ({\it i.e.}, polynomial) to perform the exact contraction of a PEPS in the general case. However, several {\em approximate} methods have been developed to contract PEPS, which allow expectation values and correlation functions to be computed with good accuracy in many cases~\cite{lubaschAlgorithmsFiniteProjected2014,lubaschUnifyingProjectedEntangled2014}.
Although the cost of these approximate PEPS contraction and optimization algorithms typically scales polynomially with the PEPS bond dimension \(\chi\), the degree of this polynomial can be large (typically 6 or more). For this reason the numerically accessible values of \(\chi\) are much smaller than for MPS. 

Among the approximate contraction methods that have been proposed, 
we can mention
the boundary MPS method~\cite{VERSTRAETE_RenormalizationAlgorithmsQuantumMany_2004},
the corner transfer matrix method~\cite{NISHINO_CornerTransferMatrix_1996,orusSimulationTwodimensionalQuantum2009},
the stochastic Monte Carlo sampling method~\cite{WANG_MonteCarloSimulation_2011},
the tensor renormalization group method~\cite{levinTensorRenormalizationGroup2007,jiangAccurateDeterminationTensor2008a,XIE_CoarsegrainingRenormalizationHigherorder_2012},
or methods based on belief propagation~\cite{ALKABETZ_TensorNetworksContraction_2021,tindallGaugingTensorNetworks2023,tindallDynamicsDisorderedQuantum2026}.
These methods differ mainly in the way they approximate the environment of a local tensor or of a finite region of the PEPS. In corner-transfer-matrix methods, this environment is represented by a finite set of corner and edge tensors surrounding the region of interest, which are updated and truncated iteratively. Tensor-renormalization methods instead coarse-grain the two-dimensional TN step by step, replacing blocks of tensors by effective tensors of controlled bond dimension.

\begin{figure}
  \begin{center}
    \includegraphics[width=1.0\textwidth]{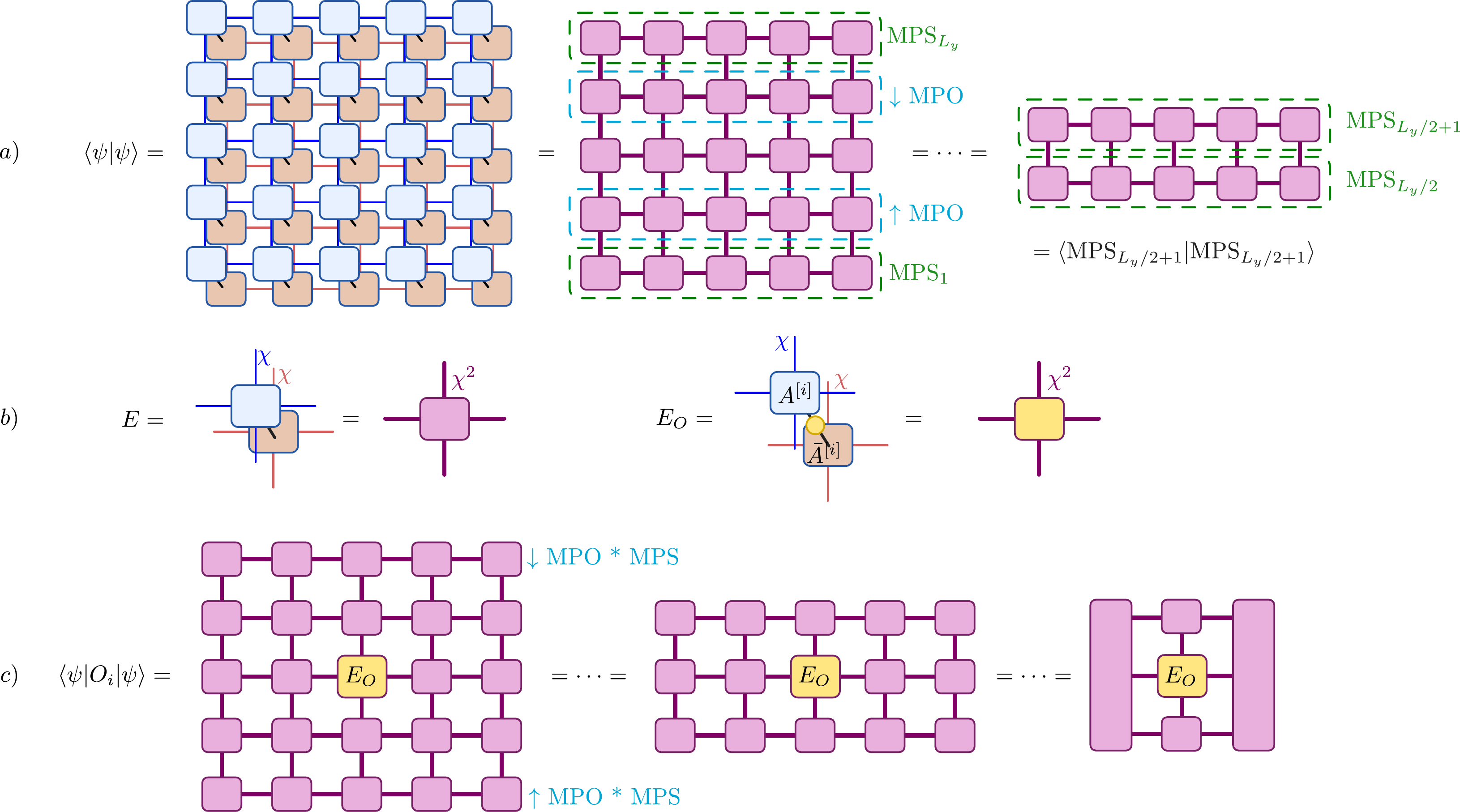}
    \caption{Contraction of a finite PEPS using double-layer tensors. The norm $\langle\psi|\psi\rangle$ and the expectation value $\langle\psi|O_i|\psi\rangle$ are represented as 2d TNs. After the physical indices are contracted, each row can be viewed as an MPO acting on a boundary MPS. The local tensor $E_O$ differs from the double-layer tensor $E$ only at the site where the operator $O_i$ is inserted.}
    \label{fig:peps_contraction}
  \end{center}
\end{figure}

Below we briefly describe the boundary MPS method, which is one of the simplest of these contraction schemes. 
For a finite PEPS, the contraction can be performed by grouping the tensors into rows and contracting them sequentially using MPS and MPO techniques.
We will sketch the method for a square-lattice PEPS and a local operator $O_i$ acting on a single site $i$, but the method can be adapted to other lattices and to more general operators.
To compute the norm $\langle \psi|\psi\rangle$ and the expectation value $\langle \psi| O_i |\psi\rangle$, we need to compute the contractions of the TNs depicted in \cref{fig:peps_contraction}.

Away from the site $i$, the physical index can be contracted locally, giving so-called double-layer tensors
\begin{equation}
E^{[j]}_{\ell\ell',rr',uu',dd'} =
\sum_s \overline{A^{[j]\,s}_{\ell' r' u' d'}}
A^{[j]\,s}_{\ell r u d},
\end{equation}
whose four virtual indices (left, right, up and down) have dimension $\chi^2$ instead of $\chi$ (\cref{fig:peps_contraction}b). 
At the site where the operator is inserted, one uses instead $\sum_{s,s'} \overline{A^{[i]\,s'} }\, O_{s's}\, A^{[i]\,s}$.
The resulting 2d network can then be contracted row by row.
The first boundary row is viewed as an MPS. Its physical index is the vertical virtual index of the double-layer tensors (dimension $\chi^2$), and its virtual indices are the horizontal virtual indices of the double-layer tensors (dimension $\chi^2$). The second row is viewed as an MPO acting on this boundary MPS, and the contraction of the two rows gives a new boundary MPS. The same procedure is carried out from the top and from the bottom, until the two boundary MPS meet at the row where the operator $O_i$ is inserted.
Unless one performs some truncation, the bond dimension of the boundary MPS will grow exponentially with the number of rows that have been absorbed.
Therefore, each MPS--MPO multiplication needs to be accompanied by a truncation step to a bond dimension $\chi_{\rm env}$, as explained in \cref{sec:zipup}. The approximation is controlled by the boundary bond dimension $\chi_{\rm env}$: increasing it improves the representation of the effective environment, at the price of a larger computational cost.
At the last step, one is left with the calculation of the scalar product of two MPS, which can be done efficiently (contraction from left to right, see \cref{sec:MPS_scalar_product}).
This approach can also be used to construct the environment tensors required for the optimization of PEPS,
as sketched in \cref{fig:peps_contraction}c.

\section{Mixed states and open systems}
\label{sec:mixed_states_open_systems}

In an open quantum system, states are generally mixed states described by density matrices $\rho$ rather than pure states $|\psi\rangle$~\cite{weimerSimulationMethodsOpen2021}. In this section we discuss two common ways of treating such states. The first, purification, represents a density matrix as the partial trace of a pure state in an enlarged Hilbert space. This pure state may be encoded as a TN, and in particular as an MPS. 
We illustrate this approach in \cref{sec:thermal_purification} in the context of thermal equilibrium states. The second, vectorization, views the density matrix as a pure state (a vector) in a doubled local Hilbert space (but without partial trace). Here again, the pure state may be encoded as an MPS, and operators acting on the density matrix may be encoded as MPOs.
The vectorization approach is presented in \cref{sec:mps_mixed_states,sec:superoperators_as_MPO,sec:lindblad_MPS_form}.
These sections contain a few code examples based on the \texttt{TensorMixedStates} library~\cite{HOUDAYER_TensorMixedStatesJuliaLibrary_2026}, a Julia package built on \texttt{ITensors.jl} for the simulation of mixed states and open quantum systems where the density matrices are vectorized and represented by MPS.

We note that 2d TNs have also been used to describe mixed states and open quantum systems~\cite{kshetrimayumSimpleTensorNetwork2017a,mckeeverStableIPEPOTensorNetwork2021}, but in these notes we focus on MPS and MPO representations.

\subsection{Thermal equilibrium states by purification}
\label{sec:thermal_purification}


The idea of purification, which is not specific to TNs, is a general way of representing a mixed state as a pure state in a larger Hilbert space.
This has been intensively exploited in the context of TNs and MPS in particular~\cite{VERSTRAETE_MatrixProductDensity_2004,FEIGUIN_FinitetemperatureDensityMatrix_2005,SCHOLLWOCK_DensitymatrixRenormalizationGroup_2011,PAECKEL_TimeevolutionMethodsMatrixproduct_2019}.

Consider a density matrix $\rho$ acting on a (physical) Hilbert space $\mathcal{H}_P$, and write its spectral decomposition as
\begin{equation}
  \rho = \sum_a p_a \ket{a}_P \bra{a}_P,
  \qquad p_a \geq 0,
  \qquad \sum_a p_a = 1.
\end{equation}
One then introduces an auxiliary Hilbert space $\mathcal{H}_Q$, which is a copy of $\mathcal{H}_P$, with an orthonormal basis $\{\ket{a}_Q\}$. One defines a {\em pure} state
\begin{equation}
  \ket{\Psi_\rho}
  =
  \sum_a \sqrt{p_a}\,\ket{a}_P \otimes \ket{a}_Q
\end{equation}
such that the original density matrix is recovered by tracing out the auxiliary space,
\begin{equation}
  \rho = \mathrm{Tr}_Q \ket{\Psi_\rho}\bra{\Psi_\rho}.
\end{equation}
Note that the purification is not unique: applying any unitary transformation acting only on the auxiliary space leaves the reduced density matrix on $\mathcal{H}_P$ unchanged.
This can sometimes be exploited to reduce the entanglement between the auxiliary degrees of freedom.

For the thermal equilibrium state associated with a Hamiltonian $H$ at inverse temperature $\beta$, 
\begin{equation}
  \rho_\beta = \frac{1}{Z(\beta)} \,e^{-\beta H},
  \qquad
  Z(\beta)=\mathrm{Tr}\, e^{-\beta H},
\end{equation}
the natural purification is the thermofield-double state
\begin{equation}
  \ket{\Psi_\beta}
  =
  \frac{1}{\sqrt{Z(\beta)}}
  \sum_n e^{-\beta E_n/2}\ket{n}_P\otimes\ket{n}_Q,
\end{equation}
where $\ket{n}_P$ are eigenstates of $H$, and $\ket{n}_Q$ are the corresponding states in the auxiliary space. 
By construction tracing out the auxiliary space gives the desired thermal density matrix for the physical system
\begin{equation}
  \mathrm{Tr}_Q \ket{\Psi_\beta}\bra{\Psi_\beta}
  =
  \frac{1}{Z(\beta)} \sum_n e^{-\beta E_n} \ket{n}_P\bra{n}_P
  =
  \rho_\beta.
\end{equation}

To construct $\ket{\Psi_\beta}$, one may start from a maximally entangled purification of the infinite-temperature state ($\beta=0$) and evolve the physical degrees of freedom in imaginary time by $\beta/2$:
\begin{equation}
  \ket{\Psi_\beta}
  =
  \sqrt{\frac{Z(0)}{Z(\beta)}}   \left(e^{-\beta H/2}\otimes \mathbbm{1}_Q\right)   \ket{\Psi_0}
\label{eq:thermal_purification_from_beta0}
\end{equation}
with
\begin{equation}  
  \ket{\Psi_0} = \frac{1}{\sqrt{Z(0)}}\sum_a \ket{a}_P \otimes \ket{a}_Q ,
  \label{eq:thermal_purification_infinite_temperature}
\end{equation}
where $Z(0)=\dim\mathcal{H}_P$.

In an MPS implementation, one convenient ordering is to alternate physical and auxiliary sites,
$P_1$, $Q_1$, $P_2$, $Q_2$, $\dots$, $P_N$, $Q_N$.
For a two-level system, a possible infinite-temperature purification
\cref{eq:thermal_purification_infinite_temperature}
is then a product of maximally entangled states on each pair $(P_i,Q_i)$,
\begin{equation}
  \ket{\Psi_0}
  =
  \bigotimes_{i=1}^N
  \left(\ket{0}_{P_i}\ket{0}_{Q_i}+\ket{1}_{P_i}\ket{1}_{Q_i}\right) /\sqrt{2} .
  \label{eq:infinite_temperature_entangled_pairs}
\end{equation}
Starting from this state, a finite-temperature purification is obtained
with \cref{eq:thermal_purification_from_beta0}. Such imaginary-time evolution can be performed using one of the time-evolution methods described in \cref{sec:time_evol_MPS}, such as TDVP or a W method. Importantly, the imaginary-time evolution operator $e^{-\beta H/2}$ acts only on the physical sites. However, because of the initial entanglement between physical and auxiliary sites, $\ket{\Psi_\beta}$ is an entangled state of the physical and auxiliary degrees of freedom.
After normalization, thermal expectation values of physical observables are ordinary pure-state expectation values,
\begin{equation}
  \langle O\rangle_\beta
  =
  \frac{\bra{\Psi_\beta}O_P\ket{\Psi_\beta}}
       {\braket{\Psi_\beta}{\Psi_\beta}} .
\end{equation}

\Cref{lst:thermal_xx_purification} illustrates this construction for an open XX spin-$\frac{1}{2}$ chain. In this example, the imaginary-time evolution is performed by TDVP on an MPS representing \cref{eq:infinite_temperature_entangled_pairs}.

\begin{longlisting}
  \jcode{thermal_xx_purification.jl}
  \caption[Thermal equilibrium state of an XX chain by purification and TDVP]{\label{lst:thermal_xx_purification}
    Finite-temperature calculation for an open XX spin-$\frac{1}{2}$ chain
    ($H=\frac{J}{2}\sum_i(S_i^+S_{i+1}^-+S_i^-S_{i+1}^+)=J\sum_i(S_i^xS_{i+1}^x+S_i^yS_{i+1}^y)$)
    using a purified MPS.
    The code prepares the infinite-temperature state as Bell pairs between physical and auxiliary sites (\cref{eq:infinite_temperature_entangled_pairs}), evolves the physical sites in imaginary time by TDVP (\texttt{ITensor} implementation), computes the mean energy, and compares it with the exact (free-fermion) expression:
   $E(\beta)=\sum_{m=1}^N \frac{\epsilon_m}{1+\exp(\beta\epsilon_m)}$
  with the fermionic mode energies $\epsilon_m=J\cos\left(\frac{\pi m}{N+1}\right)$.
    The script sweeps over a few inverse temperatures and prints a comparison table containing the TDVP energy, the exact free-fermion energy, the absolute error, and the maximum bond dimension of the purified MPS.
    \jcodegithublink
  }
\end{longlisting}

\subsection{Vectorization and TN for a mixed state}
\label{sec:mps_mixed_states}

Vectorized density matrices have also been used to simulate mixed-state dynamics in 1d quantum lattice systems~\cite{ZWOLAK_MixedStateDynamicsOneDimensional_2004}.
The vectorization of a matrix of dimension $D_1\times D_2$ consists of reshaping it into a {\em vector} of dimension $D_1 D_2$.
If a matrix $M$ represents an operator $\hat M$ acting on a Hilbert space of dimension $D$
it can be written as $\hat {M} = \sum_{i,j=1}^{D} M_{ij} \ket{i}\bra{j}$, where $\ket{i}$ are basis states of the Hilbert space.
The vectorization of the operator $\hat M$ is then given by
\begin{equation}
  |M\rangle\rangle = \sum_{i,j} M_{ij} \ket{i} \otimes \ket{j},
  \label{eq:operator_vectorization}
\end{equation}
where the two copies of the basis states $\ket{i}$ and $\ket{j}$ are now viewed as basis states of an enlarged Hilbert space of dimension $D^2$. 
This vectorization is closely related to the Choi-Jamiolkowski isomorphism~\cite{choiCompletelyPositiveLinear1975,jamiolkowskiLinearMapsStates1972}.

\begin{figure}
  \begin{center}
    \includegraphics[width=1.0\textwidth]{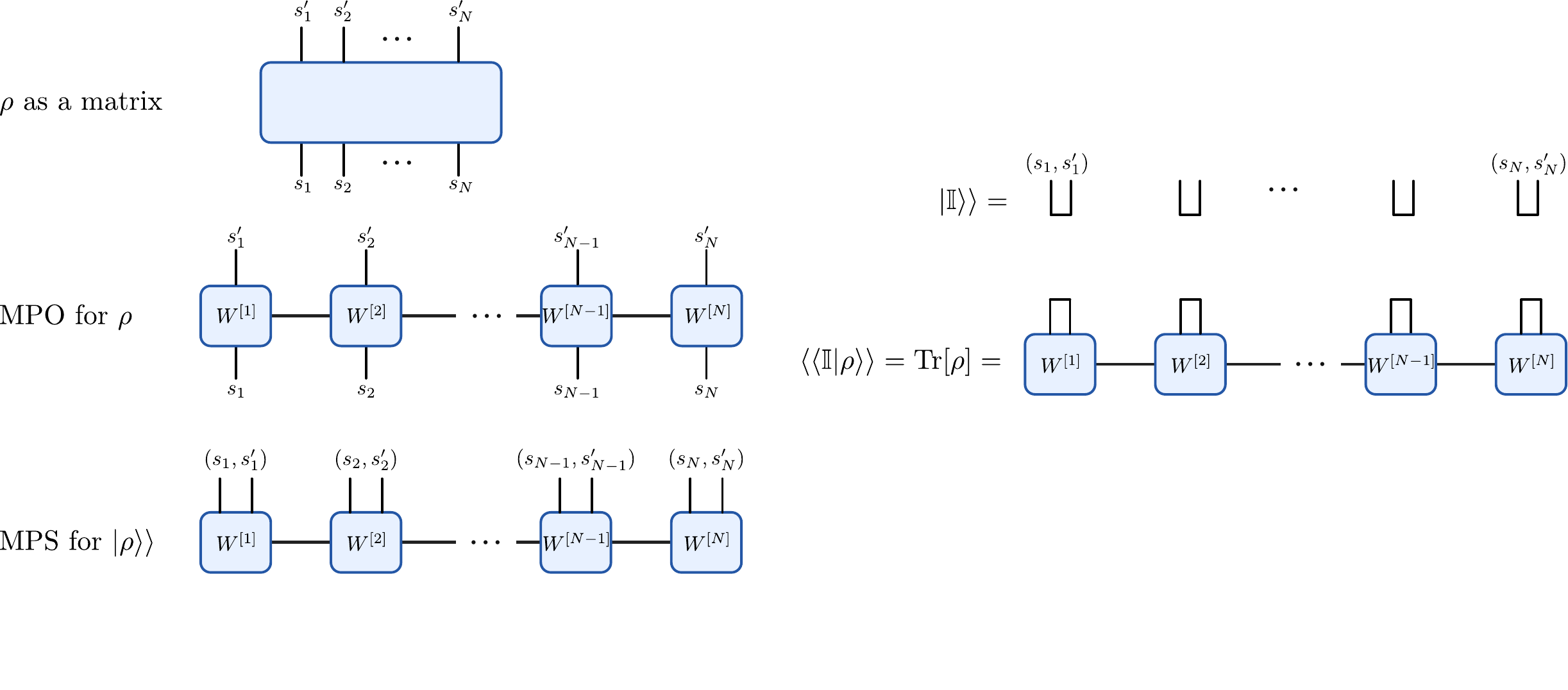}
    \caption{Left: three representations of a density matrix. The matrix $\rho$ of an $N$-site system can first be viewed as a single tensor with $N$ physical indices on the top and $N$ physical indices on the bottom. It can then be factorized as a matrix product operator (MPO), where each local tensor has two physical indices and, except at the boundaries, two virtual indices. By vectorizing the operator, or equivalently by folding the lower physical legs upward, $\rho$ can be represented as a super-ket $|\rho\rangle\rangle$ in MPS form, with a doubled local physical index $(\sigma_i,\sigma_i')$ on each site.
    Right: Graphical representation for the fully mixed state $|\mathbbm{1}\rangle\rangle$ (identity density matrix) and for the scalar product with $|\rho\rangle\rangle$, which gives the trace of $\rho$.
    }
    \label{fig:rho_mpo_superket}
  \end{center}
\end{figure}

Now consider the case of a density matrix $\rho$ of a quantum system with $N$ sites and a local Hilbert space of dimension $d$ at each site. $\rho$ can be decomposed in a basis of local operators as
\begin{equation}
  \rho = \sum_{\sigma_1,\dots,\sigma_N} R_{\sigma_1 \dots \sigma_N} O^{[1]\,\sigma_1} \otimes \cdots \otimes O^{[N]\,\sigma_N}
\end{equation}
where $O^{[k]\,\sigma_k}$ are local operators acting on site $k$ and $\sigma_k$ runs from $1$ to $d^2$.
As in \cref{eq:operator_vectorization}, the coefficients $R_{\sigma_1 \dots \sigma_N}$ can be viewed as the components of a {\em pure} state in a Hilbert space of dimension $d^{2N}$.
With this point of view, we may denote the product of local operators $O^{[1]\,\sigma_1}\otimes \cdots \otimes O^{[N]\,\sigma_N}$ as $|\sigma_1 \dots \sigma_N\rangle\rangle$ and write the mixed state $\rho$ as
\begin{equation}
  |\rho\rangle\rangle= \sum_{\sigma_1,\dots,\sigma_N} R_{\sigma_1\dots \sigma_N} |\sigma_1 \dots \sigma_N\rangle\rangle.
  \label{eq:rho_superket_full}
\end{equation}
Here the ``super-ket'' notation $|\cdots\rangle\rangle$ is used to distinguish the pure state $|\rho\rangle\rangle$ in the enlarged space from a usual pure state $|\psi\rangle$ in the original Hilbert space. In turn, the coefficients $R_{\sigma_1 \dots \sigma_N}$ can be decomposed as an MPS,
\begin{equation}
  R_{\sigma_1 \dots \sigma_N}
  =
  R^{[1]\,\sigma_1}
  \cdots
  R^{[N]\,\sigma_N}.
\end{equation}
The vectorized density matrix then takes the (super) MPS form
\begin{equation}
  |\rho\rangle\rangle
  =
  \sum_{\sigma_1,\dots,\sigma_N}
  R^{[1]\,\sigma_1}
  \cdots
  R^{[N]\,\sigma_N}
  |\sigma_1\dots \sigma_N\rangle\rangle.
  \label{eq:rho_superket_MPS}
\end{equation}
The main difference from an ordinary MPS is that the local physical index now labels a local operator, or equivalently a pair of local ket/bra indices, and therefore has dimension $d^2$ instead of $d$. Note that one may also view $\rho$ as an MPO with two physical indices per site. Folding the lower physical leg upward turns this MPO representation of $\rho$ into the MPS representation of $|\rho\rangle\rangle$, as illustrated in \cref{fig:rho_mpo_superket}.

Thanks to the representation of $|\rho\rangle\rangle$ as an MPS, most of the tools introduced in the previous sections for pure states can be used to manipulate mixed states. 

\begin{remarkblock}
This representation does not automatically enforce the positive-semidefinite character of the density matrix. 
In particular, after truncating the MPS representation of $|\rho\rangle\rangle$ there is no guarantee that the reconstructed $\rho$ is positive semidefinite. In practice, in numerical simulations, one monitors the trace, Hermiticity, and the convergence of observables with the MPS bond dimension.
It is also possible to use representations that automatically enforce the positivity of $\rho$, such as the purification representation $\rho = X X^\dagger$ with $X$ represented as an MPO~\cite{wernerPositiveTensorNetwork2016a}. See \cref{sec:thermal_purification}.
Such TNs, with a double-layer structure (one for $X$ and one for $X^\dagger$), are called matrix product density operators (MPDOs).
Even though it does not guarantee the positivity of $\rho$, the simple vectorized form is often used in practice since purification-based descriptions can be numerically quite expensive~\cite{cuevasPurificationsMultipartiteStates2013}.
As always in MPS simulations, the convergence of the results with the bond dimension should be checked.
\end{remarkblock}

The scalar product in the vectorized space is the Hilbert-Schmidt scalar product,
\begin{equation}
  \langle\langle A|B\rangle\rangle
  =
  \mathrm{Tr}(A^\dagger B).
\end{equation}
For instance,
\begin{equation}
  \mathrm{Tr}(\rho)=\langle\langle \mathbbm{1}|\rho\rangle\rangle,
  \qquad
  \mathrm{Tr}(\rho^2)=\langle\langle \rho|\rho\rangle\rangle
  \qquad
  \langle O\rangle = \mathrm{Tr}(O\rho)    =   \langle\langle O^\dagger|\rho\rangle\rangle
\end{equation}
(in $\mathrm{Tr}(\rho^2)$ we have used the fact that for a density matrix $\rho=\rho^\dagger$).

\subsection{Operator-space entanglement entropy}
\label{sec:OSEE}

The entanglement entropy of the vectorized state $|\rho\rangle\rangle$ across a spatial bipartition is called the {\em operator-space entanglement entropy} (OSEE)~\cite{zanardiEntanglementQuantumEvolutions2001a,prosenOperatorSpaceEntanglement2007,znidaricComplexityThermalStates2008a,pizornOperatorSpaceEntanglement2009,dubailEntanglementScalingOperators2017a,wellnitzRiseFallSlow2022a}.\footnote{When the density matrix is normalized by $\mathrm{Tr}(\rho)=1$, the vectorized state $|\rho\rangle\rangle$ is not in general normalized in the Hilbert-Schmidt scalar product: $\langle\langle \rho|\rho\rangle\rangle=\mathrm{Tr}(\rho^\dagger\rho)$, which is the purity of the density matrix. Thus, if the Schmidt decomposition
of $|\rho\rangle\rangle$ across a cut has singular values $\lambda_\alpha$, the weights that should be used to compute the OSEE are $p_\alpha=\lambda_\alpha^2/\sum_\beta \lambda_\beta^2$; equivalently, one first normalizes $|\rho\rangle\rangle$ in Hilbert-Schmidt norm.
Then ${\rm OSEE}=-\sum_\alpha p_\alpha \log p_\alpha$.}
The OSEE quantifies the amount of correlations, both quantum and classical, between the two subsystems in the mixed state $\rho$. The OSEE can also be used to characterize a unitary operator (rather than a mixed state).

Just as the entanglement entropy of a pure state controls the MPS bond dimension needed to represent it, the OSEE controls the bond dimension needed to represent a density matrix in the vectorized MPS form.
The OSEE is therefore an important quantity in the context of numerical simulations since 
it quantifies how expensive it is to represent $|\rho\rangle\rangle$ as an MPS.
Roughly speaking, a bond dimension $\chi$ can describe an OSEE up to order $\sim \log\chi$.
In the same way as entanglement entropy growth in pure-state dynamics may limit the use of TN methods, and MPS in particular, a fast growth of the OSEE during mixed-state or open-system dynamics may require large bond dimensions.

Two simple limits are useful to keep in mind.
If $\rho=|\psi\rangle\langle\psi|$ is a pure-state density matrix, then the OSEE of $|\rho\rangle\rangle$ is twice the von Neumann entanglement entropy of $|\psi\rangle$ for the same spatial bipartition.
On the other hand, if $\rho=\rho_1 \otimes \dots \otimes \rho_N$ is a product state, then its vectorized form is also a product state, and the OSEE vanishes (a special limit is the fully mixed case, $\rho\propto \mathbbm{1}$).

\begin{figure}
  \begin{center}
    \includegraphics[width=0.65\textwidth]{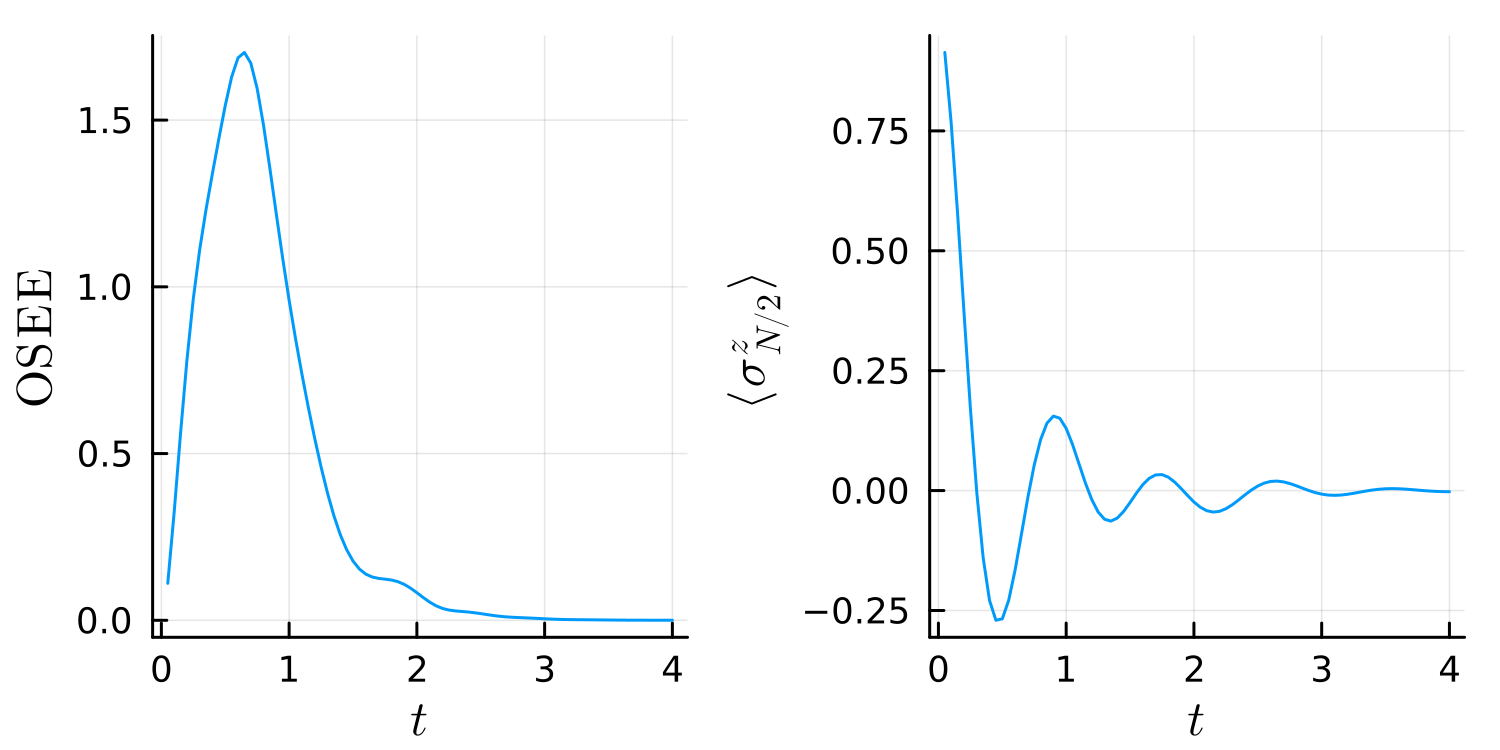}
  \end{center}
  \caption{Left: Operator-space entanglement entropy (OSEE) across the central bond plotted as a function of time during the Lindblad evolution of the dissipative XX chain simulated in \cref{lst:lindblad_XX_spin_chain_OSEE}.
  Right: Magnetization \(\langle \sigma^z_{N/2}\rangle\) at the central site.
  In this example, the
  Hamiltonian is the XX spin-1/2 Hamiltonian and the
  jump operators in the Lindbladian are local decay and excitation operators \(L_i=\sqrt{\Gamma}\sigma_i^\pm\). The dissipation drives the system toward the fully mixed state $\rho_{\rm ss}=\mathbbm{1}/2^N$.}
  \label{fig:lindblad_xx_OSEE}
\end{figure}

\Cref{lst:lindblad_XX_spin_chain_OSEE} is a code that computes the OSEE as a function of time for a dissipative spin chain. The results are shown in \cref{fig:lindblad_xx_OSEE}.
The initial growth of the OSEE is due to the buildup of correlations in the system under the action of the Hamiltonian, while the later decrease is due to dissipation, which drives the system toward a fully mixed product state (hence with zero OSEE).
For a discussion of the OSEE in the context of dissipative 1d systems, see for instance~\cite{wellnitzRiseFallSlow2022a}.

\subsection{Superoperators as MPOs}
\label{sec:superoperators_as_MPO}

\begin{figure}
  \begin{center}
    \includegraphics[width=0.6\textwidth]{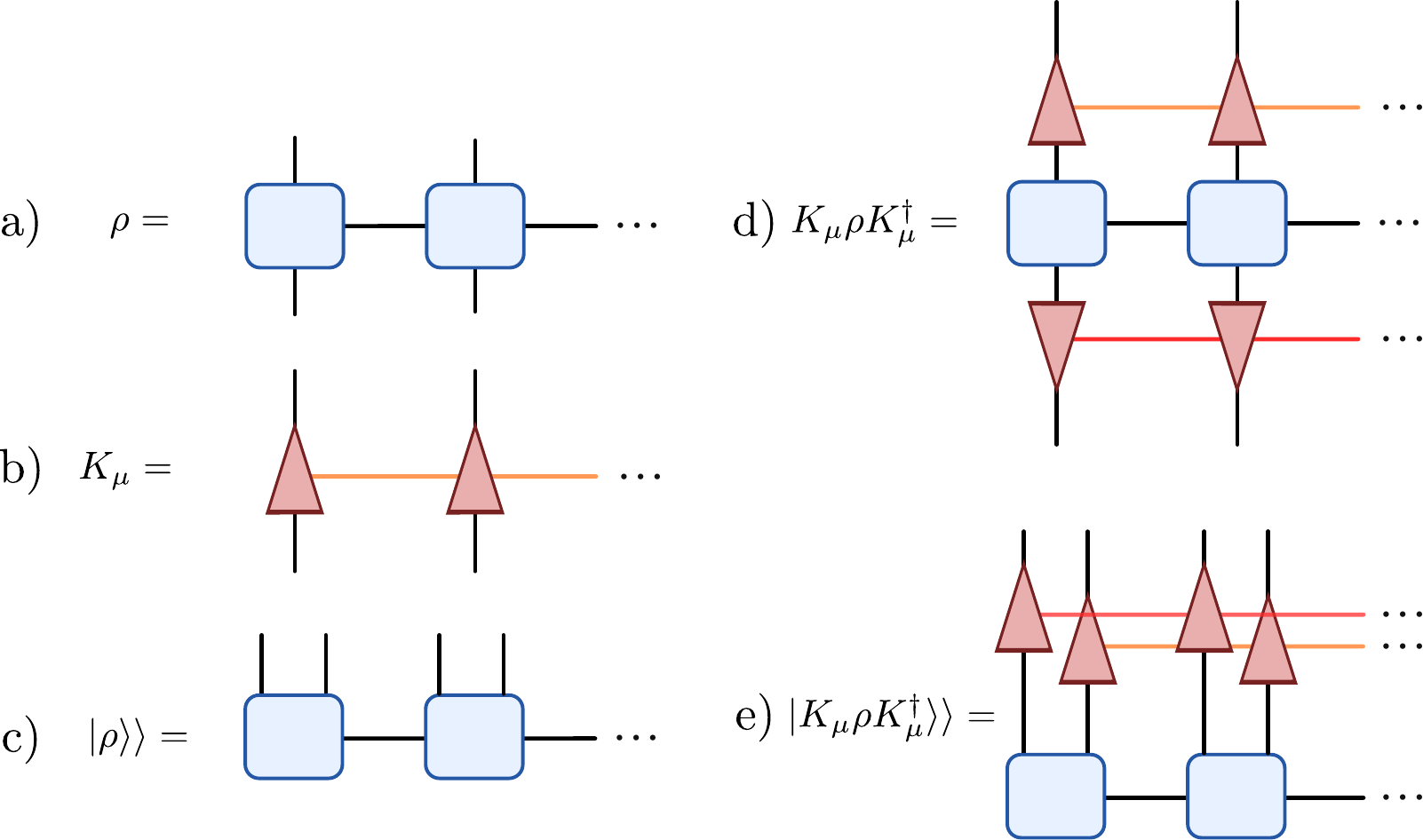}
    \caption{Diagrammatic representation of the vectorization procedure for a quantum channel in Kraus form.
    a) Density matrix $\rho$ as an MPO.
    b) Jump/Kraus operator $K_\mu$ as an MPO.
    c) Vectorized density matrix $|\rho\rangle\rangle$ as an MPS in the doubled Hilbert space
    (see also \cref{fig:rho_mpo_superket}).
    d) The result of the action of $K_\mu$ on $\rho$ is a product of three MPOs.
    e) After vectorization, the action of $K_\mu$ on $\rho$ becomes the application of an MPO on an MPS in the doubled Hilbert space.
    }
    \label{fig:MPO_kraus_superoperators}
  \end{center}
\end{figure}

With vectorization, linear maps acting on density matrices become ordinary linear operators acting on the super-ket $|\rho\rangle\rangle$.
With the convention of \cref{eq:operator_vectorization}, one has the identity 
\begin{equation}
  |A\rho B^\dagger\rangle\rangle
  =
  (A\otimes \overline{B})|\rho\rangle\rangle,
  \label{eq:vectorized_left_right_action}
\end{equation}
where the overline denotes complex conjugation in the local basis used for the vectorization.
In other words, acting on the left of the density matrix is equivalent to acting on the first copy of the doubled space, while acting on the right is equivalent to acting on the second copy.
Thus, left and right multiplications of $\rho$ become ordinary operators in the doubled Hilbert space. The idea is to express the operator $A\otimes \overline{B}$ as an MPO (acting on the doubled Hilbert space), so that $|A\rho B^\dagger\rangle\rangle$ can be computed by an MPO--MPS multiplication, as in \cref{sec:MPO}.
In the special case of a unitary gate acting on a mixed state we have 
\begin{equation}
  \rho \mapsto U \rho U^\dagger,
  \qquad
  |\rho\rangle\rangle \mapsto
  (U\otimes\overline{U})|\rho\rangle\rangle.
  \label{eq:unitary_gate_superoperator}
\end{equation}
Again, this can be viewed as the application of an MPO (representing $U\otimes\overline{U}$) on an MPS. 
More generally, a quantum channel is a completely positive map that can be written in Kraus form~\cite{breuer_theory_2007}
\begin{equation}
  \mathcal{E}(\rho)
  =
  \sum_\mu K_\mu \rho K_\mu^\dagger,
  \qquad
  \sum_\mu K_\mu^\dagger K_\mu = \mathbbm{1}
\end{equation}
for a trace-preserving channel. In vectorized form this becomes
\begin{equation}
  |\mathcal{E}(\rho)\rangle\rangle
  =
  \left(
  \sum_\mu K_\mu\otimes\overline{K_\mu}
  \right)
  |\rho\rangle\rangle.
  \label{eq:kraus_channel_superoperator}
\end{equation}
The operator in parentheses is a superoperator and can be represented as an MPO.
This is illustrated in \cref{fig:MPO_kraus_superoperators}.

\begin{figure}
  \begin{center}
    \includegraphics[width=0.55\textwidth]{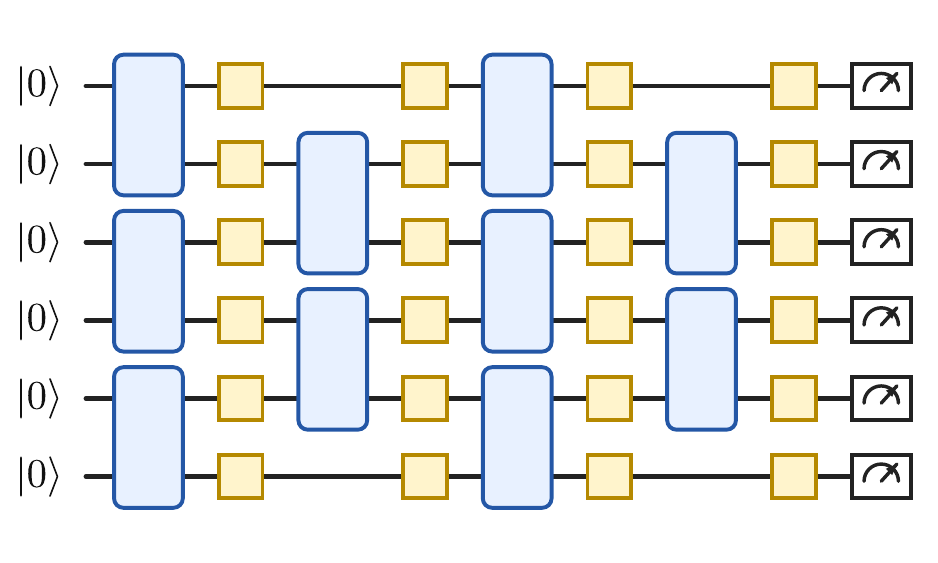}
    \caption{Brick-wall random circuit with Haar-random 2-qubit gates (blue) and single-qubit depolarizing channels (yellow), shown here for $N=6$ qubits and depth $D=4$.
    An MPS-based code that simulates this circuit is given in \cref{lst:noisy_random_circuit_xeb}.
    }
    \label{fig:noisy_random_circuit_xeb}
  \end{center}
\end{figure}

A code example that simulates a quantum circuit containing unitary gates
(as in \cref{eq:unitary_gate_superoperator})
and noise channels
(as in \cref{eq:kraus_channel_superoperator})
is given in \cref{lst:noisy_random_circuit_xeb}.
The code provides an example of a noisy random quantum circuit simulated with MPS~\cite{zhouWhatLimitsSimulation2020,ayral_density-matrix_2023}.
The initial state is the product state $\ket{0}^{\otimes N}$, represented as a mixed state. 
The circuit (shown in \cref{fig:noisy_random_circuit_xeb}) consists of layers of Haar-random two-qubit gates arranged in a brick-wall pattern, followed by a single-qubit depolarizing channel with error probability $p$. Because of these error gates, the output state of the circuit is a mixed state represented by a density matrix $\rho$.
In parallel the same circuit is also simulated with $p=0$.
In that case the circuit is unitary and the output $\rho_0=|\psi\rangle\langle\psi|$ is a pure state. The code compares the ideal output probability distribution (from $\rho_0$) with the noisy output probability distribution (from $\rho$) via the linear cross-entropy benchmarking (XEB) estimator that was popularized by the Google quantum supremacy experiment of 2019~\cite{aruteQuantumSupremacyUsing2019}.
The XEB estimator is defined as
\begin{equation}
  \mathcal{F}_{\rm XEB} = 2^N \sum_{\mathbf \sigma} p_{\rm noisy}(\mathbf \sigma) p_{\rm ideal}(\mathbf \sigma) - 1
  \label{eq:FXEB}
\end{equation}
where $p_{\rm noisy}(\mathbf \sigma)={\rm Tr}[\rho |\mathbf \sigma\rangle\langle \mathbf \sigma|]$
and $p_{\rm ideal}(\mathbf \sigma)={\rm Tr}[\rho_0 |\mathbf \sigma\rangle\langle \mathbf \sigma|]=|\langle \psi | \mathbf \sigma\rangle|^2$ 
are the probabilities of the computational-basis bitstring $\mathbf \sigma=(\sigma_1,\dots,\sigma_N)$ in the noisy and ideal output distributions, respectively. $\mathcal{F}_{\rm XEB}$ is an approximation of the fidelity between the noisy and ideal output states, and it is expected to decay exponentially with the number of qubits $N$ and the depth $D$ of the circuit, as $\mathcal{F}_{\rm XEB} \sim \exp(-\alpha N D)$, where $\alpha$ is a constant that depends on the error probability $p$ of the noise channel.

\begin{figure}
  \begin{center}
    \includegraphics[width=9cm]{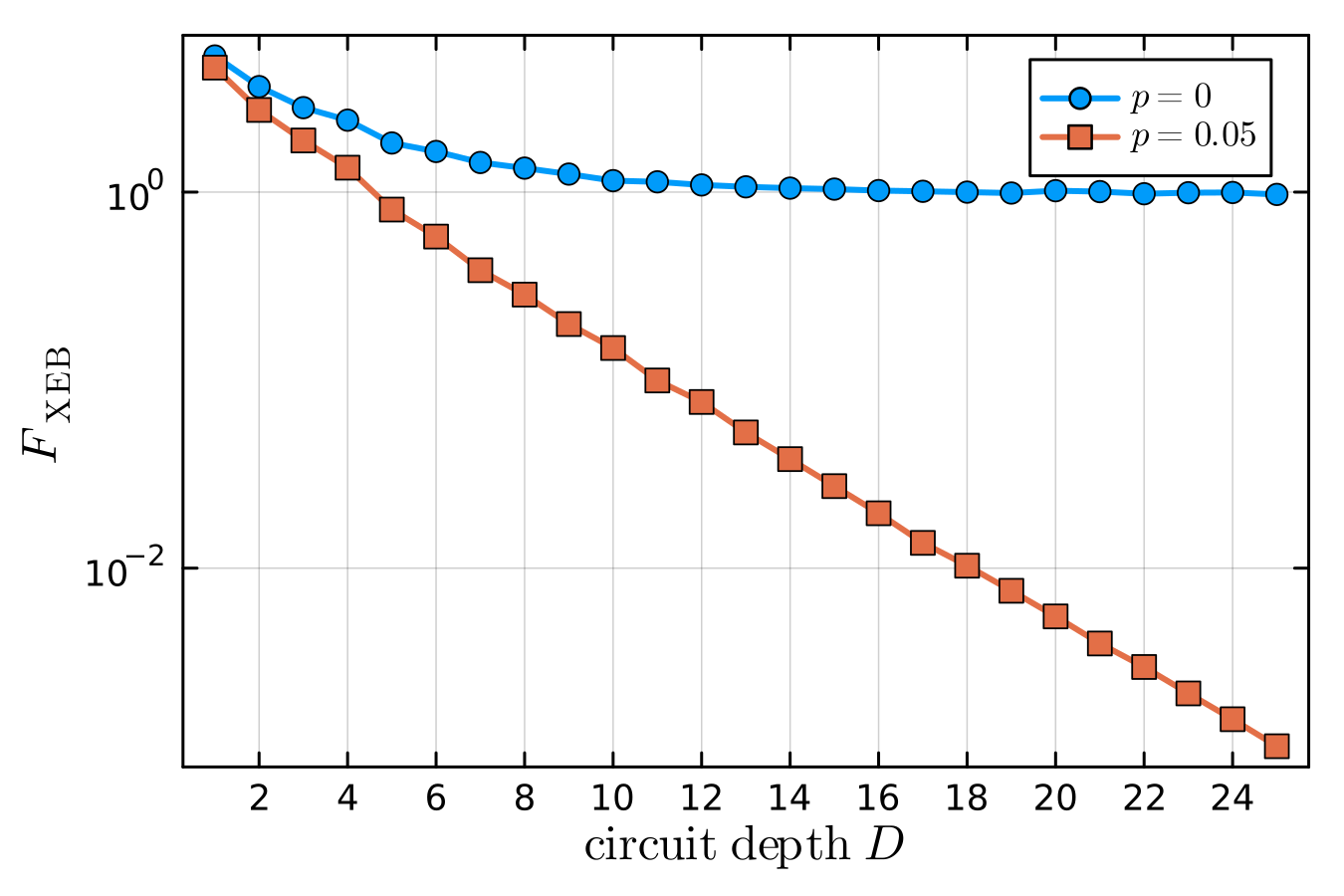}
    \caption{$\mathcal{F}_{\rm XEB}$ for random brick-wall circuits with $N=8$ qubits as a function of the circuit depth $D$. The two curves correspond to the ideal case, $p=0$, and to a single-qubit depolarizing channel with error probability $p=0.05$ applied after each unitary layer (see \cref{fig:noisy_random_circuit_xeb}).
    A code example for such a simulation is given in \cref{lst:noisy_random_circuit_xeb}.}
    \label{fig:noisy_random_circuit_xeb_depth_sweep}
  \end{center}
\end{figure}

Suppose that a quantum device is a noisy implementation of the ideal random circuit. If $N$ is large one cannot compute {\em all} the probabilities $p_{\rm ideal}(\mathbf \sigma)$ of the ideal output distribution, and one cannot run enough repetitions of the device circuit to estimate all the probabilities $p_{\rm noisy}(\mathbf \sigma)$. In that case, one can estimate $\mathcal{F}_{\rm XEB}$ by sampling a subset of bitstrings $\mathbf \sigma$ from the noisy distribution:
\begin{equation}
\mathcal{F}_{\rm XEB} \approx 2^N \frac{1}{M}\sum_{i=1}^M p_{\rm ideal}(\mathbf \sigma_i) - 1
\end{equation}
where $\mathbf \sigma_1,\dots,\mathbf \sigma_M$ are $M$ bitstrings sampled from the noisy distribution.\footnote{Even with such sampling, the estimation of $\mathcal{F}_{\rm XEB}$ remains computationally hard because one has to compute $p_{\rm ideal}(\mathbf \sigma_i)$ for each sampled bitstring, which requires simulating the ideal circuit. Specialized TN methods can target this task for many bitstrings~\cite{panSimulationQuantumCircuits2022}; in the present code this simulation is done with MPS. Due to the growth of the entanglement in the ideal state as a function of the circuit depth, the MPS bond dimension needed to accurately represent the ideal state grows exponentially with $D$. However, this is not an issue for small values of $N$.}
The \cref{lst:noisy_random_circuit_xeb} performs the sampling above (variable \texttt{NSAMPLES}). It also performs an average over several realizations of the random circuit (variable \texttt{NCIRCUITS}).
\Cref{fig:noisy_random_circuit_xeb_depth_sweep} shows the XEB estimator as a function of the circuit depth $D$ for $N=8$ qubits, in the ideal case $p=0$ and in the noisy case $p=0.05$. In the ideal case, the circuit approaches a Haar-random unitary as $D$ increases, and the XEB estimator approaches the value $\mathcal{F}_{\rm XEB}=1$. Such behavior is expected for a Haar-random state, as a consequence of the Porter-Thomas distribution of the output probabilities for a Haar-random state. In the noisy case, the XEB estimator decays exponentially with $D$. Note that, by construction, the XEB estimator is equal to $\mathcal{F}_{\rm XEB}=0$ for completely random outputs with $p_{\rm noisy}(\mathbf \sigma) = 2^{-N}$.

\begin{longlisting}
  \jcode{noisy_random_circuit_xeb.jl}
  \caption[Noisy random-circuit simulation]{\label{lst:noisy_random_circuit_xeb}
    Noisy random-circuit simulation with \texttt{TensorMixedStates}, including sampling of computational-basis bitstrings and evaluation of the linear XEB estimator.
    Internally, the density matrix $\rho$ is represented as a vectorized MPS, and the quantum channels (unitary gates and noise channels) as MPOs. 
    Haar-random two-qubit gates are generated by QR decomposition of random complex matrices and arranged in a brick-wall circuit. After each unitary layer, the code applies a one-qubit depolarizing channel with error probability \texttt{PNOISE}, samples bitstrings from the noisy output state, evaluates their ideal probabilities using the corresponding \(p=0\) circuit, and averages the resulting XEB estimate over several circuit realizations. It also prints the trace and purity, maximum bond dimensions, OSEE of \(\rho\), and the entanglement entropy of the ideal pure state.
    \jcodegithublink
  }
\end{longlisting}

\subsection{Lindblad dynamics in MPS form}
\label{sec:lindblad_MPS_form}

For Markovian open-system dynamics, the density matrix often obeys a Gorini-Kossakowski-Sudarshan-Lindblad master equation (called simply the Lindblad master equation below)~\cite{breuer_theory_2007}
\begin{equation}
  \frac{d\rho}{dt}
  =
  \mathcal{L}(\rho),
\end{equation}
with
\begin{equation}
  \mathcal{L}(\rho)
  =
  -i[H,\rho]
  +
  \sum_\mu
  \left(
  L_\mu \rho L_\mu^\dagger
  -
  \frac{1}{2}\{L_\mu^\dagger L_\mu,\rho\}
  \right),
  \label{eq:lindblad_equation}
\end{equation}
where the operators $L_\mu$ are the jump operators describing the non-unitary part of the dynamics.
In the vectorized representation this becomes an ordinary linear differential equation
\begin{equation}
  \frac{d}{dt}|\rho\rangle\rangle
  =
  \mathbb{L}|\rho\rangle\rangle,
  \label{eq:vectorized_lindblad_equation}
\end{equation}
where $\mathbb{L}$ is the Lindbladian operator acting in the doubled Hilbert space.
Using \cref{eq:vectorized_left_right_action}, the Hamiltonian part is
\begin{equation}
  -i[H,\rho]
  \quad \longrightarrow \quad
  -i\left(H\otimes \mathbbm{1} - \mathbbm{1}\otimes H^T\right)|\rho\rangle\rangle,
  \label{eq:lindblad_hamiltonian_superoperator}
\end{equation}
and a dissipative term gives
\begin{equation}
  L_\mu \rho L_\mu^\dagger
  -
  \frac{1}{2}\{L_\mu^\dagger L_\mu,\rho\}
  \quad \longrightarrow \quad
  \left[
  L_\mu\otimes\overline{L_\mu}
  -
  \frac{1}{2}L_\mu^\dagger L_\mu\otimes \mathbbm{1}
  -
  \frac{1}{2}\mathbbm{1}\otimes (L_\mu^\dagger L_\mu)^T
  \right]|\rho\rangle\rangle.
  \label{eq:lindblad_dissipative_superoperator}
\end{equation}


Like any linear operator acting on an MPS, the Hamiltonian contribution \cref{eq:lindblad_hamiltonian_superoperator}, the dissipative contributions \cref{eq:lindblad_dissipative_superoperator}, or the whole Lindbladian $\mathbb{L}$ can be represented as MPOs.
To construct such MPOs, one can use the finite-state automaton picture described in \cref{sec:MPO_finite_state_automata}.
Consider for instance the Hamiltonian part of the Lindbladian. Assume that $H$
contains $k$ terms acting across a given bond.
According to the finite-state automaton picture of \cref{sec:MPO_finite_state_automata}, the MPO representation of $H$ (as an operator acting on the original Hilbert space with local physical dimension $d$)
has a bond dimension $k+2$ across that bond.
When acting on a density matrix through $[H,\cdot]$, each term gives rise to {\em two} terms in the Lindbladian, one acting on the left and one acting on the right of $\rho$. So, when acting on $\vert\rho\rangle\rangle$, the Hamiltonian part of the Lindbladian contains $2k$ terms acting across the same bond.
The operator $[H,\cdot]$ can thus be represented as an MPO with bond dimension $2k+2$ and local physical dimension $d^2$.
Now consider the jump operators $L_\mu$ (\cref{eq:lindblad_equation}). Each jump operator gives rise to {\em three} terms in the Lindbladian, one of the form $L_\mu\otimes\overline{L_\mu}$ and two of the form $L_\mu^\dagger L_\mu\otimes \mathbbm{1}$ and $\mathbbm{1}\otimes (L_\mu^\dagger L_\mu)^T$. It is common
to have dissipative models where the jump operators are local (each $L_\mu$ acts on a single site).\footnote{
For a spin-$\frac{1}{2}$ system, typical examples of local jump operators are the so-called dephasing $L_\mu=\sqrt{\gamma}\sigma_i^z$
or amplitude damping or local relaxation $L_\mu=\sqrt{\gamma}\sigma_i^-$.}
In such cases, they do not increase the bond dimension of the Lindbladian MPO compared to the Hamiltonian part (they are included in the blocks $D^{[i]}$ of the MPO of $\mathbb{L}$; see \cref{eq:general_H_MPO}).

As before, the bond dimension of these MPOs will depend on the range of the interactions. 
The Lindblad evolution can therefore be viewed as an MPS time-evolution problem in operator space:
\begin{equation}
  |\rho(t)\rangle\rangle
  =
  e^{t\mathbb{L}}|\rho(0)\rangle\rangle.
\end{equation}
The TDVP method discussed in \cref{sec:TDVP} and the W$^{\rm I}$/W$^{\rm II}$ MPO methods discussed in \cref{sec:W} can thus be used 
in this context. They provide different ways
to approximate the dissipative evolution.
The short code in \cref{lst:MPO_bond_dim} illustrates how to inspect the MPO bond dimensions for an XX Hamiltonian, for the corresponding Hamiltonian-only Lindbladian, and for Lindbladians with local or two-site dephasing terms.

\begin{longlisting}
  \jcode{MPO_bond_dim.jl}
  \caption[MPO bond dimensions for Lindbladians]{\label{lst:MPO_bond_dim}
      The code builds MPO representations for an XX spin-$\frac{1}{2}$ Hamiltonian acting on a pure state, for the corresponding Hamiltonian-only Lindbladian acting on a density matrix, for local dephasing, and for XX Lindbladians with local or two-site dephasing. The bond dimensions of the resulting MPOs are printed.
    \jcodegithublink
  }
\end{longlisting}

Code examples based on the \(W^{\mathrm{II}}\) method to solve a Lindblad problem in a dissipative spin chain are given in \cref{lst:lindblad_XX_spin_chain,lst:lindblad_XX_spin_chain_OSEE}.

\begin{longlisting}
  \jcode{lindblad_XX_spin_chain.jl}
  \caption[Lindblad dynamics of a dissipative XX chain]{\label{lst:lindblad_XX_spin_chain}
    Lindblad time evolution of a dissipative XX spin-$\frac{1}{2}$ chain.
    The model is the boundary-driven XX chain with dephasing. The boundary reservoirs are parametrized by a driving strength $\Gamma$ 
    and a magnetization bias $\mu$
    (jump operators $L_{\pm,1}=\sqrt{2\Gamma(1\mp\mu)}\sigma_1^\pm$ and $L_{\pm,N}=\sqrt{2\Gamma(1\pm\mu)}\sigma_N^\pm$ at the left and right boundaries) 
    while dephasing is controlled by the rate $\gamma$ (jump operators of the form $L_i=\sqrt{\gamma}\sigma_i^z$ for $i=1,\dots,N$).
    The code builds the Lindbladian as an MPO (\texttt{make\_mpo}), and,
    starting from the infinite-temperature density matrix
    (\texttt{rho0 = State\{Mixed\}(System(N, Qubit()), "FullyMixed")}), evolves the density matrix up to a finite time $t$, and computes the
    magnetization profile $\langle \sigma_i^z\rangle$. These results are compared with the exact nonequilibrium steady-state result derived in Ref.~\cite{znidaricExactSolutionDiffusive2010}.
    The time evolution is computed by approximating the exponential $\exp(\delta t\,\mathbb{L})$ of the Lindbladian by a product of MPOs, using a fourth-order version of the W$^{\rm II}$ algorithm (\texttt{approx\_W} and \texttt{order=4}) (see \cref{sec:W}). With this scheme the time-step error is of order $\delta t^5$.
    \jcodegithublink
  }
\end{longlisting}

The high-level interface of \texttt{TensorMixedStates} can take care of the time-step loop and of the output files. A compact example is shown in \cref{lst:lindblad_XX_spin_chain_OSEE}.

\begin{longlisting}
  \jcode{lindblad_XX_spin_chain_OSEE.jl}
  \caption[Lindblad dynamics and OSEE in a dissipative XX chain]{\label{lst:lindblad_XX_spin_chain_OSEE}
    Lindblad time evolution of an XX spin chain with local ``decay'' jump operators \(L_i=\sqrt{\Gamma}\sigma_i^-\)
    and local ``excitation'' jump operators \(L_i=\sqrt{\Gamma}\sigma_i^+\).
    The combination of these dissipators drives the system toward the fully mixed steady state $\rho_{\rm ss}=\mathbbm{1}/2^N$.
    The initial state is the N{\'e}el state \(\ket{\uparrow\downarrow\uparrow\downarrow\cdots}\).
    The simulation uses the high-level \texttt{SimData}/\texttt{runTMS} interface of \texttt{TensorMixedStates}, which makes it easy to set up the time evolution and write multiple observables to files. 
    The code writes the OSEE across the central bond, \(\langle \sigma^z_{N/2}\rangle\), and the trace error as functions of time to files, then reads back the first two data files and saves the corresponding figure.
    The saved figure is displayed in \cref{fig:lindblad_xx_OSEE}.
    \jcodegithublink
  }
\end{longlisting}

The steady state satisfies
\begin{equation}
  \mathbb{L}|\rho_{\rm ss}\rangle\rangle=0.
\end{equation}
One variational strategy to approximate the steady state is then to search, within the MPS manifold, for a normalized vectorized density matrix minimizing
\(\|\mathbb{L}|\rho\rangle\rangle\|^2\), or equivalently for the ground state of the Hermitian positive-semidefinite operator
\(\mathbb{L}^\dagger\mathbb{L}\)~\cite{cuiVariationalMatrixProduct2015a}. One can construct an MPO representation of
\(\mathbb{L}^\dagger\mathbb{L}\) and use DMRG to find its ground state.\footnote{Note that if the bond dimension of the MPO for \(\mathbb{L}\) is \(\chi_{\mathbb L}\), then the bond dimension of the MPO for \(\mathbb{L}^\dagger\mathbb{L}\) is at most \(\chi_{\mathbb L}^2\).}
This approach is implemented in \texttt{TensorMixedStates}
(as the \texttt{SteadyState} phase in the high-level interface). Note, however, that the convergence of this DMRG method may be slow when
\(\mathbb{L}^\dagger\mathbb{L}\) has a small spectral gap above its zero eigenvalue, or if it has many low-lying eigenvalues.\footnote{These eigenvalues are the squares of the singular values of \(\mathbb{L}\).}
This difficulty is expected to become more pronounced if the gap closes in the thermodynamic limit. Our experience on several simple spin-chain models suggests that this situation is not uncommon.
See also the closely related method proposed in \cite{mascarenhasMatrixproductoperatorApproachNonequilibrium2015a},
which targets the steady state with DMRG, working directly with $\mathbb{L}$ rather than with $\mathbb{L}^\dagger\mathbb{L}$.

\section*{Acknowledgements}
\addcontentsline{toc}{section}{Acknowledgements}

I am grateful to the organizers of the 9$^{\rm th}$ Les Houches School on Computational Physics -- Open Quantum Systems:
Markus Holzmann, Cécile Repellin, Tommaso Roscilde and Marco Schir\`o.
I also thank Jérôme Houdayer for discussions and collaboration on the \texttt{TensorMixedStates} library~\cite{HOUDAYER_TensorMixedStatesJuliaLibrary_2026}.
This work is supported by France 2030 under French National Research Agency grant No. ANR-22-PETQ-0007 (PEPR integrated project EPiQ).

\clearpage
\appendix
\phantomsection
\addcontentsline{toc}{section}{Appendices}


\section{Singular-value decomposition (SVD) and QR factorization}
\label{sec:SVD_QR}

\subsection{SVD and Schmidt decompositions}
\label{sec:SVD_Schmidt}
The singular value decomposition (SVD) is a fundamental tool in the manipulation of TN states and MPS in particular. It allows us to decompose a matrix into a product of three matrices: $M = U \Lambda V^\dagger$, where $U$ and $V$ are unitary matrices and $\Lambda$ is a real and nonnegative diagonal matrix containing the singular values.
By truncating the small singular values in $\Lambda$, one can obtain an approximation of the original matrix $M$ with a smaller rank. This is a key step in many TN algorithms. The faster the singular values decay, the better the approximation obtained by keeping only a few singular values.

A key property of the SVD is that it gives the optimal low-rank approximation of a matrix. If
\begin{equation}
  M = U \Lambda V^\dagger
\end{equation}
is the SVD of $M$, with singular values ordered as $\Lambda_1\geq \Lambda_2\geq \cdots$, then the matrix
\begin{equation}
  M_r = U_r \Lambda_r V_r^\dagger
\end{equation}
obtained by keeping only the $r$ largest singular values and the corresponding columns of $U$ and $V$ is the best approximation to $M$ among all matrices $M'$ of rank at most $r$, in the sense that it minimizes the Frobenius norm
\begin{equation}
  \delta^2 =
  \mathrm{Tr}\left[(M-M')^\dagger(M-M')\right].
\end{equation}
The minimal error is
\begin{equation}
  \delta^2_{\rm min} = \sum_{\alpha>r} \Lambda_\alpha^2 .
\end{equation}
This result is known as the Eckart--Young theorem~\cite{ECKART_ApproximationOneMatrix_1936}.

\begin{figure}
  \begin{center}
    \includegraphics[width=0.9\textwidth]{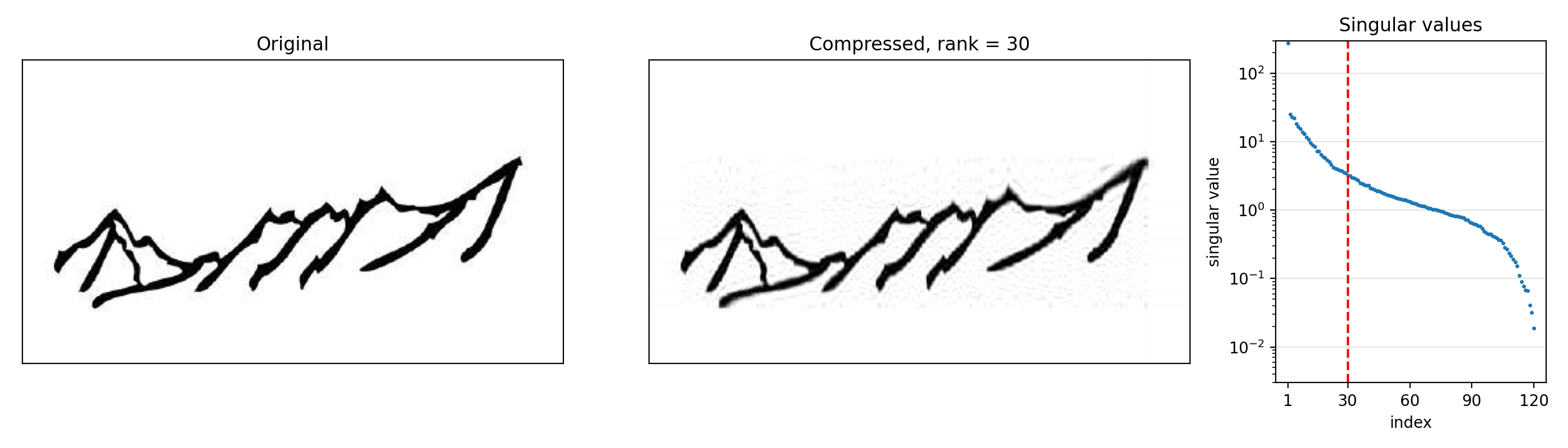}
  \end{center}
  \caption{Example of image compression by keeping only the 30 largest singular values of a grayscale image (Les Houches School of Physics logo). The right panel shows the singular values in descending order and the vertical dotted line indicates the cutoff for keeping only the 30 largest singular values (and associated singular vectors).
  The image size is $394 \times 221$ pixels and only the first 120 singular values (out of 221) are shown in the right panel. 
  }
  \label{fig:svd_image_compression}
\end{figure}

An image can be considered as a matrix, and \cref{fig:svd_image_compression} shows an example of image compression by keeping only a small fraction of the largest singular values of a grayscale image. Numerically, the SVD of a matrix of size $n\times m$ takes a time of the order $\mathcal{O}\left(n m\cdot \min(n,m)\right)$. 

The SVD is also important for the Schmidt decomposition of a bipartite state $|\psi\rangle = \sum_{a,b} M_{ab} |a\rangle \otimes |b\rangle$
defined in the Hilbert space $\mathcal{H} = \mathcal{H}_A \otimes \mathcal{H}_B$ (of dimension $d_A\times d_B$), where $|a\rangle$ and $|b\rangle$ are orthonormal bases of the two subsystems $A$ and $B$, respectively.
The SVD of $M$ allows one to express $|\psi\rangle$ as
\begin{equation}
  |\psi\rangle = \sum_i \Lambda_i |u_i\rangle \otimes |v_i\rangle
\end{equation}
where $\Lambda_i$ are the singular values (real and nonnegative), and
$|u_i\rangle = \sum_{a=1}^{d_A} U_{ai}|a\rangle$
and
$|v_i\rangle= \sum_{b=1}^{d_B} \overline{V_{bi}}|b\rangle$
form orthonormal families in the two subsystems $A$ and $B$, respectively.
The normalization of $|\psi\rangle$ implies that $\sum_i \Lambda_i^2 = 1$.

This decomposition is important in quantum information theory and is closely related to the entanglement properties of quantum states. Indeed, the Schmidt decomposition provides the diagonal forms of the reduced density matrices of the two subsystems:
\begin{equation}
  \rho_A = \mathrm{Tr}_B |\psi\rangle\langle\psi| = \sum_i \Lambda_i^2 |u_i\rangle\langle u_i|
\end{equation}
\begin{equation}
  \rho_B = \mathrm{Tr}_A |\psi\rangle\langle\psi| = \sum_i \Lambda_i^2 |v_i\rangle\langle v_i|.
\end{equation}
From the expression above we see that the singular values $\Lambda_i$ are directly related to the entanglement spectrum of the state $|\psi\rangle$ with respect to the A/B partition. In particular, the von Neumann entanglement entropy of the state $|\psi\rangle$ with respect to this partition is given by
\begin{equation}
  S_{\rm vN} = -\sum_i \Lambda_i^2 \log \Lambda_i^2.
\end{equation}

\begin{figure}
  \begin{center}
    \includegraphics[width=0.324\textwidth]{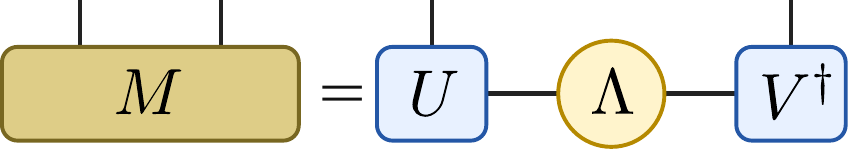}
  \end{center}
  \caption{
    Graphical representation of the singular value decomposition (SVD) of a matrix $M$ into $U \Lambda V^\dagger$.
    This can also be viewed as the Schmidt decomposition of a bipartite state $|\psi\rangle$ into $|\psi\rangle = \sum_\alpha \Lambda_\alpha |u_\alpha\rangle \otimes |v_\alpha\rangle$, where $\Lambda_\alpha$ are the singular values, and $|u_\alpha\rangle$ and $|v_\alpha\rangle$ are the left and right singular vectors, respectively.
  }
  \label{fig:svd}
\end{figure}

The graphical representation of the SVD is shown in \cref{fig:svd}. The SVD can be used to perform various operations on MPS, such as truncating the state by approximating it with an MPS with lower bond dimension or bringing the MPS into a canonical form.
Concerning truncation, the Eckart--Young theorem guarantees that truncating the SVD spectrum is the optimal way to approximate a bipartite state by a state with a smaller Schmidt rank (and hence a smaller bond dimension). 

\subsection{QR decomposition}
\label{sec:QR}

The QR decomposition of an $m\times n$ complex rectangular matrix $A$ is a factorization of the form $A = QR$ where $Q$ is an $m\times m$ unitary matrix ($Q^\dagger Q=1$) and $R$ is an $m\times n$ upper-triangular (more generally, upper-trapezoidal) matrix.\footnote{If $A$ is real then $Q$ is orthogonal, $Q^tQ=1$.}
One way to construct it is to apply the Gram--Schmidt orthogonalization procedure to the columns of $A$ and, if necessary, complete the resulting orthonormal family into a basis of $\mathbb C^m$. The columns of $A$ are then expressed in this new basis, and the corresponding coefficients form the entries of $R$. For a wide matrix ($n>m$), there are at most $m$ orthonormal vectors in $\mathbb C^m$; the remaining columns of $A$ are still represented by the corresponding columns of the upper-trapezoidal matrix $R$.

When $m>n$ the matrix $A$ is ``tall'' (more rows than columns). In such a case it is easy to see that the last $m-n$ rows of $R$ are zero by construction, so that the last $m-n$ columns of $Q$ do not contribute to the product $QR$. For this reason, one can consider instead the ``thin'' or ``reduced'' QR decomposition, where $Q$ is an $m\times n$ matrix with orthonormal columns ($Q^\dagger Q=1$) and $R$ is an $n\times n$ upper triangular matrix. Such a reduced QR decomposition can be obtained from the standard QR decomposition by keeping only the first $n$ columns of $Q$ and the first $n$ rows of $R$.

\begin{figure}
  \begin{center}
    \includegraphics[width=0.544\textwidth]{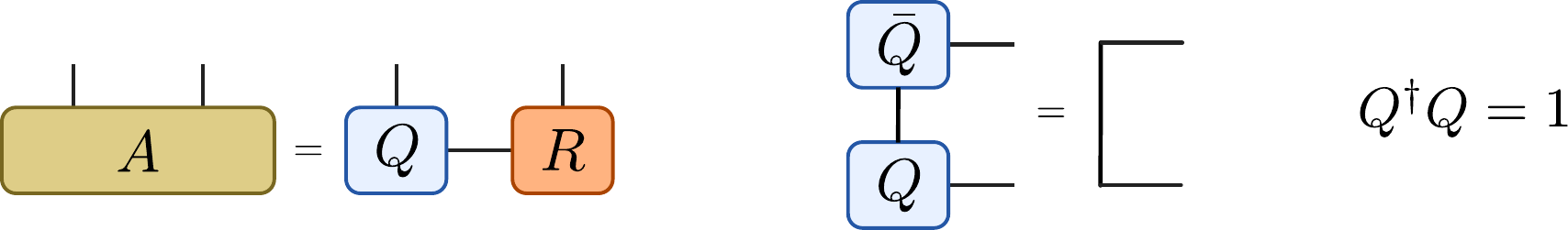}
  \end{center}
  \caption{
    Graphical representation of the QR decomposition. The matrix $Q$ is unitary and the bottom part is the graphical representation for $Q^\dagger Q=1$. The matrix $R$ is upper triangular.
    Why do we choose to write $\bar Q$ (and not $Q^\dagger$) in the above graphical representation? The reason is that the information about which index is contracted with which index is already included in the representation: we see that the ``vertical'' index of the lower tensor is contracted with the same vertical index (turned down to simplify the drawing) of the upper tensor (see also \cref{fig:tensor_notation_basics}).
  }
  \label{fig:QR}
\end{figure}


\section{Injective MPS}
\label{sec:injective_MPS}

\subsection{Map from the virtual space to the physical space}
\label{sec:injective_MPS_def1}
The notion of injective MPS was introduced in \cite{PEREZ-GARCIA_MatrixProductState_2007}.
Injective MPS are generic and they enjoy several useful mathematical properties.
Intuitively, injectivity means that a sufficiently long block of physical sites contains enough information to distinguish all possible virtual boundary conditions. In a non-injective MPS, different virtual boundary matrices may lead to the same physical block state. 

\begin{figure}[htbp]
  \begin{center}
    \includegraphics[width=0.467\textwidth]{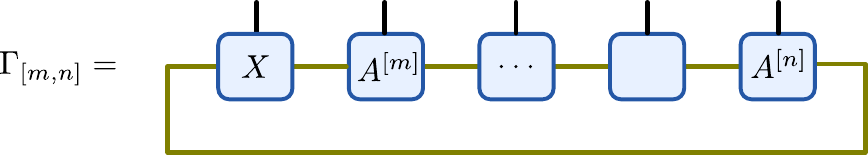}
  \end{center}
  \caption{Graphical representation of the map \(\Gamma_{[m,n]}\), which contracts a boundary matrix \(X\) with the MPS tensors on the block from site \(m\) to site \(n\), leaving the physical indices of the block ``open'' (\cref{eq:Gamma_map0}).}
  \label{fig:map_Gamma}
\end{figure}

Consider the following MPS
\begin{equation}
  |\psi\rangle   =   \sum_{s_1,\dots,s_N}   A^{[1]\,s_1}A^{[2]\,s_2}\cdots A^{[N]\,s_N}  \,|s_1\dots s_N\rangle
\end{equation}
where the matrices have dimension $\chi\times \chi$ (except for the first and last ones, which are $1\times \chi$ and $\chi\times 1$, respectively)
and the local Hilbert space has dimension $d$.
Next, consider a block of consecutive sites extending from site $m>1$ to site $n<N$.
We can define a linear map $\Gamma_{[m,n]}:\mathrm{Mat}_{\chi}(\mathbb C)\to (\mathbb C^d)^{\otimes (n-m+1)}$ from the doubled virtual space (of dimension $\chi^2$) to the physical Hilbert space of dimension $d^{(n-m+1)}$ of the block as follows:\footnote{The trace below is a compact way of contracting the two virtual boundary indices of the block with the boundary matrix \(X\). It should not be
confused with the trace used to close a periodic MPS.}
\begin{equation}
  \Gamma_{[m,n]} : X \mapsto \sum_{s_m,\dots,s_n}
  \operatorname{Tr}\!\left(
  X\,A^{[m]\,s_m}A^{[m+1]\,s_{m+1}}\cdots A^{[n]\,s_n}
  \right)
  \,|s_m\dots s_n\rangle .
  \label{eq:Gamma_map0}
\end{equation}
This can also be written using a more compact notation as
\begin{equation}
  \Gamma_{[m,n]}(X) = \sum_{\mathbf s} \operatorname{Tr}\!\left(X A^{\mathbf s}\right) \ket{\mathbf s}
  \label{eq:Gamma_map}
\end{equation}
where $\mathbf s=(s_m,\dots,s_n)$.
This map is represented graphically in \cref{fig:map_Gamma}.

The block is called injective if the map $\Gamma_{[m,n]}$ is injective.
Injectivity implies that the space spanned by the states $\Gamma_{[m,n]}(X)$ has dimension $\chi^2$.
Generic MPS are injective in the sense that non-injective MPS form a set of measure zero in the space of MPS. Injective MPS have several important properties, which make them particularly useful for both analytical and numerical purposes. For instance, they are the unique ground states of their local gapped parent Hamiltonians~\cite{FANNES_FinitelyCorrelatedStates_1992,PEREZ-GARCIA_MatrixProductState_2007,CIRAC_MatrixProductStates_2021}.

If the MPS is such that $A^{[k]}=A$ for all $k$, then the MPS is said to be injective if there is some length $r$ such that the map $\Gamma_{[1,r]}$ is injective. In the following we will focus, for simplicity, on this case of translationally invariant MPS where all the matrices are identical, $A^{[k]}=A$.

\subsection{Space of matrices spanned by products of MPS matrices}
\label{sec:injective_MPS_matrix_span}

Another definition of injectivity is the following: a block of $r$ consecutive sites is injective if the matrix space spanned by the products $A^{s_1}A^{s_2}\cdots A^{s_r}$ has dimension $\chi^2$ (and thus spans the full space of $\mathrm{Mat}_{\chi}(\mathbb C)$).
This second definition is in fact equivalent to the first one, as we show below. 

We define $A^{\mathbf s}=A^{s_1}\cdots A^{s_r}$ with $\mathbf s=(s_1,\dots,s_r)$ and consider the block map \cref{eq:Gamma_map}
$\Gamma_r(X)=\sum_{\mathbf s} \operatorname{Tr}\!\left(X A^{\mathbf s}\right) \ket{\mathbf s}$. We will prove that
\[
\Gamma_r \text{ is injective}
\quad\Longleftrightarrow\quad
\operatorname{span}\{A^{\mathbf s}\}_{\mathbf s}
=
\mathrm{Mat}_{\chi}(\mathbb C).
\]

First, suppose that $\operatorname{span}\{A^{\mathbf s}\}_{\mathbf s} = \mathrm{Mat}_{\chi}(\mathbb C)$.
If \(\Gamma_r(X)=0\), then $\operatorname{Tr}\!\left(X A^{\mathbf s}\right)=0$ $\forall \mathbf s$.
Since the matrices \(A^{\mathbf s}\) span \(\mathrm{Mat}_{\chi}(\mathbb C)\), it follows that
$\forall Y\in \mathrm{Mat}_{\chi}(\mathbb C)$ we have $\operatorname{Tr}(X Y)=0$. Choosing $Y=X^\dagger$ we get $X=0$.
Hence, \(\Gamma_r\) is injective.

Conversely, suppose that \(\Gamma_r\) is injective. Assume, for contradiction, that
$S=\operatorname{span}\{A^{\mathbf s}\}_{\mathbf s} \neq \mathrm{Mat}_{\chi}(\mathbb C)$.
Since the bilinear pairing $(X,Y)\mapsto \operatorname{Tr}(XY)$ is nondegenerate, we can pick a nonzero matrix $X$ such that $\operatorname{Tr}(XY)=0$ for all $Y\in S$.
In particular, $\operatorname{Tr}\!\left(X A^{\mathbf s}\right)=0$ $\forall \mathbf s$.
Therefore, $\Gamma_r(X) = 0$. This contradicts the injectivity of $\Gamma_r$.
Hence, $\operatorname{span}\{A^{\mathbf s}\}_{\mathbf s} = \mathrm{Mat}_{\chi}(\mathbb C)$.

\subsection{Parent Hamiltonian of an injective MPS}
\label{sec:injective_MPS_parent_Hamiltonian}

Consider a translation-invariant MPS on a chain of \(N\) sites, with bond dimension \(\chi\):
\[ |\psi\rangle = \sum_{s_1,\dots,s_N} \operatorname{Tr}\left(A^{s_1}A^{s_2}\cdots A^{s_N}\right)
|s_1 s_2\dots s_N\rangle ,
\]
where each physical index \(s\) takes values in \(\{1,\dots,d\}\), and each matrix \(A^s\) has dimension \(\chi\times \chi\).
We will construct a local Hamiltonian
\[ H=\sum_n h_n \]
such that \(|\psi\rangle\) is a ground state.


Take a block of \(r\) consecutive sites and consider, as in \cref{sec:injective_MPS_def1}, the linear map
\[
\Gamma_r : X\in \mathrm{Mat}_{\chi}(\mathbb C)
\longmapsto
\sum_{s_1,\dots,s_r}
\operatorname{Tr}\left(X A^{s_1}\cdots A^{s_r}\right)
|s_1,\dots,s_r\rangle .
\]
Equivalently, one can write the block states
\[
|\phi_{\alpha\beta}^{[r]}\rangle
=
\sum_{s_1,\dots,s_r}
\left(A^{s_1}\cdots A^{s_r}\right)_{\alpha\beta}
|s_1,\dots,s_r\rangle ,
\]
where \(\alpha,\beta=1,\dots,\chi\).
These \(\chi^2\) states span a subspace
\[
  {\rm Im}(\Gamma_r) = \operatorname{span}
\left\{
|\phi_{\alpha\beta}^{[r]}\rangle
\right\}.
\]
This is the set of all possible physical states that can appear on a block of \(r\) sites when the virtual boundary indices of that block are left open.

For an injective MPS, for sufficiently large \(r\), the map \(\Gamma_r\) is injective. Therefore
$\dim {\rm Im}(\Gamma_r)  = \chi^2$. 
Since the full block Hilbert space has dimension \(d^r\), for large enough \(r\) one has $\chi^2 < d^r$,
so \({\rm Im}(\Gamma_r)\) is a proper subspace of the block Hilbert space.

We want to construct a Hamiltonian that penalizes local states that do not belong to \({\rm Im}(\Gamma_r)\).
For this we introduce the projector $\Pi_r$ onto \({\rm Im}(\Gamma_r)\) and define
\[
h = \mathbb I - \Pi_r.
\]
\(h\) is a positive-semidefinite operator acting on \(r\) consecutive sites, and
\[
h|\varphi\rangle = 0
\]
if and only if
\[
|\varphi\rangle \in {\rm Im}(\Gamma_r).
\]
The parent Hamiltonian is then defined as the sum of these local projectors:
\[
H = \sum_{n=1}^N h_{n,n+1,\dots,n+r-1},
\]
where \(h_{n,n+1,\dots,n+r-1}\) acts as \(h\) on the block \((n,\dots,n+r-1)\).
For periodic boundary conditions, the site labels are understood modulo \(N\).

\begin{figure}[htbp]
  \begin{center}
    \includegraphics[width=0.673\textwidth]{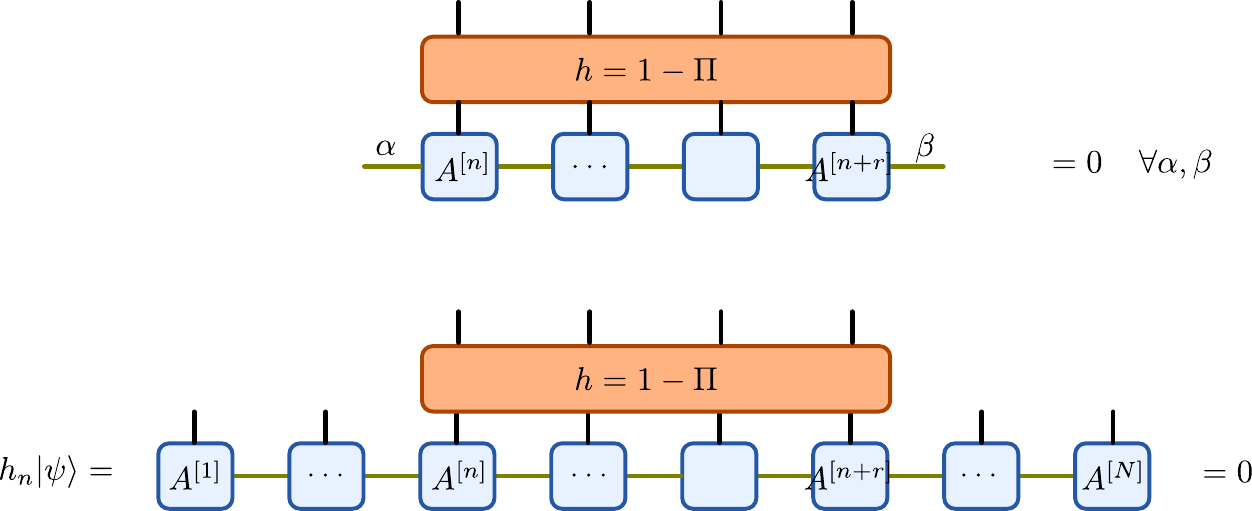}
  \end{center}
  \caption{Parent Hamiltonian of an injective MPS. Each local term \(h=\mathbb I-\Pi\) penalizes the orthogonal complement of the image of the block map on \(r\) consecutive sites.}
  \label{fig:H_parent}
\end{figure}

Now consider the full MPS and isolate a block of \(r\) sites. The rest of the chain supplies some effective virtual boundary matrix \(X\) and the wavefunction restricted to that block is always a vector in
\(
  {\rm Im}(\Gamma_r)
\)
and, as illustrated in \cref{fig:H_parent}, 
 \[
h_{n,n+1,\dots,n+r-1}|\psi\rangle=0
\]
for every \(n\).
Hence
\[
H|\psi\rangle=0.
\]
Since each term \(h_n\) is positive semidefinite,
\[
H\geq 0,
\]
so \(|\psi\rangle\) is a ground state with ground-state energy zero.
This type of Hamiltonian is called frustration-free:
\[
h_n|\psi\rangle=0
\qquad \text{for all } n.
\]

For an injective MPS, the corresponding parent Hamiltonian has \(|\psi\rangle\) as its unique ground state on a sufficiently large periodic chain, and the spectral gap remains open in the thermodynamic limit~\cite{FANNES_FinitelyCorrelatedStates_1992,PEREZ-GARCIA_MatrixProductState_2007} (proof not given here). 

We also mention another property of injective MPS: their transfer matrix has a unique eigenvalue of largest magnitude~\cite{CIRAC_MatrixProductStates_2021} (proof not given here). As discussed in \cref{sec:MPS_correlation_function}, such nondegeneracy implies that the correlation functions of an injective MPS decay exponentially with distance. Finally, we mention that the concept of injectivity can be generalized to higher-dimensional tensor networks, such as PEPS~\cite{CIRAC_MatrixProductStates_2021}.

\section{MPS representation of the AKLT state}
\label{sec:AKLT_MPS}

\begin{figure}
  \begin{center}
  \includegraphics[width=0.398\textwidth]{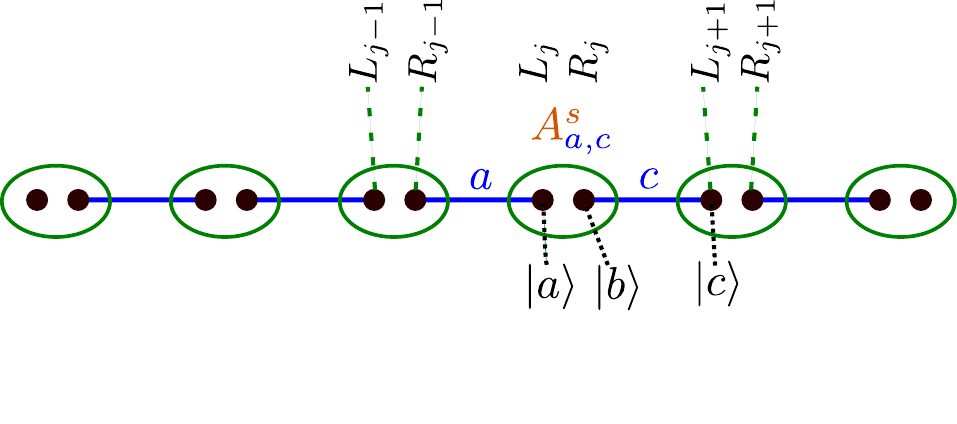}
  \caption{MPS representation of the AKLT valence-bond solid state
  (see also \cref{fig:AKLT}). The figure represents the index convention used in \cref{eq:AKLT_As}.
  }
  \label{fig:AKLT_MPS}
  \end{center}
\end{figure}

The construction of the AKLT state given in \cref{sec:AKLT} can be used to construct an explicit MPS representation. To do so, one associates 
the virtual spin-\(\frac12\) degrees of freedom to {\em bond} degrees of freedom in the MPS, as explained below.
At each site \(j\), we have two virtual spin-\(\frac12\)'s, denoted by \(L_j\) and \(R_j\). The
right virtual spin of site \(j\) is paired into a singlet with the left virtual
spin of site \(j+1\) (\cref{fig:AKLT_MPS}) 
\[
|\omega\rangle_{R_j,L_{j+1}}
=
\frac{1}{\sqrt{2}}
\left(
|\uparrow\rangle_{R_j}|\downarrow\rangle_{L_{j+1}}
-
|\downarrow\rangle_{R_j}|\uparrow\rangle_{L_{j+1}}
\right).
\]
At each site, the two virtual spins \(L_j\) and \(R_j\) need to be projected onto the spin-1 subspace and a natural basis for this subspace is
\[
|+\rangle = |\uparrow\uparrow\rangle,
\qquad
|0\rangle =
\frac{1}{\sqrt{2}}
\left(
|\uparrow\downarrow\rangle
+
|\downarrow\uparrow\rangle
\right),
\qquad
|-\rangle = |\downarrow\downarrow\rangle .
\]
The projection operator onto the $S=1$ subspace reads
\[
P
=
|+\rangle\langle\uparrow\uparrow|
+
\frac{|0\rangle}{\sqrt{2}}
\left(
\langle\uparrow\downarrow|
+
\langle\downarrow\uparrow|
\right)
+
|-\rangle\langle\downarrow\downarrow| .
\]
We now derive the MPS matrices.
The basis states of the virtual spin-1/2 are denoted by \(a,b\in\{\uparrow,\downarrow\}\). Thus, from the above expression for the projector we get its matrix elements
\[
P^s_{ab}
=
\langle s|P|a,b\rangle ,
\qquad
s\in\{+,0,-\}.
\]
Thus,
\[
P^+_{ab} = \delta_{a,\uparrow}\delta_{b,\uparrow},
\]
\[
P^0_{ab}
=
\frac{1}{\sqrt{2}}
\left(
\delta_{a,\uparrow}\delta_{b,\downarrow}
+
\delta_{a,\downarrow}\delta_{b,\uparrow}
\right),
\]
and
\[
P^-_{ab} = \delta_{a,\downarrow}\delta_{b,\downarrow}.
\]

To construct the MPS tensors we have to take into account the fact that in each bond the virtual spins $\frac12$ are paired into a singlet. 
To do this, we introduce an MPS bond index connecting site $j$ and $j+1$ that takes two values (bond dimension $\chi=2$).
We choose the value of this index to be the state of the virtual spin $L_{j+1}$ on the {\em right} side of the bond.
If $R_j$ is in the basis state $b$, then $L_{j+1}$ is in state $c=\bar b$. Including the sign of the singlet wave function,
the wave function of the singlet is simply
$|\omega\rangle_{R_j,L_{j+1}}=\sum_{b,c=\uparrow,\downarrow}\epsilon_{bc}|b\rangle_{R_j}|c\rangle_{L_{j+1}}$, where $\epsilon_{bc}$ is the antisymmetric tensor defined by
\[
\epsilon
  =
  \frac{1}{\sqrt{2}}
  \begin{pmatrix}
    0 & 1 \\
    -1 & 0
  \end{pmatrix},
\]
The MPS matrices can then be chosen as
\begin{equation}  
A^s_{a c}
=
\sum_{b=\uparrow,\downarrow}
P^s_{ab}\,\epsilon_{bc}.
\label{eq:AKLT_As}
\end{equation}
$s\in\{+,0,-\}$ is the physical index in the MPS,
$a$ is at the same time the MPS index of the bond $(j-1,j)$ and the state of the virtual spin $L_j$,
and $c$ is the MPS index of the bond $(j,j+1)$ and the state of the virtual spin $L_{j+1}$.
The index $b$ is the state of the virtual spin $R_j$, which is related to $c$ by $\epsilon$.
This index convention is illustrated in \cref{fig:AKLT_MPS}.

The final step is to explicitly compute the three matrices. For \(s=+\), only
\(P^+_{\uparrow\uparrow}=1\) is nonzero. Therefore
\[
A^+_{ac}
=
\sum_b P^+_{ab}\epsilon_{bc}
=
\delta_{a,\uparrow}\epsilon_{\uparrow c},
\]
so that
\[
A^+
=
\frac{1}{\sqrt{2}}
\begin{pmatrix}
0 & 1 \\
0 & 0
\end{pmatrix} = \frac{1}{\sqrt{2}} \sigma^+.
\]
For \(s=0\), the only nonzero coefficients are
\[
P^0_{\uparrow\downarrow}
=
P^0_{\downarrow\uparrow}
=
\frac{1}{\sqrt{2}} .
\]
Hence,
\[
A^0_{ac}
=
\frac{1}{\sqrt{2}}
\left(
\delta_{a,\uparrow}\epsilon_{\downarrow c}
+
\delta_{a,\downarrow}\epsilon_{\uparrow c}
\right),
\]
which gives
\[
A^0
=
\frac{1}{2}
\begin{pmatrix}
-1 & 0 \\
0 & 1
\end{pmatrix} = -\frac{1}{2}\sigma^z.
\]
Finally, for \(s=-\), only \(P^-_{\downarrow\downarrow}=1\) is nonzero, and
therefore
\[
A^-_{ac}
=
\delta_{a,\downarrow}\epsilon_{\downarrow c}
\]
and
\[
A^-
=
\frac{1}{\sqrt{2}}
\begin{pmatrix}
0 & 0 \\
-1 & 0
\end{pmatrix} = -\frac{1}{\sqrt{2}} \sigma^-.
\]

Equivalently, one can group the physical spin-1 states into the entries of a state-valued matrix
\begin{equation}
  \mathcal{A}
  =
  \sum_{s\in\{+,0,-\}} A^s\ket{s}
  =
  \begin{pmatrix}
    -\frac{1}{2}\ket{0} & \frac{1}{\sqrt{2}}\ket{+}\\[1mm]
    -\frac{1}{\sqrt{2}}\ket{-} & \frac{1}{2}\ket{0}
  \end{pmatrix}.
  \label{eq:AKLT_state_valued_A}
\end{equation}
There is no remaining physical index on $\mathcal{A}$; each matrix entry is now a ket in the local spin-1 Hilbert space. The AKLT state on a periodic chain is obtained by taking the virtual trace of the product of $N$ copies of this state-valued matrix, with tensor products of the local kets understood in the multiplication.

These matrices give an MPS representation of the AKLT state. Note that other normalizations and other gauges can be found in the literature.
On a periodic chain the AKLT state can then be written as
\[
|\Psi_{\mathrm{AKLT}}\rangle
=
\sum_{s_1,\ldots,s_N\in\{+,0,-\}}
\mathrm{Tr}
\left[
A^{s_1}A^{s_2}\cdots A^{s_N}
\right]
|s_1s_2\dots s_N\rangle .
\]
For open boundary conditions, the trace is replaced by boundary vectors,
\[
|\Psi_{\mathrm{AKLT}(\mathbf{v}_{\mathrm L},\mathbf{v}_{\mathrm R})}\rangle
=
\sum_{s_1,\ldots,s_N\in\{+,0,-\}}
\mathbf{v}_{\mathrm L}^{\dagger}
A^{s_1}A^{s_2}\cdots A^{s_N}
\mathbf{v}_{\mathrm R}
|s_1s_2\dots s_N\rangle.
\]
These additional degrees of freedom in the choice of the boundary vectors $\mathbf{v}_{\mathrm L}$ and $\mathbf{v}_{\mathrm R}$ are related to the edge states of the AKLT model in the presence of open boundary conditions. 

\printbibliography[heading=bibintoc]

\end{document}